\documentclass[12pt]{article}
\usepackage[utf8]{inputenc}
\usepackage[svgnames,dvipsnames]{xcolor}
\usepackage[a4paper,left=1.5cm,right=1.5cm,top=1.5cm,bottom=1.5cm]{geometry}
\usepackage[bottom, multiple]{footmisc}
\usepackage{booktabs, multirow, nicefrac, float, xspace}
\usepackage{graphicx}
\usepackage{tikz}
\usetikzlibrary{decorations.markings}
\usetikzlibrary{decorations.pathmorphing}
\usepackage{framed}
\usepackage{csquotes}
\definecolor{shadecolor}{rgb}{0.90,0.90,0.90}
\usepackage{cite}
\usepackage{hyperref}
\usepackage{subcaption}
\usepackage{pifont}
\usepackage{mathtools}
\usepackage{setspace}
\usepackage{amsmath, amssymb, amsthm, float, graphicx,amsfonts,braket}
\numberwithin{equation}{section}
\usepackage{amsthm}

\theoremstyle{definition}

\usepackage{float}
\allowdisplaybreaks

\hypersetup{colorlinks=true, linkcolor=Maroon, citecolor=FireBrick,urlcolor=Green,linktocpage}
\definecolor{amber(sae/ece)}{rgb}{1.0, 0.49, 0.0}
\definecolor{deepsaffron}{rgb}{1.0, 0.6, 0.2}

\usepackage{multirow}

\usepackage{amstext} 
\usepackage{array}   
\newcolumntype{L}{>{$}l<{$}}
\newcolumntype{C}{>{$}c<{$}}

\def\beq{\begin{eqnarray}}\def\eeq{\end{eqnarray}}
\def\be{\begin{equation}}\def\ee{\end{equation}}

\def\D{\Delta}

\usepackage{bm}
\usepackage{bm}

\begin{document}

\title{\bf Bootstrapping string models\\with entanglement minimization\\ and Machine-Learning}
\author{Faizan Bhat$^{a}$\footnote{faizanbhat@iisc.ac.in}, Debapriyo Chowdhury$^{a}$\footnote{debapriyoc@iisc.ac.in},\\ Arnab Priya Saha$^{a}$\footnote{arnabsaha@iisc.ac.in}~~and Aninda Sinha$^{a,b}$\footnote{asinha@iisc.ac.in}\\
\it ${^a}$Centre for High Energy Physics,
\it Indian Institute of Science,\\ \it C.V. Raman Avenue, Bangalore 560012, India.\\ \it ${^b}$Department of Physics and Astronomy, University of Calgary,\\ \it Alberta T2N 1N4, Canada.}

\maketitle

\abstract{We present a new approach to bootstrapping string-like theories by exploiting a local crossing symmetric dispersion relation and field redefinition ambiguities. This approach enables us to use mass-level truncation and to go beyond the dual resonance hypothesis. We consider both open and closed strings, focusing mainly on open tree-level amplitudes with integer-spaced spectrum, and two leading Wilson coefficients as inputs. Using entanglement minimization in the form of the minimum of the first finite moment of linear entropy or entangling power, we get an excellent approximation to the superstring amplitudes, including the leading and sub-leading Regge trajectories. We find other interesting S-matrices which do not obey the duality hypothesis, but exhibit a transition from Regge behaviour to power law behaviour in the high energy limit. Finally, we also examine Machine-Learning techniques to do bootstrap and discuss potential advantages over the present approach.} 

\newpage
\tableofcontents
\onehalfspacing

\section{Introduction}

{\it ``The garbage of the past often becomes the treasure of the present (and vice versa)."}\\---Alexander Polyakov.

\subsection{A historical perspective:}
Dual resonance models, describing pion-nucleon scattering, posit that the sum over $s$-channel poles should be reproduced by the sum over $t$-channel poles \cite{DHS}. This led to the birth of string theory, as explained in the beautiful introduction chapter by Green, Schwarz, and Witten \cite{Green:1987sp}. In the same chapter, however the authors are careful in admitting that\\
\begin{center}
{\it The duality hypothesis never had more than slender experimental support, and the Veneziano model was merely an ad hoc way of satisfying this not-so-well motivated hypothesis.}\\
\end{center}

They question: {\it Is duality\footnote{In modern literature, this is referred to as world-sheet duality to distinguish from other dualities in string theory. In this paper, when we mention duality, we always mean world-sheet duality.} an approximation or a principle?} By assuming the latter, we get the world-sheet picture of string theory, which appears to be an important ingredient in the ultraviolet finite properties of string theory, especially when reinterpreted as a consistent theory of quantum gravity. However, let us continue to examine the older literature. This will nicely set up much of the motivation behind the way we will re-examine string-like theories using the bootstrap. Our approach, while having some similarities with several recent papers \cite{Huang:2020nqy, Maity:2021obe, Arkani-Hamed:2022gsa, Cheung:2023adk,Cheung:2023uwn, Rigatos:2023asb, Haring:2023zwu,Eckner:2024ggx,Eckner:2024pqt,Berman:2024wyt,Cheung:2024uhn, Albert:2024yap, Chiang:2023quf,Huang:2024ihm}, differs substantially in spirit.

Dolen, Horn, and Schmid (see around eq.(20) in their seminal work \cite{DHS} as well as the recollection by Nambu in \cite{birth}) mention that the experimental evidence suggested that the correct prescription for the amplitude describing pion-nucleon scattering data should be:

\begin{equation}\label{DHS1}
M=M_{Regge}+M_{Res}-\langle M_{Res}\rangle
\end{equation}
where the first two terms are the sum over the $t$-channel and the $s$-channel respectively, as one would expect in usual quantum field theory. The last term is needed to make the amplitude not-too-big and avoid ``double-counting", while still allowing for the resonance structures--without the last term, the representation was historically called the ``interference model".  Dolen, Horn, Schmid observe that if the resonances overlap, then $M_{Res}\approx \langle M_{Res}\rangle$ and only then the duality hypothesis emerges as
\begin{equation}\label{DHS2}
M=M_{Regge}=M_{Res}\,,
\end{equation}
\textit{i.e.} the crossing symmetric amplitude in terms of a single channel emerges (the second equality follows from crossing symmetry). This indirectly implies that the high-energy behaviour is compatible with the unsubtracted fixed-$t$ dispersion relation.
However, we should emphasise that one cannot take eq.(\ref{DHS1}) literally as it is not entirely clear how to calculate the last term. Rather, we interpret eq.(\ref{DHS1}) as a phenomenological possibility of adding some extra piece that would add both channels, avoid double-counting while being able to explain experimental results. In light of our recent work \cite{Saha:2024qpt}, we will make eq.(\ref{DHS1}) precise.

 Although most of the recent attempts have focused on having duality ala eq.(\ref{DHS2}) in-built, it is worthwhile to expand the framework by allowing for theories which deviate from duality--in addition to the caveats in \cite{Green:1987sp}, see e.g. \cite{yellin} for an old review motivating this. In this paper, we take a step in this direction. A related motivation is the fact that glue-ball scattering in gauge theories exhibits power-law behavior in the high-energy limit and not the Regge behavior observed in 10-dimensional string theory. The beautiful paper by Polchinski and Strassler \cite{polchinskistrassler} uses AdS/CFT insights to argue that this may indeed happen, specifically hinting that there should be a cross-over from the Regge behavior to the power-law behavior. They consider glueball scattering and argue that at low values of negative $t$, one would have $s^{2+ct}$ for some $c>0$ while for higher negative $t$'s this would change to $s^2$. The bootstrap program must be able to investigate such behavior in amplitudes non-perturbatively (we will have only modest observations, which will hopefully spur more interest in this direction). Note that the hypergeometric deformation of the bosonic string amplitude in \cite{Cheung:2023adk, Haring:2023zwu} behaves as $s^t+c/s$; therefore, it exhibits a similar kind of transition, while respecting duality.

 Is there a representation of the amplitude that resembles eq.(\ref{DHS1})? String field theory suggests that the answer must be yes. Simply put, thinking about perturbative string theory in the framework of string field theory suggests that there must exist a representation of the amplitude that is exactly as in quantum field theory \cite{Polchinski:1998rq, Sen:2019jpm}. The phenomenological expectation of eq.(\ref{DHS1}) then can be reinterpreted by positing that the third term $\langle M_{Res}\rangle$ arises from the contact terms that arise in string field theory. We should point out here that a split in terms of channels and contact terms is subject to the usual field-redefinition ambiguities in quantum field theories. We will exploit this in our work.

\subsection{Recent developments}
 In \cite{Saha:2024qpt}, using a local crossing symmetric dispersion relation \cite{ak,Sinha:2020win,Gopakumar:2021dvg,bieb,Raman:2021pkf,Chowdhury:2021ynh,az1, az2, ak1, deRham:2022gfe,bissime,Song:2023quv, Bhat:2023ekh, bhat}, a field-theory representation of tree-level string theory amplitudes was found. The advantage of this representation is that it converges everywhere except at the poles, which is what is expected from string field theory arguments. An independent mathematical proof based on partial fractions for the representations in \cite{Saha:2024qpt} and its generalizations was found by the mathematician Hjalmar Rosengren in\footnote{In the same paper, it was appreciated that the series corresponds to generalizations of the hypergeometric series, which have not been studied in the mathematics literature.} \cite{rosengren}. In this paper, we will use fresh insights gained from the analysis in \cite{Saha:2024qpt} to set up the bootstrap of tree-level string theory in a new way. First, let us use the results of \cite{Saha:2024qpt} and correlate with eq.(\ref{DHS1}). 

 Using the local two-channel symmetric dispersion relation, we can obtain a one-parameter family of representations for tree-level open string amplitude, given by 
 \begin{equation}\label{op-st}
     \frac{\Gamma\left(-s_{1}\right)\Gamma\left(-s_{2}\right)}{\Gamma\left(1-s_{1}-s_{2}\right)}  =  \frac{1}{s_{1}s_{2}} + \sum_{n=1}^{\infty}\frac{1}{n!}\left(\frac{1}{s_{1}-n}+\frac{1}{s_{2}-n}+\frac{1}{\lambda+n}\right)\left(1-\lambda+\frac{\left(s_{1}+\lambda\right)\left(s_{2}+\lambda\right)}{\lambda+n}\right)_{n-1},
 \end{equation}
 where $\lambda$ is a free parameter and $\text{Re}(\lambda)>-1$ for convergence of the series; while $\lambda$ can be complex, we will just focus on real values in this paper. An important feature of this series representation is that not only does the sum converge for any values of $s_{1}$ and $s_{2}$ away from the poles, but each of the three pieces on the rhs converges on its own. This is reminiscent of the proposal by DHS in eq.(\ref{DHS1}). To understand how this is still compatible with eq.(\ref{DHS2}), we need to work a bit harder. Let us call the term containing $\frac{1}{s_{1}-n}$ inside the sum in \eqref{op-st} as $s_{1}$-channel sum, similarly the term with $\frac{1}{s_{2}-n}$ as $s_{2}$-channel sum and the remaining term with $\frac{1}{\lambda+n}$ as the contact sum. It can be shown that each of these sums is individually convergent.
 \begin{table}[H]
     \centering
     \begin{tabular}{|c|c|c|c|c|c|}
     \hline
          $s_{1}$& $s_{2}$ & $s_{1}$\text{-channel sum}& $s_{2}$\text{-channel sum} & \text{contact sum} & $\widehat{\mathcal{M}}\left(s_{1},s_{2}\right)$\\
          \hline
          5.5 & -7.2 & 7.4306 & -0.0088 & 0.02351 & 7.4454\\
          5.5 & 4.7 & 2.4908 & -85.2407 & 20.346 & -62.404\\
          -5.5 & -3.6 & -0.1013 & -0.147 & 0.1982 & -0.0501\\
          \hline
     \end{tabular}
     \caption{Convergence of individual sum. Here, sum is performed with $1000$ terms and $\lambda=2.5$. The last column gives the actual value of $\widehat{\mathcal{M}}\left(s_{1},s_{2}\right)= \frac{\Gamma\left(-s_{1}\right)\left(-s_{2}\right)}{\Gamma\left(1-s_{1}-s_{2}\right)}-\frac{1}{s_{1}s_{2}}$. The values obtained from the numerical sum match the actual values in the given decimal places.}
     \label{tab:ch-conv}
 \end{table}

Table \ref{tab:ch-conv} also indicates the ambiguity in splitting into channels. What we have identified as the $s_1$-channel in table 1 also includes contact terms, which are vital for the convergence of the sum. A different decomposition is discussed in the Appendix (\ref{indiv-ch-conv}). That the Veneziano amplitude respects both (\ref{DHS1}) and (\ref{DHS2}) serves as encouragement to use similar ideas in numerics. An immediate advantage over the fixed-$t$ dispersion relation is that we will be able to impose constraints on a wider domain, since our representation converges everywhere except at the poles. We will next summarize the salient features of the new approach we propose in this paper.

 \subsection{A summary of our approach}
 Let us consider tree-level, four-point, color-ordered scattering amplitude of massless external states, belonging to the Yang-Mills supermultiplet, in open string theory \cite{Green:2019tpt},
 \begin{equation}
     \mathcal{M}_{\text{op}}\left(s_{1},s_{2}\right) = g^2 P_{4}\frac{\Gamma\left(-s_{1}\right)\Gamma\left(-s_{2}\right)}{\Gamma\left(1-s_{1}-s_{2}\right)}.
 \end{equation}
 Various projections of the prefactor $P_{4}$ correspond to different component fields contained in the SYM multiplet. For example, for the four-gluon amplitude, $P_{4}$ becomes $\mathcal{F}^{4}$. $\frac{\Gamma\left(-s_{1}\right)\Gamma\left(-s_{2}\right)}{\Gamma\left(1-s_{1}-s_{2}\right)}$ is the universal part of the amplitude, and we refer to it as the form factor. This form factor can be thought of as four-point amplitude of identical scalars. 
 
In this work, we are interested in studying the analogue of the form factor---by abuse of terminology we henceforth call the form factor as the amplitude---in putative theories. The key assumptions in our analysis are the following:
 \begin{enumerate}
     \item The amplitude is crossing symmetric in $s_{1}$ and $s_{2}$,
		\begin{equation}
			\mathcal{M}\left(s_{1},s_{2}\right) = \mathcal{M}\left(s_{2},s_{1}\right).
		\end{equation}
  At tree-level, $\mathcal{M}\left(s_{1},s_{2}\right)$ is analytic for all values of $s_{1}$ and $s_{2}$, except at the locations of physical singularities. Furthermore, since we are interested in tree-level amplitudes, we will consider $\mathcal{M}\left(s_{1},s_{2}\right)$ to be a meromorphic function with simple poles in $s_{1}$ and $s_{2}$. The last consideration follows from locality.

  \item We consider an equidistant spectrum with poles at integer values of $s_{1}$ and $s_{2}$, \textit{i.e.} $m_{n}^{2}=n$, for $n=\mathbb{Z}^{+}$. At any given mass-level only a finite number of particles are exchanged. This rules out any accumulation points in the spectrum. This assumption is not compatible with the Coon amplitude \cite{COON1969669, Maldacena:2022ckr}, which has infinitely many states below a particular energy level. Although one could accommodate this amplitude by considering delta functions in the absorptive part having support at $q$-deformed integer values, we will not look into it here. 

  \item We assume spins of the exchanged particles at any given mass-level are bounded from above. In the case of open string amplitude, at the $n$-th mass level, spins are $0\le \ell\le n$, with the condition that only even (odd) spins contribute when $n$ is odd (even). We use this as an input to find a class of amplitudes; we study amplitudes relaxing the even-odd constraint in the appendix. This assumption can also be relaxed to include all spins from $0$ to $n$ contribute as is the case for the Veneziano amplitude describing tachyon scattering. 

  \item The residue of the amplitude at a pole in a particular channel is given by a polynomial in the other channel. Unitarity at tree level implies that the polynomial can be expanded in terms of Gegenbauer polynomials with positive coefficients. This is because at the pole the four-point amplitude factorizes to two three-point amplitudes, and the four-point coupling constant is the three-point coupling squared. The positivity of the coefficients follows from the fact that four-point couplings should be real. Therefore, the absorptive part of the amplitudes in the $s_{1}$-channel can be given by\footnote{$\mathcal{A}^{(s_{1})}$ is the discontinuity of the amplitude in $s_{1}$ channel and is given by 
\begin{equation}
	\mathcal{A}^{(s_{1})}\left(s_{1},s_{2}\right) = \frac{1}{2i}\lim_{\epsilon\rightarrow 0}\left[\mathcal{M}\left(s_{1}+i\epsilon,s_{2}\right)-\mathcal{M}\left(s_{1}-i\epsilon,s_{2}\right)\right].
\end{equation}
Overall negative sign in \eqref{gb-exp} appears because 
\begin{equation}
    \lim_{\epsilon\rightarrow0}\frac{1}{x\pm i\epsilon} = \text{Principle value}\left(\frac{1}{x}\right)\mp i\pi\delta\left(x\right).
\end{equation} }
  \begin{equation}\label{gb-exp}
	\mathcal{A}^{(s_{1})}\left(s_{1}, s_{2}\right) = -\pi\sum_{n=1}^{\infty}\sum_{\substack{\ell=0\\\ell\in\text{spectrun}}}^{n}c^{(n)}_{\ell}\mathcal{G}^{\left(\frac{D-3}{2}\right)}_{\ell}\left(1+\frac{2s_{2}}{s_{1}}\right)\delta\left(s_{1}-n\right),
\end{equation}
    with $c^{(n)}_{\ell}\ge 0$. Here $\mathcal{G}^{\left(\frac{D-3}{2}\right)}_{\ell}\left(z\right)$ is the Gegenbauer polynomial in the $D$ dimension and $c_{\ell}^{(n)}$ are the partial-wave coefficients (we will mainly be interested in $D=10$). While writing \eqref{gb-exp} we have only considered poles corresponding to massive exchanges. To justify \eqref{gb-exp} we can first choose to work with amplitudes with the pole at $s_{1}=0$, $s_{2}=0$ removed and then add the massless pole separately\footnote{As an example consider an amplitude, $\mathcal{M}\left(s_{1},s_{2}\right)= \frac{1}{s_{1}s_{2}}+\widehat{\mathcal{M}}\left(s_{1},s_{2}\right)$. The left side of \eqref{gb-exp} then corresponds to $\widehat{\mathcal{M}}\left(s_{1},s_{2}\right)$. The reason for distinguishing between massless and massive poles is that residues over massless poles are not always polynomial in nature and hence cannot be expanded in terms of a finite number of Gegenbauer polynomials.}.  

    \item Rather than the locality constraints or the equivalent null constraints, we will use the independence of the amplitude from the parameter in the representation. This independence is a consequence of the field redefinition ambiguity as discussed in \cite{Saha:2024qpt}.  The resulting constraints are given in eq.(\ref{lambdaconstraints}). These constraints are non-trivially related to the locality/null constraints. This is because the difference between the local crossing symmetric dispersion relation and the nonlocal one are the locality/null constraints \cite{Sinha:2020win, Song:2023quv} and the parametric dependence arises due to the freedom to add a linear combination of these constraints to the representation \cite{Sinha:2020win}. For reasons we explain in the appendix, setting up the constraints in this manner provides us with certain numerical advantages.

 \end{enumerate}

 With the given ansatz in \eqref{gb-exp}, we find a one-parameter family of local, crossing symmetric representations of amplitudes given by \eqref{gen-disp-1}. The motivation to look for such a parametric representation comes from \eqref{op-st}. Details of the derivation of this expression are present in section \ref{sec:lambda-amp-deriv}. Our goal is to obtain the partial wave coefficients $c_{\ell}^{(n)}$ using numerical analysis subject to certain extremization constraints. We will carry out a similar analysis for the closed-string in section \ref{sec:clst}. The immediate advantages of this new representation for the bootstrap are obvious: (a) With truncation, the representation is expected to capture all the features of the amplitude \cite{Saha:2024qpt}---this is not possible with a truncated fixed-$t$ representation, which only converges when $t<0$\,. (b) One does not have to worry about analytic continuation since the representation is expected to converge everywhere except at the poles. 

 \subsection{What quantity to minimize?}
 What principles do we use to select interesting theories using bootstrap? In all investigations carried out so far, one has tried to bound ratios of Wilson coefficients. This is a perfectly sensible approach, since presumably the first few Wilson coefficients are precisely measurable in experiments and can be considered as inputs. Is there any other way to zoom into the space of interesting theories? There is yet no sharp answer to this question \cite{bose1,bose2,bell}. Investigations \cite{PhysRevLett.125.181602, Cheung:2023hkq, Beane:2021zvo, Aoude:2024xpx, Kowalska:2024kbs, Low:2024mrk}  related to entanglement minimization have suggested the possibility that this could be a physics principle to use. The rationale behind this is that generating entanglement is resource intensive (for sure entangling photons in the laboratory needs non-linear crystals and it is hard to make two photons interact!) so it is possible that natural processes will try to lower complexity by using minimum entanglement. We will discuss this further in section 3. To use this for bootstrap, we need to carefully specify the important parameters that we keep fixed. There is an implicit coupling constant that sits in front of the tree-level amplitude. We will assume that it is non-zero. We will also fix two normalized low-energy Wilson coefficients (to be specified below) to the superstring values. At this stage there is a lot of freedom left, and the theory that minimizes entanglement does not necessarily have to be close to string theory. However, crossing symmetry and locality, together with a mild assumption of high-energy behavior, appear to pick out an S-matrix, which is very close to superstring theory! For other Wilson coefficients, we get a wide variety of theories. We will examine one in particular, which exhibits the Polchinski-Strassler-type transition.

 \subsection{Help from Machine Learning}
 The numerical approach described above only involves constraints that are linear in the unknowns $c_{\ell}^{(n)}$s. The constraints are the independence of the amplitude from the field redefinition parameter $\lambda$ and need to be linear in the unknowns. However, we can only impose the constraints up to some tolerance as we truncate in the mass-levels. The choice of this tolerance is somewhat ad hoc as we do not a priori have knowledge of how well the level-truncation works. As we increase the number of mass-levels, we should be able to reduce the tolerance. However, implementing this is a painful exercise. Instead, what is desirable is to demand that the ratio of the amplitudes for two different values of $\lambda$ is close to unity. However, since we do not know the sign of the amplitude at different $s_1, s_2$-values (barring some general statements in specific domains), we cannot convert this into a linear programming problem. 

 We examine Machine Learning techniques to get a handle on this problem. Our analysis of ML techniques is not exhaustive; rather, our hope is that our preliminary attempts will spur further research in this direction. Indeed, the problem that we analyse is one of the simplest one to investigate from the ML and bootstrap points of view. 

 We organize our paper as follows. In section \ref{sec:lambda-amp-deriv}, we review the local crossing-symmetric representations and the usefulness of the new parametric representations in a general set-up for 2-2 scattering of identical particles, having 2-channel symmetry. In section \ref{sec:EP}, we turn to linear entropy or entangling power in perturbation theory, keeping in mind gluon scattering in open superstring theory. In section \ref{sec:bootstrap}, we set up the bootstrap problem and discuss our findings. Section \ref{sec:clst} briefly examines the closed string bootstrap, leaving a more detailed study for future work. In section \ref{sec:ML}, we set up a Machine Learning approach and explain the advantages of considering such an approach. In section \ref{sec:parametricLCSDR}, we give a unified dispersion relation whose various limits go over to known dispersion relations. We conclude in section \ref{sec:conclusion}.

\section{Crossing-symmetric parametric representations}  
\label{sec:lambda-amp-deriv}
Let us consider a tree-level amplitude $\mathcal{M}\left(s_{1},s_{2}\right)$ that has only simple poles in the $s_{1}$ and $s_{2}$ variables. We start with the local two-channel symmetric dispersion relation obtained in \cite{Saha:2024qpt},
\begin{equation}\label{localDisp}
	\mathcal{M}\left(s_{1}, s_{2}\right) = \mathcal{M}\left(0,0\right)+ \frac{1}{\pi}\int_{s_{0}}^{\infty}\mathrm{d}\sigma\left[\frac{1}{\sigma-s_{1}}+\frac{1}{\sigma-s_{2}}-\frac{1}{\sigma}\right]\mathcal{A}^{(s_{1})}\left(\sigma,\frac{y}{\sigma}\right) - \frac{1}{\pi}\int_{s_{0}}^{\infty}\frac{\mathrm{d}\sigma}{\sigma}\mathcal{A}^{(s_{1})}\left(\sigma,0\right),
\end{equation}
where $s_{0}$ is the location of pole at lowest massive exchange.  
Here we have defined 
\begin{equation}
    x = s_{1} + s_{2}, \qquad y = s_{1}s_{2}.
\end{equation}
While writing \eqref{localDisp} we implicitly assume that there are no massless poles and that $\mathcal{M}\left(0,0\right)$ is finite. If there is pole at $s_{1}=0$, $s_{2}=0$ then we can separate the singular term and replace 
\begin{equation}
    \mathcal{M}\left(0,0\right) \rightarrow \text{Massless pole} + W_{00}\:.
\end{equation}
$W_{00}$ is finite. Note that the first and second integrals in \eqref{localDisp} cancel each other out at $s_{1}=0, s_{2}=0$. 
As an example, let us examine how the dispersion relation works for the amplitude $1/s_1 s_2$. Here $\mathcal{M}(0,0)$ blows up so let us shift the variables by $\alpha$ (which can be thought of as a regulator in this example). Then using the dispersion relation (assuming $\alpha>s_0$) we have:
\begin{eqnarray}
\frac{1}{(s_1-\alpha)(s_2-\alpha)}&=&\frac{1}{\alpha^2}-\frac{1}{\pi}\int_{s_0}^\infty \mathrm{d}\sigma \left[\frac{1}{\sigma-s_{1}}+\frac{1}{\sigma-s_{2}}-\frac{1}{\sigma}\right]\pi \frac{\delta(\sigma-\alpha)}{\frac{s_1 s_2}{\sigma}-\alpha}-\frac{1}{\pi}\int_{s_0}^\infty \frac{\mathrm{d}\sigma}{\sigma}\frac{\pi \delta(\sigma-\alpha)}{\alpha}\,,\nonumber\\
&=& -\left[\frac{1}{\alpha-s_1}+\frac{1}{\alpha-s_2}-\frac{1}{\alpha}\right]\frac{\alpha}{s_1 s_2-\alpha^2}\,,\\
&=&\frac{1}{(s_1-\alpha)(s_2-\alpha)}\,,\nonumber
\end{eqnarray}
Using shifted variables in the arguments of the amplitude we can obtain a one-parameter family of dispersive representation from \eqref{localDisp}, 
\begin{equation}
\label{Gen_Sing_LCSDR}
\begin{split}
    \mathcal{M}(s_1,s_2)=&\:\mathcal{M}(-\lambda,-\lambda) - \frac{1}{\pi} \int_{s_0}^{\infty} \mathrm{d}\sigma\frac{\mathcal{A}^{(s_{1})}\left(\sigma,- \lambda \right)}{\sigma + \lambda} \\
    &+\frac{1}{\pi} \int_{s_0}^{\infty} \mathrm{d}\sigma  \left[\frac{1}{\sigma -s_1}+\frac{1}{\sigma - s_2}- \frac{1}{\sigma + \lambda}\right] \mathcal{A}^{(s_{1})}\left(\sigma,\frac{(s_1+\lambda)(s_2 + \lambda)}{\sigma + \lambda} - \lambda \right)
\end{split}
\end{equation}
A derivation of the above representation is given in (\ref{sec:parametricLCSDR}).
\\
Let us consider the partial wave expansion of the amplitude in the $s_{1}$-channel, which is given by
\begin{equation}
	\mathcal{M}\left(s_{1}, s_{2}\right) = \sum_{\ell}f_{\ell}\left(s_{1}\right)\mathcal{G}_{\ell}^{\left(\frac{D-3}{2}\right)}\left(1+\frac{2s_{2}}{s_{1}}\right).
\end{equation}
Note that $s_{1}$-channel discontinuity of the amplitude is given by 
\begin{equation}
	\mathcal{A}^{(s_{1})}\left(s_{1}, s_{2}\right) = \frac{1}{2i}\left[\mathcal{M}\left(s_{1}+i\epsilon, s_{2}\right)- \mathcal{M}\left(s_{1}-i\epsilon,s_{2}\right)\right],
\end{equation}
which leads us to 
\begin{equation}\label{gen-disc}
	\mathcal{A}^{(s_{1})}\left(s_{1},s_{2}\right) = \sum_{\ell}a_{\ell}\left(s_{1}\right)\mathcal{G}_{\ell}^{\left(\frac{D-3}{2}\right)}\left(1+\frac{2s_{2}}{s_{1}}\right).
\end{equation}
Here we have defined 
\begin{equation}
	a_{\ell}\left(s\right) := \frac{1}{2i}\left[f_{\ell}\left(s+i\epsilon\right)-f_{\ell}\left(s-i\epsilon\right)\right].
\end{equation}
Now plugging \eqref{gen-disc} in the parametric representation of \eqref{Gen_Sing_LCSDR}, we obtain
\begin{eqnarray}
	&& \mathcal{M}\left(s_{1}, s_{2}\right)\nonumber\\
	& = & \mathcal{M}\left(-\lambda,-\lambda\right) \nonumber\\
	&& -\frac{1}{\pi}\int_{s_{0}}^{\infty}\mathrm{d}\sigma\left[\frac{1}{s_{1}-\sigma}+\frac{1}{s_{2}-\sigma}+\frac{1}{\sigma+\lambda}\right]\sum_{\ell}a_{\ell}\left(\sigma\right)\mathcal{G}^{\left(\frac{D-3}{2}\right)}_{\ell}\left[1+\frac{2}{\sigma}\left(\frac{\left(s_{1}+\lambda\right)\left(s_{2}+\lambda\right)}{\sigma+\lambda}-\lambda\right)\right]\nonumber\\
	&& -\frac{1}{\pi}\int_{s_{0}}^{\infty}\frac{\mathrm{d}\sigma}{\sigma+\lambda}\sum_{\ell}a_{\ell}\left(\sigma\right)\mathcal{G}_{\ell}^{\left(\frac{D-3}{2}\right)}\left(1-\frac{2\lambda}{\sigma}\right). 
\end{eqnarray}
In case of tree-level amplitude which contains only poles, we can write
\begin{equation}\label{tree-disc}
	a^{(0)}_{\ell}\left(s\right) = -\pi\sum_{n=1}^{\infty}c^{(n)}_{\ell}\delta\left(s-n\right).
\end{equation}
$c_{\ell}^{(n)}\ge 0$ due to unitarity. $\ell$ runs over spins of the exchanged states. We specify the spectrum of the states and this acts as an input in our set up. 
\begin{itemize}\item For open superstring, at mass-level $n$ we have  $0\le \ell\le n-1$ with $n-\ell \equiv 1, \mod 2$. \item In case of the bosonic string and its deformations, the condition is $0\le\ell\le n$ at mass-level $n$. 
\end{itemize}
Finally we obtain the following parametric representation for the tree-level amplitude
\begin{eqnarray}\label{gen-disp-1}
	&&\mathcal{M}\left(s_{1}, s_{2}\right) \nonumber\\
 & = & P\left(\frac{1}{s_1},\frac{1}{s_2}\right) + W_{00}\left(\lambda\right)\nonumber\\
	&& + \sum_{n=1}^{\infty}\sum_{\ell}\left[\frac{1}{s_{1}-n}+ \frac{1}{s_{2}-n}+ \frac{1}{\lambda+n}\right] c_{\ell}^{(n)}\mathcal{G}_{\ell}^{\left(\frac{D-3}{2}\right)}\left[1+\frac{2}{n}\left(\frac{\left(s_{1}+\lambda\right)\left(s_{2}+\lambda\right)}{\lambda+n}-\lambda\right)\right] \nonumber\\
	&& + \sum_{n=1}^{\infty}\sum_{\ell}\frac{1}{\lambda+n}c_{\ell}^{(n)}\mathcal{G}_{\ell}^{\left(\frac{D-3}{2}\right)}\left(1-\frac{2\lambda}{n}\right),
\end{eqnarray}
where, $W_{00}\left(\lambda\right)=\mathcal{M}\left(-\lambda,-\lambda\right) - P\left(-\frac{1}{\lambda},-\frac{1}{\lambda}\right)$. $P\left(\frac{1}{s_1},\frac{1}{s_2}\right)$ is the expression for massless pole. $W_{00}\left(\lambda\right)$ depends on the details of the specific amplitude being considered. As an example, we take $\mathcal{M}(s_1,s_2) = \left(\frac{1}{s_{1}-1} + \frac{1}{s_{2}-1}\right)$. In this case, we have $W_{00}\left(\lambda\right)=-\frac{2}{\lambda+1}$ and the only non-zero partial-wave coefficient, $c_{0}^{(1)}=1$. There is no massless pole for this amplitude.  As examples for amplitudes with massless pole, $W_{00}\left(\lambda\right)$ for the following two cases are:
\begin{align}
	&\text{Open superstring}: \: \mathcal{M}(s_1,s_2)=\frac{\Gamma\left(-s_{1}\right)\Gamma\left(-s_{2}\right)}{\Gamma\left(1-s_{1}-s_{2}\right)} & \Rightarrow&  & W_{00}\left(\lambda\right) =& \frac{\Gamma\left(\lambda\right)^{2}}{\Gamma\left(1+2\lambda\right)}-\frac{1}{\lambda^{2}},\nonumber\\
	&\text{Bosonic string}: \: \mathcal{M}(s_1,s_2)=\frac{\Gamma\left(-s_{1}\right)\Gamma\left(-s_{2}\right)}{\Gamma\left(-s_{1}-s_{2}\right)} & \Rightarrow&  & W_{00}\left(\lambda\right)  =& \frac{\Gamma\left(\lambda\right)^{2}}{\Gamma\left(2\lambda\right)} - \frac{2}{\lambda}.
\end{align}

The goal of our work is to solve for the partial wave coefficients, $c_{\ell}^{(n)}$ using physical constraints. $W_{00}\left(\lambda\right)$ can be determined by fixing Wilson coefficients that are part of the data used as inputs. Since both sides of \eqref{gen-disp-1} are independent of $\lambda$, we can use constraint $\frac{\partial^k}{\partial\lambda^k}\mathcal{M}\left(s_{1},s_{2}\right)=0$, for any positive integer values of $k$ as constraints in our analysis. 

Henceforth, we will be interested in gluon scattering, keeping open superstring theory in mind, and will use 
\begin{equation}
    P\left(\frac{1}{s_1},\frac{1}{s_2}\right)=\frac{1}{s_1 s_2}\,.
\end{equation}

\subsection{Locality/Null constraints}

A two-channel symmetric dispersion relation was obtained in \cite{Raman:2021pkf},
\begin{equation}\label{non-localDisp}
	\mathcal{M}\left(s_{1}, s_{2}\right) = \mathcal{M}\left(0,0\right) + \frac{1}{\pi}\int_{a}^{\infty}\frac{\mathrm{d}\sigma}{\sigma}\left[\frac{x\sigma-2y}{\sigma^{2}-x\sigma+y}\right]\mathcal{A}\left(\sigma, \frac{a\sigma}{\sigma-a}\right).
\end{equation}
Here $a=y/x$. The above dispersion relation contains non-local terms, which eventually drop out after performing an infinite sum. 

By following a similar analysis as before, we can obtain a one-parameter family of representations from \eqref{non-localDisp} for two-channel symmetric amplitudes,
\begin{eqnarray}\label{gen-nonlocalDisp}
	\mathcal{M}\left(s_{1},s_{2}\right) & = & P\left(\frac{1}{s_1},\frac{1}{s_2}\right) + W_{00}\left(\lambda\right) \nonumber\\
	&& + \sum_{n=1}^{\infty}\sum_{\ell}\left[\frac{1}{s_{1}-n} + \frac{1}{s_{2}-n} + \frac{2}{\lambda+n}\right]\nonumber\\
	&& \hspace{0.5cm}
	c_{\ell}^{(n)}\mathcal{G}_{\ell}^{\left(\frac{D-3}{2}\right)}\left[1+\frac{2}{n}\Biggl\{\frac{\left(s_{1}+\lambda\right)\left(s_{2}+\lambda\right)}{s_{1}+s_{2}+2\lambda}\frac{\lambda+n}{\lambda+n-\frac{\left(s_{1}+\lambda\right)\left(s_{2}+\lambda\right)}{s_{1}+s_{2}+2\lambda}}-\lambda\Biggr\}\right].
\end{eqnarray}
The difference between \eqref{gen-nonlocalDisp} and \eqref{gen-disp-1} gives null constraints at different values of $\left(s_{1}, s_{2}\right)$. We have checked that the open string amplitude indeed obeys the resulting constraints. {We could have used these locality/null constraints to set up our bootstrap analysis, but employing $\frac{\partial^{k}}{\partial\lambda^{k}}\mathcal{M}\left(s_{1}, s_{2}\right)=0$ proves to be more efficient in our work.}

\subsection{Advantage of using $\lambda$-parametrization}
Here we briefly discuss why we prefer using $\lambda$-parametrization than the naive situation where $\lambda=0$. Let us examine the open-string solution for this. For \eqref{gen-nonlocalDisp} to converge, we will need \cite{Saha:2024qpt}
\begin{equation}\label{CSDRconvOS}
	\text{Re}\left(\frac{s_{1} s_{2}-\lambda ^{2} }{s_{1}+s_{2}+2
		\lambda}\right) <2\,,
\end{equation}
with $\text{Re}\left(\lambda\right)>-1$. Thus, one can impose the locality/null constraints only when this condition holds. 

\begin{figure}[h]
    \centering
    \subfloat[\centering Allowed region for $\lambda=0.01$.]{{\includegraphics[width=5cm]{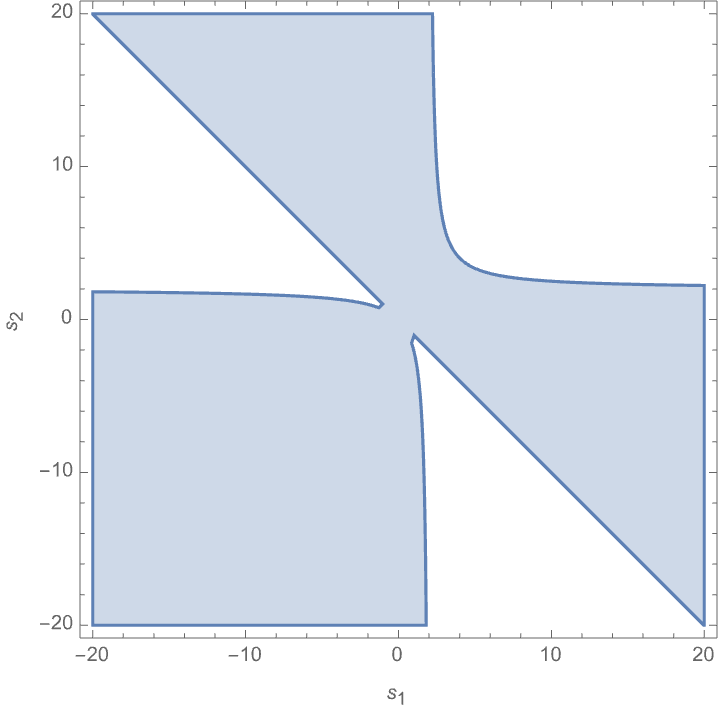} }}%
    \qquad
    \subfloat[\centering Allowed region for $\lambda=14.6$.]{{\includegraphics[width=5cm]{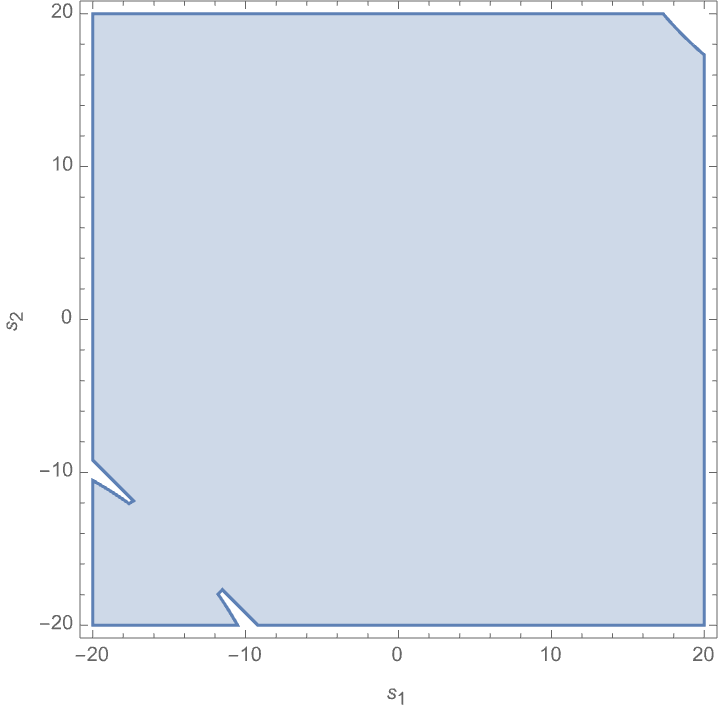} }}%
    \caption{Plot showing that for non-zero $\lambda$, a larger $s_1,s_2$ region opens up where the open-string locality/null constraints hold. }%
   \label{w11w02}
\end{figure}

As the plot in Fig.(\ref{CSDRconvOS}) shows, by exploiting non-zero $\lambda$, a larger region in the $s_1, s_2$ plane opens up where one can impose the constraints. With the fixed-$t$ representation, the constraints come from equating the s-channel to the $t$-channel---this can only be done in the 3rd quadrant. Thus, one can expect the numerical bootstrap to exhibit faster convergence with non-zero $\lambda$. This also motivates us to use the independence of $\lambda$ instead of locality constraints, since clearly the choice of $\lambda$ should not affect the outcome, and it is prudent to impose this independence as constraints.

\section{Entanglement minimization}
\label{sec:EP}
Now we have to specify what physical quantity we extremise to get bootstrap solutions. For this, we turn to quantifying the entanglement generated in scattering of gluons. See \cite{Aoude:2024xpx, bose1, bell, PhysRevLett.125.181602, bose2, Cheung:2023hkq, Beane:2021zvo,Kowalska:2024kbs, Low:2024mrk} for recent developments. 
Given two subsystems, $\mathcal{H}_{A}$ and $\mathcal{H}_{B}$, we can define linearized entropy as
\begin{equation}
    \mathcal{E}\left[\Omega\right] = 1-\text{tr}_{A}\left[\rho^{2}_{A}\right].
\end{equation}
Here, $\rho_{AB}$ is the density matrix constructed from the state $|\Omega\rangle = \sum_{a,b}\Omega_{ab}|p_{1},a\rangle_{A}\otimes|p_{2},b\rangle_{B}$ and $\rho_{A}=\text{tr}_{B}\left[\rho_{AB}\right]$. In \cite{Aoude:2024xpx} linearized entropy has been studied in the context of perturbative scattering. Let us consider $2\rightarrow 2$ scattering, where $|p_{1},a\rangle$ and $|p_{2},b\rangle$ are the corresponding asymptotic scattering states. These states are labeled by momenta, $p_{i}$ of the particles and internal quantum numbers, if any, denoted by $a, b$. We are interested in the difference in linearized entropy between final and initial states in scattering, which is given by \cite{Aoude:2024xpx}
\begin{equation}
    \Delta\mathcal{E}\left[\Omega\right] = 4\mathcal{N}\text{Im}\left[\sum_{ij}\Omega_{ij}\mathcal{M}_{ij}^{ab}\left(p_{1},p_{2}\rightarrow p_{1},p_{2}\right)\left(\Omega\cdot\Omega^{\dagger}\cdot\Omega\right)_{ab}\right].
\end{equation}
Here $\mathcal{M}_{ij}^{ab}$ is the S-matrix element with initial states denoted by $i,j$ and final states are $a,b$.
In case $\text{tr}\left[\left(\Omega^{\dagger}\cdot\Omega\right)^{2}\right] = 1$, the initial linearized entropy vanishes and this implies that $|\Omega^{i}\rangle$ can be written as a product state. Then the above equation reduces to 
\begin{equation}\label{EP}
    \Delta\mathcal{E}\left[\Omega^{\text{prod}}\right] = 4\mathcal{N}\text{Im}\mathcal{M}_{\alpha\beta}^{\alpha\beta}\left(p_{1},p_{2}\rightarrow p_{1}, p_{2}\right).
\end{equation}
Here, the normalization is $\mathcal{N} = \frac{T}{4E_{\Vec{p}_{1}}E_{\Vec{p}_{2}}V}$, with $T=\delta\left(0\right)$ coming from the energy conservation and $V=\delta^{(3)}\left(\Vec{0}\right)$ coming from the 3-momenta conservation. 
It is to be noted that in the above expression the scattering amplitude is in the forward limit in kinematic space and elastic in the internal quantum numbers. The entangling power is obtained by integrating over all initial state configurations\footnote{In \cite{PhysRevLett.125.181602, bose1, bose2} entangling power is defined as $\mathcal{E}=1-\frac{1}{16\pi^{2}}\int\mathrm{d}\Omega_{A}\mathrm{d}\Omega_{B}\text{tr}_{A}\rho_{A}^{2}$. Here, $\Omega$ represents the phase space. In the perturbative set up we are working at leading order in coupling constant where the amplitude in the forward limit appears. The difference in linearized entropy given in \eqref{EP} is sign semidefinite, so the phase-space integration does not make any qualitative difference in our analysis}. In a slight abuse of nomenclature, we will continue using the terminology ``entangling power" when referring to $\Delta\mathcal{E}$. The entangling power is related to the absorptive part in the forward limit, which is proportional to the total forward scattering cross-section. In weakly coupled theories, this can be expected to be small \cite{mandelstam}. We can examine the S-matrices in theoretical space that would minimize this. Alternatively, the heuristic considerations in the Introduction about minimizing the complexity associated with entangling gates in a circuit picture give us another motivation to examine the same question.

Consider the four-point color-ordered scattering amplitude of gluons, $\mathcal{M}\left(1^{h_{1}},2^{h_{2}},3^{h_{3}},4^{h_{4}}\right)$. $h_{i}$ is the helicity of $i$th particle. To apply \eqref{EP}, it will be convenient to take linear combinations of helicity amplitudes and work with $\mathcal{M}_{\varepsilon_{1}\varepsilon_{2}}^{\varepsilon_{4}\varepsilon_{3}}\left(p_{1},p_{2},p_{3},p_{4}\right)$. We follow the convention where all outgoing momenta are positive, such that 
\begin{equation}
	s = -\left(p_{1}+p_{2}\right)^{2}, \qquad t=-\left(p_{1}+p_{4}\right)^{2}, \qquad u=-\left(p_{1}+p_{3}\right)^{2}.
\end{equation}
We want to calculate the entanglement between two gluons in a scattering process. In open superstring theory color ordered four-gluon scattering amplitude is given by 
\begin{equation}
    \mathcal{M}_{4}\left(1234\right) = g^2\mathcal{F}^{4}\frac{\Gamma\left(-s\right)\Gamma\left(-t\right)}{\Gamma\left(1-s-t\right)},
\end{equation}
where the prefactor is given by (trace over gauge indices is implied)
\begin{eqnarray}
	\mathcal{F}^{4} & = & F_{1,\mu\nu}F_{2}^{\mu\nu}F_{3,\alpha\beta}F_{4}^{\alpha\beta} + F_{1,\mu\nu}F_{3}^{\mu\nu}F_{4,\alpha\beta}F_{2}^{\alpha\beta} + F_{1,\mu\nu}F_{4}^{\mu\nu}F_{2,\alpha\beta}F_{3}^{\alpha\beta} \nonumber\\
	&& -4\biggl\{F_{1\mu\nu}F_{2}^{\nu\alpha}F_{3,\alpha\beta}F_{4}^{\beta\mu} + F_{1\mu\nu}F_{3}^{\nu\alpha}F_{4,\alpha\beta}F_{2}^{\beta\mu} + F_{1\mu\nu}F_{4}^{\nu\alpha}F_{2,\alpha\beta}F_{3}^{\beta\mu}\biggr\},
\end{eqnarray}
with $F_{i,\mu\nu} = p_{i,\mu}\varepsilon_{i,\nu}-p_{i,\nu}\varepsilon_{i,\mu}$ for the $i$th particle. 

In the forward limit, we can choose a frame such that the momenta of the incoming and outgoing particles lie along the $z$ axis and the polarization vectors are in the $x\;y$ plane. Therefore, $\varepsilon_{i}\cdot p_{j} =0$, for all $i,j$. In this limit, the prefactor reduces to
\begin{equation}
	\lim_{t\rightarrow 0}\mathcal{F}^{4} = -2s^{2}\varepsilon_{1}\cdot\varepsilon_{4}\; \varepsilon_{2}\cdot\varepsilon_{3}.
\end{equation}
If we had considered instead a tensor structure of the form 
\begin{equation}
	\widetilde{\mathcal{F}}^{4} = c_{1}\biggl\{F_{1,\mu\nu}F_{2}^{\mu\nu}F_{3,\alpha\beta}F_{4}^{\alpha\beta} + \text{cyclic}\left(2\rightarrow3\rightarrow4\right)\biggr\} - c_{2}\biggl\{F_{1\mu\nu}F_{2}^{\nu\alpha}F_{3,\alpha\beta}F_{4}^{\beta\mu}+  \text{cyclic}\left(2\rightarrow3\rightarrow4\right)\biggr\},
\end{equation}
then the leading order difference in linear entropy would be
\begin{equation}
	\Delta\mathcal{E}\left[\Omega\right] = -4\mathcal{N}s^{2}\biggl\{\left(c_{1}-\frac{c_{2}}{4}\right)\left(\varepsilon_{1}\cdot\varepsilon_{2}\;\varepsilon_{1}^{\ast}\cdot\varepsilon_{2}^{\ast} + \varepsilon_{1}\cdot\varepsilon_{2}^{\ast}\;\varepsilon_{2}\cdot\varepsilon_{1}^{\ast}\right) -\frac{c_{2}}{2}\varepsilon_{1}\cdot\varepsilon_{1}^{\ast}\; \varepsilon_{2}\cdot\varepsilon_{2}^{\ast}\biggr\}\sum_{n=1}^{\infty}\text{Res}_{s=n}M\left(s,0\right)\delta\left(s-n\right)\,,
\end{equation}
and so for the superstring case the first term vanishes---we will focus on the superstring case.
\normalcolor
To implement \eqref{EP} in bootstrap, we retain the tensor structure of four-gluon scattering and consider an arbitrary form factor, $M\left(s,t\right)$, with the assumption that $M\left(s,t\right)$ has simple poles at positive integer values of $s$ and $t$. Then we obtain from the massive poles,
\begin{equation}
    \Delta\mathcal{E}_{massive} = 8\pi\mathcal{N}s^{2}\varepsilon_{1}\cdot\varepsilon_{4}\; \varepsilon_{2}\cdot\varepsilon_{3}\sum_{n=1}^{\infty}\text{Res}_{s=n}M\left(s,0\right)\delta\left(s-n\right).
\end{equation}
The massless pole contributes a term proportional to $s/t\sim s/\epsilon$ as $t\rightarrow \epsilon$ and is the same for all theories. Thus, this will not affect our results.
Since the right-hand side of the above equation is non-negative, our goal is to minimize this expression. However, note that for tree-level amplitudes, we have a sum over delta functions, and hence to make sense of this expression, we will need to do an averaging over $s$. We will be interested in cases where $|M(s,0)|\sim o(s^0)$ and hence \begin{equation}\label{minEp}\int_1^\Lambda ds\, \frac{\Delta\mathcal{E}}{s^2}\end{equation} 
will be finite. This is also the first finite moment of the entangling power. Alternatively, we can also think of this expression as an average w.r.t. $1/s$. This is the quantity that we will minimize. We can also consider $\int_1^\Lambda ds\, \frac{\Delta\mathcal{E}}{s^n}$ for some real $n$ with $n>1$. We will discuss this case in the appendix.

\section{Bootstrap}
\label{sec:bootstrap}
We will use \eqref{gen-disp-1} to set up the bootstrap analysis. We remind the reader that while the unsubtracted fixed-$t$ dispersion relation holds for theories for which the amplitude at fixed negative $s_2$ and large $s_1$ behaves as\footnote{It is possible to consider higher subtracted dispersion relations such that the amplitude behaves as $o(s_1^n)$ for $n>0$, but we will refrain from considering these theories.} $O(s_1^{-\epsilon})$ with $\epsilon>0$, \eqref{gen-disp-1} holds more generally, even for positive $s_2$. 

In the low-energy expansion around $s_1=0, \; s_2=0$, let us denote the constant part of the amplitude as $W_{00}$. This is obtained by subtracting the massless pole and setting $s_1=s_2=0$ in \eqref{gen-disp-1}. 
Specifically, the low energy expansion takes the form:
\begin{eqnarray}
\mathcal{M}_{low}\left(s_{1},s_{2}\right) &= &g^2 
\left(\frac{1}{s_1 s_2}+W_{00}+W_{10}(s_1+s_2)+W_{01} s_1 s_2 +\cdots\right) \nonumber\\
& =& {g^{2}\left(\frac{1}{s_{1}s_{2}} +  \sum_{m,n =0}^{\infty}W_{mn}x^{m}y^{n}\right)}\,,
\end{eqnarray}
where $x=s_1+s_2, y= s_1 s_2$.
In what follows, we will absorb the $g^2$ factor inside the Wilson coefficients so that the massless pole is $1/(s_1 s_2)$. To emphasize again, unlike most attempts in the literature, we will not bound ratios of Wilson coefficients. Rather, we will minimize the linear entropy/entangling power as in \eqref{minEp} with string coupling,  $W_{10}$ and $W_{01}$ held fixed.

We can find $W_{00}\left(\lambda\right)$ in \eqref{gen-disp-1} by fixing the Wilson coefficient, $W_{00}$ from the following relation,
\begin{eqnarray}\label{W00lambda}
	W_{00} & = & W_{00}\left(\lambda\right) + \sum_{n=1}^{\infty}\sum_{\ell}\left[-\frac{2}{n}+\frac{1}{\lambda+n}\right]c_{\ell}^{(n)}\mathcal{G}_{\ell}^{\left(\frac{D-3}{2}\right)}\left(\frac{n-\lambda}{n+\lambda}\right) \nonumber\\
	&& + \sum_{n=1}^{\infty}\sum_{\ell}\frac{1}{\lambda+n}c_{\ell}^{(n)}\mathcal{G}_{\ell}^{\left(\frac{D-3}{2}\right)}\left(1-\frac{2\lambda}{n}\right).
\end{eqnarray}
This gives the representation:
\begin{eqnarray}\label{ansatzfull}
	&&\mathcal{M}\left(s_{1}, s_{2}\right) \nonumber\\
 & = & P\left(\frac{1}{s_1},\frac{1}{s_2}\right)+W_{00} + \sum_{n=1}^{\infty}\sum_{\ell}c_{\ell}^{(n)}\Biggl\{ \left(\frac{2}{n}-\frac{1}{\lambda+n}\right)\mathcal{G}_{\ell}^{\left(\frac{D-3}{2}\right)}\left(\frac{n-\lambda}{n+\lambda}\right)\nonumber\\
	&& + \left[\frac{1}{s_{1}-n}+ \frac{1}{s_{2}-n}+ \frac{1}{\lambda+n}\right] \mathcal{G}_{\ell}^{\left(\frac{D-3}{2}\right)}\left[1+\frac{2}{n}\left(\frac{\left(s_{1}+\lambda\right)\left(s_{2}+\lambda\right)}{\lambda+n}-\lambda\right)\right]\Biggr\}. 
\end{eqnarray}
\eqref{ansatzfull} will play a crucial role in our analysis. As we explain below, $W_{00}$ is related to entangling power.

We wish to minimize $\int_{s_0}^\infty \mathrm{d}\sigma {\mathcal A}^{(s_{1})}(\sigma,0)/\sigma$ as this is the quantity appearing in the first finite moment of the entangling power\footnote{Higher finite moments would allow for $o(s_1)$ behaviour; we will deal with this in the appendix.}. Explicitly, we have (the $Im$ part is $\lambda$-independent):

\begin{equation}\label{epowerexp}
\int_{1}^\Lambda ds\, \frac{\Delta\mathcal{E}}{s^2}=\sum_{n=1}^{N_{max}<\Lambda}\sum_{\ell}\frac{1}{n}c_{\ell}^{(n)}\mathcal{G}_{\ell}^{\left(\frac{D-3}{2}\right)}\left(1\right)\,.
\end{equation}

To make sense, this must be finite (which can be checked in the numerics by increasing the number of modes) as $\Lambda\rightarrow \infty$. Hence, we need to figure out when this can happen. In other words, what high-energy behavior is compatible with the finiteness of this moment of the entangling power.
Let us first note the unsubtracted single channel (fixed-$s_2$) dispersion relation for an amplitude given by 
\begin{equation}\label{t-fix}
	\mathcal{M}\left(s_{1},s_{2}\right) = P\left(\frac{1}{s_1},\frac{1}{s_2}\right)+\frac{1}{\pi}\int_{s_{0}}^{\infty}\frac{\mathrm{d}\sigma}{\sigma-s_{1}}\mathcal{A}\left(\sigma,s_{2}\right) + \mathcal{B}(s_2),
\end{equation}
where $\mathcal{B}\left(s_{2}\right)$ is the boundary contribution coming from the arc at infinity.

There are two interesting cases which for brevity we will refer to as Category I and II:
\begin{itemize}
\item {\bf Category I}: The boundary term $\mathcal{B}(s_2)$ vanishes if the amplitude decays at large $s_{1}$, \textit{i.e.,} $\lim_{s_{1}\rightarrow \infty}\mathcal{M}\left(s_{1},s_{2}\right) \sim O(s_{1}^{-\epsilon})$, for $\epsilon>0$. This condition can hold in certain domain of $s_{2}$, mostly for $s_{2}<0$.

 Therefore, in this case, we conclude that if $\frac{1}{\pi}\int_{s_{0}}^{\infty}\frac{\mathrm{d}\sigma}{\sigma-s_{1}}\mathcal{A}\left(\sigma,s_{2}\right)$ is finite, then by Cauchy's residue theorem it must be equal to the amplitude with the massless pole subtracted from it, \textit{i.e.,} $\mathcal{M}\left(s_{1},s_{2}\right)- P\left(\frac{1}{s_1},\frac{1}{s_2}\right)$. Further, in this case, since the unsubtracted fixed-$s_2$ dispersion relation holds, we expect these theories to respect the dual resonance condition.

\item {\bf Category II}: The amplitude goes to a constant for large $|s_1|$ and fixed $s_2$. Naively, it would appear that the integral in \eqref{t-fix} would diverge, but this does not always happen as we will explain below. In this case the boundary term would contribute a constant (since this is at fixed-$s_2$, the constant can depend on $s_2$) and the absorptive integral could still be finite, for instance, if the absorptive part went to zero faster than a constant. One could say that in this case $\mathcal{M}(s_1,s_2)-\mathcal{B}(s_2)$ has a single channel representation. Such theories would be beyond the usual duality condition.  A simple example of this category of theories, where for fixed-$s_{2}$ the boundary term goes to a constant, is as follows. Consider a deformation of the string amplitude by a single scalar at a fixed mode number. Using \eqref{ansatzfull}, for a scalar, we find that the contribution is $\lambda$-independent and contributes $c_0^{(n)}[2/n+1/(s_1-n)+1/(s_2-n)]$. For  $s_1\rightarrow \infty$, this gives the $s_2$-dependent contribution $c_0^{(n)}[2/n+1/(s_2-n)]$. Further, since the absorptive part for this additional scalar has delta function support, it would contribute as a constant to the absorptive integral. The question is if the entanglement minimisation procedure picks up such theories. The answer is yes! Note that one can adjust the unfixed $W_{00}$ in \eqref{ansatzfull} such that $\mathcal{B}(s_2=0)=0$, to yield S-matrices which satisfy the dual resonance condition.
\item {\bf Category III}: For completeness, we will also consider models where the \eqref{epowerexp} does not have a convergent {\it rhs}. In this case, we will need higher (negative) moments of the entangling power. This will enable us to locate models which can in principle behave as $o(s_1)$ for $s_{2}<0$, which is the maximum behaviour captured by \eqref{ansatzfull}. We will not consider higher subtracted dispersion relations, which would allow for $o(s_1^n)$, with $n>1$. 
\end{itemize}

Let us consider the first possibility. The last term in \eqref{localDisp} is $\frac{1}{\pi}\int_{s_{0}}^{\infty}\frac{\mathrm{d}\sigma}{\sigma}\mathcal{A}^{(s_{1})}\left(\sigma,0\right)$, which by the above argument is the amplitude evaluated at $s_{1}=0$, $s_{2}=0$ and therefore is equal to $W_{00}$. This together with \eqref{localDisp} implies that the amplitude can be simplified to 
\begin{equation}\label{intamp}
    \mathcal{M}\left(s_{1},s_{2}\right) = P\left(\frac{1}{s_1},\frac{1}{s_2}\right)-\frac{1}{\pi}\int_{s_{0}}^{\infty}\mathrm{d}\sigma\left[\frac{1}{s_{1}-\sigma}+\frac{1}{s_{2}-\sigma}+\frac{1}{\sigma}\right]\mathcal{A}^{(s_{1})}\left(\sigma,\frac{y}{\sigma}\right).
\end{equation}
This gives rise to the representation:
\begin{eqnarray}\label{ansatz}
	\mathcal{M}\left(s_{1}, s_{2}\right)
 & = & P\left(\frac{1}{s_1},\frac{1}{s_2}\right)  \nonumber\\&&+\sum_{n=1}^{\infty}\sum_{\ell}c_{\ell}^{(n)}\left[\frac{1}{s_{1}-n}+ \frac{1}{s_{2}-n}+ \frac{1}{\lambda+n}\right] \mathcal{G}_{\ell}^{\left(\frac{D-3}{2}\right)}\left[1+\frac{2}{n}\left(\frac{\left(s_{1}+\lambda\right)\left(s_{2}+\lambda\right)}{\lambda+n}-\lambda\right)\right]. \nonumber\\
\end{eqnarray}

It is clear that for the representation to work, we must have $\lambda>-1$.
The amplitudes arising from eq.(\ref{intamp}) have the feature that they coincide with the fixed-$t$ single-channel representation in the forward limit $\left(s_{2}\rightarrow 0\right)$. In this case $W_{00}$ using the fixed-$t$ form is:
\begin{equation}\label{W00}
	W_{00} = -\sum_{n=1}^{\infty}\sum_{\ell}\frac{1}{n}c_{\ell}^{(n)}\mathcal{G}_{\ell}^{\left(\frac{D-3}{2}\right)}\left(1\right).
\end{equation}
Equating this to the $W_{00}$ arising from \eqref{ansatz}, we can define a duality measure:
\begin{equation} \label{dualcheck}
    D_M\equiv\sum_{n=1}^{\infty}\sum_{\ell}c_{\ell}^{(n)}\Biggl\{\frac{1}{n}\mathcal{G}_{\ell}^{\left(\frac{D-3}{2}\right)}\left(1\right)+ \left(-\frac{2}{n}+\frac{1}{\lambda+n}\right)\mathcal{G}_{\ell}^{\left(\frac{D-3}{2}\right)}\left(\frac{n-\lambda}{n+\lambda}\right)\Biggr\} = 0.
\end{equation}

This is obviously true for $\lambda=0$.
Note that for this case, \eqref{epowerexp} is just $-W_{00}$ (truncated); therefore, minimizing the entangling power is maximizing $W_{00}$.
For the situation where the amplitude goes to a constant at large $s_{1}$ for some fixed $s_{2}$, this condition and, conversely, the dual resonance condition will not hold. Here, $W_{00}$ and \eqref{epowerexp} are not related. In the following, we also write down the formulae for $W_{10}$ and $W_{01}$ which we fix for the bootstrap. These are given by:
\begin{eqnarray}
W_{10}&=&-\sum_{n=1}^{\infty}\sum_{\ell}\frac{2c_\ell^{(n)}}{n^2}\left\{ \frac{1}{2}\mathcal{G}_{\ell}^{\left(\frac{D-3}{2}\right)}\left(\frac{n-\lambda}{n+\lambda}\right)+(D-3)\frac{\lambda(n+2\lambda)}{(n+\lambda)^2}\mathcal{G}_{\ell-1}^{\left(\frac{D-1}{2}\right)}\left(\frac{n-\lambda}{n+\lambda}\right)\right\}\nonumber \\
W_{01}&=&\sum_{n=1}^{\infty}\sum_{\ell}\frac{2c_\ell^{(n)}}{n^2}\left\{ \frac{1}{n}\mathcal{G}_{\ell}^{\left(\frac{D-3}{2}\right)}\left(\frac{n-\lambda}{n+\lambda}\right)-(D-3)\frac{(n+2\lambda)}{(n+\lambda)^2}\mathcal{G}_{\ell-1}^{\left(\frac{D-1}{2}\right)}\left(\frac{n-\lambda}{n+\lambda}\right)\right\}\,.\label{W10W01}
\end{eqnarray}
The $\mathcal{G}_{\ell-1}$ terms contribute from $\ell=1$ onward. Using \eqref{W10W01}, we notice that
\begin{equation}
	W_{10}-\lambda\, W_{01}=-\sum_{n=1}^{\infty}\sum_{\ell}\frac{c_\ell^{(n)}}{n^2}\left\{ \left(1+\frac{2\lambda}{n}\right)\mathcal{G}_{\ell}^{\left(\frac{D-3}{2}\right)}\left(\frac{n-\lambda}{n+\lambda}\right)\right\}.
\end{equation}
Now for $-1<\lambda\leq 0$, the Gegenbauers are positive and $\left(1+\frac{2\lambda}{n}\right)\geq 0$ if $-\frac{1}{2}\leq \lambda \leq 0$. The upper limit $\lambda\rightarrow 0$ gives the bound $W_{10}\leq 0$ and the lower limit $\lambda \rightarrow -\frac{1}{2}$ makes the \textit{rhs} negative and gives the rigorous upper bound
\begin{equation}\label{W01ub}
	W_{01}\leq -2W_{10} \,.
\end{equation}
This bound is the same as \cite{Raman:2021pkf}, obtained using different means. 

For numerical bootstrap, we will use \eqref{ansatz} or \eqref{ansatzfull}. Since the constraints involve derivatives w.r.t. $\lambda$, from \eqref{ansatzfull}, $W_{00}$ will remain unfixed by the numerics, which will fix $c_\ell^{(n)}$. One may have wondered how restrictive are \eqref{W10W01}. To perform numerics we always deal with a truncated series instead of keeping all terms up to $n\rightarrow\infty$. Whenever we truncate a series, say $n=N_{\text{max}}$, it starts depending on $\lambda$. Now suppose that numerically we wanted $W_{10}$ and $W_{01}$ to be $\lambda$ independent, would we get any interesting theories? The answer is no. This is easy to see, as the contribution $\ell=0$ is always independent of $\lambda$, hence if we only imposed the $\lambda$-independence of \eqref{W10W01}, we would only get terms with $\ell=0$ turned on. The best we can hope to do with level truncation is to impose $\lambda$-independence up to a tolerance.

Thus, we can proceed in two ways:

\begin{enumerate}
\item We start with \eqref{ansatz}, find solutions satisfying the bootstrap conditions (to be elaborated below) including \eqref{epowerexp}  and then verify \eqref{dualcheck}, in which case $W_{00}$ will be given by \eqref{W00}.  If this holds up to some tolerance, we can say that the resulting amplitudes satisfy the dual resonance condition. This will locate models in {\bf category-I}.
\item We start with \eqref{ansatzfull}, find solutions satisfying the bootstrap conditions (to be elaborated below) including \eqref{epowerexp}  and look for solutions which do not obey \eqref{dualcheck}. By construction, such amplitudes which generically do not obey \eqref{dualcheck}, if they exist, should go to a constant for fixed $s_2$ and large $s_1$. Here $W_{00}$ is an unfixed parameter. This will locate models in {\bf category-II}.
\item We start with \eqref{ansatzfull}, find solutions satisfying the bootstrap conditions and minimising a higher moment of the entangling power. We will relegate a study of such {\bf category-III} models to the appendix.
\end{enumerate}

To locate string-type models, we could have just proceeded with the line-of-attack in point 2 but we will find that convergence is faster to locate dual resonance models, if we proceed along point 1. We will be able to study the daughter Regge trajectories in this way. 

\begin{figure}[H]
\centering
    \includegraphics[scale=0.4]{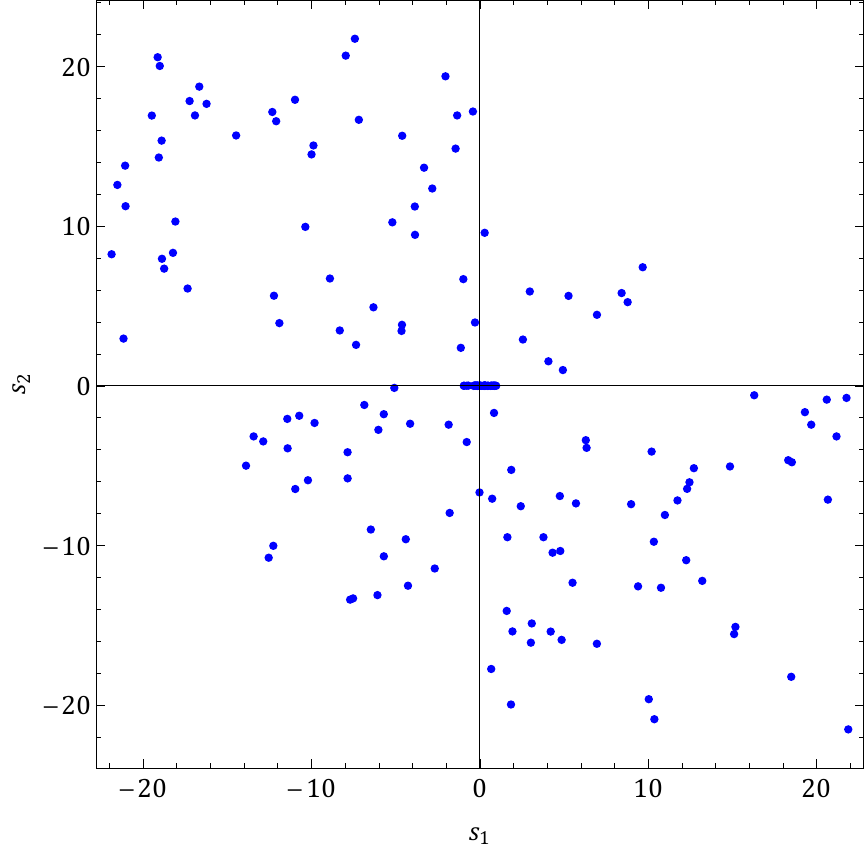}
    \caption{An example plot showing the grid of points in the $(s_1,s_2)$ plane where the field redefinition constraints are imposed. The constraints in the 1st quadrant do not appear to play a major role in our numerics.}
    \label{s1s2grid}
\end{figure}

For the bootstrap, we truncate the level in the $n$-sum in ($\ref{ansatz}$) upto $N_{\text{max}}$ and then we impose the tree-level unitarity constraints
\begin{equation}
c_{\ell}^{(n)}\geq 0\,,
\end{equation}
and the field redefinition invariance constraints:
\begin{equation}\label{lambdaconstraints}
    \frac{\partial^k\mathcal{M}\left(s_{1}, s_{2}\right)}{\partial\lambda^k}=0\, \text{ for }k=1,2,\cdots,k_{\text{max}}.
\end{equation}
Henceforth, we shall call these constraints $\lambda$-constraints. We want to impose these constraints up to some tolerance on a grid of points in the $s_1,s_2$ plane. 
We use \texttt{FindInstance} in Mathematica to make this grid, avoiding positive integer values of $s_1, s_2$ to avoid poles of the open-string amplitude with the help of the following two conditions: $\lfloor s_1\rfloor-s_1\neq 0$ and $\lfloor s_2\rfloor-s_2\neq 0$. Since there are poles when $s_1, s_2>0$, we expect the convergence rates to be different in the different quadrants in the $s_{1}, s_{2}$ plane. 
Let us fix the number of modes to be $N_{\text{max}}=30$. We will choose\footnote{Since $f(x+h)=f(x)+h f'(x)+\cdots$, the appropriate quantity to consider would have been $f'(x)/f(x)$ but we cannot handle inequalities on ratios of this sort in what we are doing. This affects how far we can go in the 1st quadrant (where the amplitude becomes large) and the 3rd quadrant (where the amplitude becomes small). }  $|s_i|<10$ in the 1st quadrant with 10 points,  $|s_i|<22$ in the 2nd and 4th quadrant with 50 points each and $|s_i|<14$ in the 3rd quadrant with 30 points. We also use a sprinkle of 40 points as additional constraints in the region $-1<s_1<1$ and $-0.01<s_2<0.01$. The points are chosen using the \texttt{RandomSeeding -> Automatic} command for $s_i\in$ \texttt{Reals} and the bootstrap is rerun for different random grids, so chosen. The reason for the restriction in the 3rd quadrant is that convergence slows down as $|s_1|\sim |s_2|$ approach $N_{\text{max}}$ \cite{Saha:2024qpt}. 
We impose
\begin{equation}\label{lambdaconstraintsN}
    \left|\frac{\partial^k\mathcal{M}\left(s_{1}, s_{2}\right)}{\partial\lambda^k}\right|\leq T,
\end{equation}
where $T$ is a suitable tolerance for $k=1$ to $k_{\text{max}}$. We use SDPB \cite{sdpb} to find an accurate solution (although Mathematica could have sufficed, SDPB gives better control on accuracy).  A plot with a random grid of points where these constraints are imposed is shown in fig.(\ref{s1s2grid}).  We use $N_{\text{max}}=30$, $k_{\text{max}}=6$,  $T=10^{-9}$. Our seed $\lambda=14.6$. We checked the independence of $\lambda$ in the solution so obtained. We perform the bootstrap in $D=10$. We could also have considered imposing the $\lambda$-independence on a combination of different $\lambda$ values. The results are similar and the choice of the approach relies on the available resources.

\subsection{Category-I theories: Dual resonance models}

\subsubsection{Negativity from crossing}
\begin{figure}[H]
    \centering
    \begin{subfigure}{0.45\textwidth}
        \includegraphics[width=7.5cm]{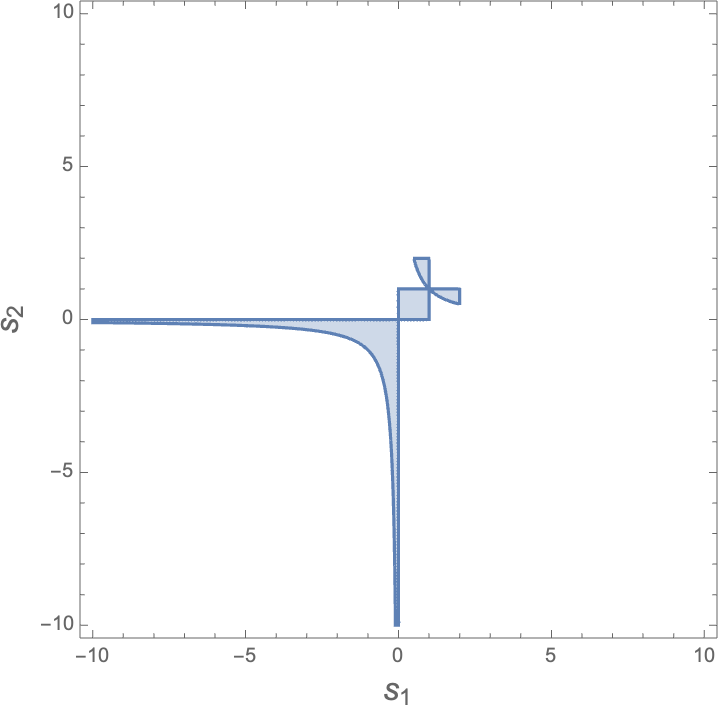}
        \caption{Negativity domain ($\lambda=0$). ``Grumpy Longhands" (tilt your head to the right!)}
        \label{lambda0}
    \end{subfigure}
    \qquad
    \begin{subfigure}{0.45\textwidth}
        \includegraphics[width=7.5cm]{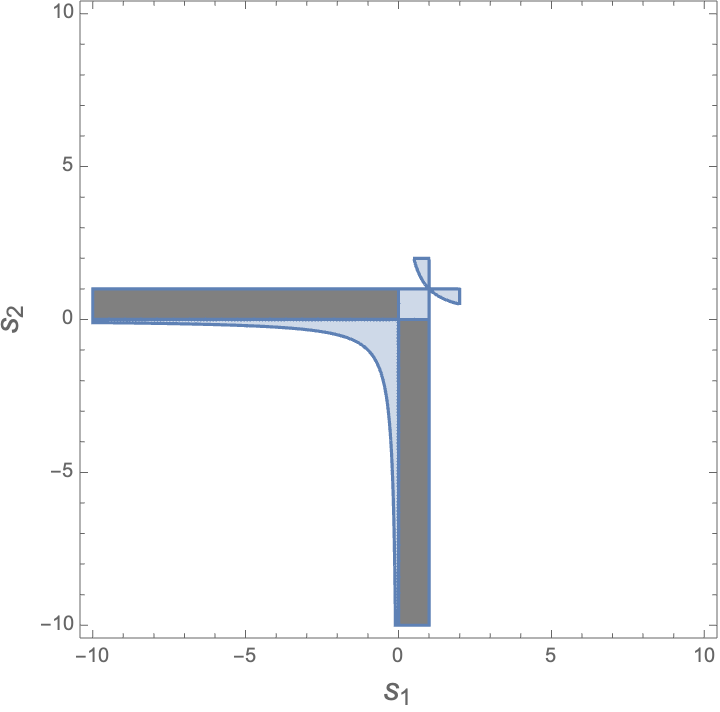}
        \caption{Extended domain obtained by dialing $-1<\lambda<0$.}
        \label{lambdan0}
    \end{subfigure}
    \caption{Negativity domains.}%
   \label{negdom}
\end{figure}

A nontrivial consequence of the representation \eqref{ansatz} is that it implies that $\mathcal{M}-1/(s_1 s_2)$ is negative in the $s_1, s_2$ region shown in the fig.(\ref{negdom}). To obtain this, we first set $\lambda=0$. This is the region where $1/(s_1-n)+1/(s_2-n)+1/n$ is negative for all $n\geq 1$ and $s_1 s_2>0$ so that the Gegenbauer argument is greater than 1. This region is given by 
\begin{equation}
    \bigl\{0<s_{1}s_{2}<1\,\cap \,s_{1}s_{2}>s_{1}+s_{2}-1\bigr\}\cup\bigl\{1<s_{1}s_{2}<s_{1}+s_{2}-1 \,\cap \, 4>s_{1}s_{2}>2\left(s_{1}+s_{2}\right)-4\bigr\}.
\end{equation}
This would imply that each term in the non-zero mode sum in \eqref{ansatz} is negative. The domain thus obtained is shown in fig.(\ref{lambda0}). Extra negativity is obtained by dialing $\lambda$. First, note that the argument of the Gegenbauer, for some large mode-number will become less than 1 if $\lambda>0$. Hence, to find extra negativity domains, we restrict ourselves to $-1<\lambda<0$---the lower limit is set by the convergence of the open string amplitude. By dialing $-1<\lambda<0$, we obtain the extra wings, indicated by gray in fig.(\ref{lambdan0}). This illustrates an advantage of the parametric representation. We have checked that the hypergeometric deformations discussed in the appendix obey this negativity. It is not possible to see these negativity conditions starting with the fixed-$t$ dispersive representation.

\subsubsection{Numerics}

\begin{figure}[H]
    \centering
    \subfloat[\centering $s_2=-5.1$]{{\includegraphics[width=8cm]{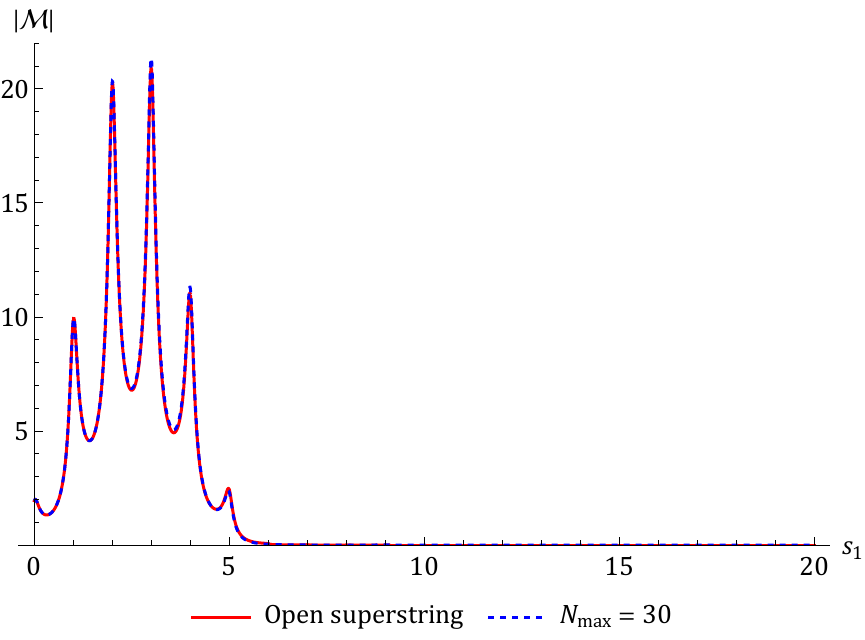} }}%
    \qquad
    \subfloat[\centering $s_2=-1.1$]{{\includegraphics[width=8cm]{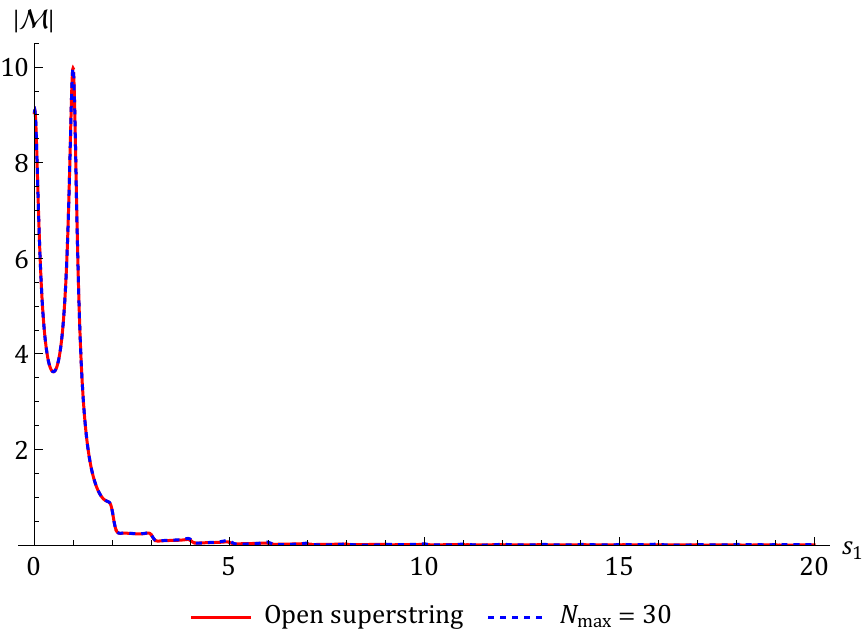} }}%
    \qquad
    \subfloat[\centering $s_2=-0.1$]{{\includegraphics[width=8cm]{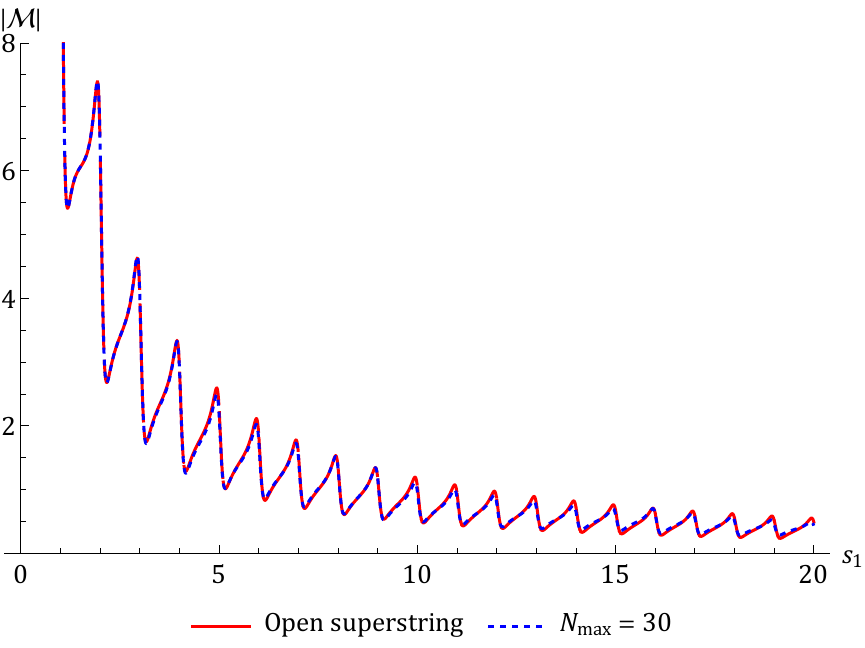} }}%
    \qquad
    \subfloat[\centering $s_2=0.5$]{{\includegraphics[width=8cm]{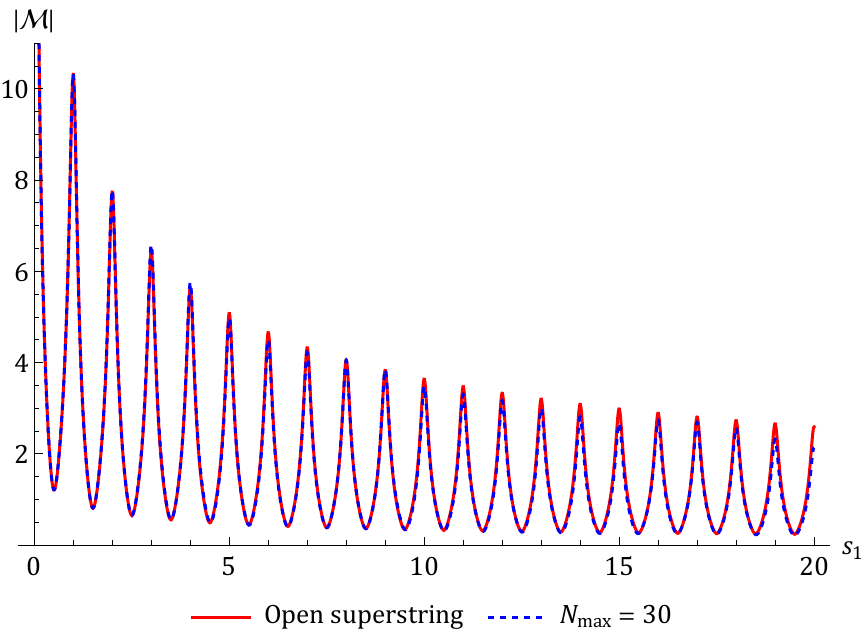} }}%
    \qquad
    \subfloat[\centering $s_2=1.1$]{{\includegraphics[width=8cm]{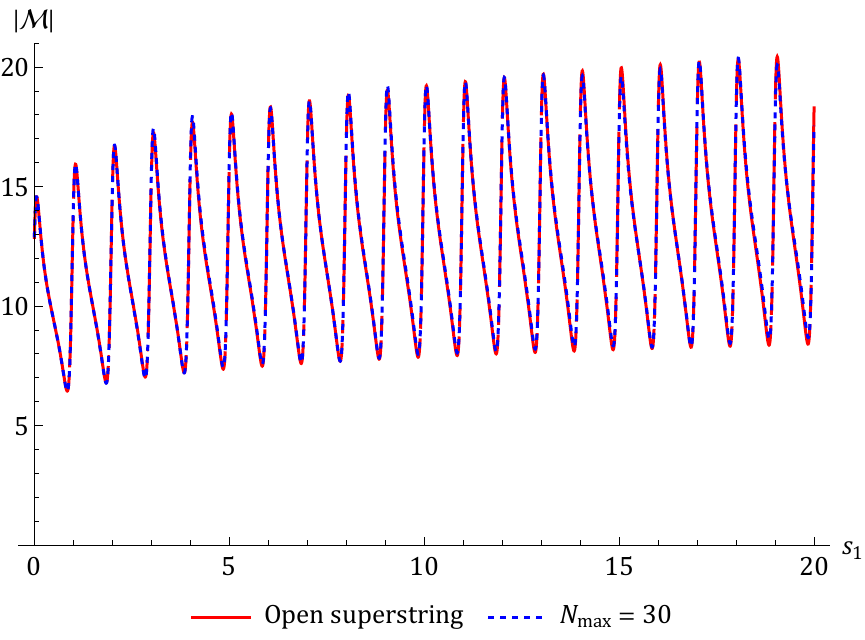} }}%
    \qquad
    \subfloat[\centering $s_2=3.2$]{{\includegraphics[width=8cm]{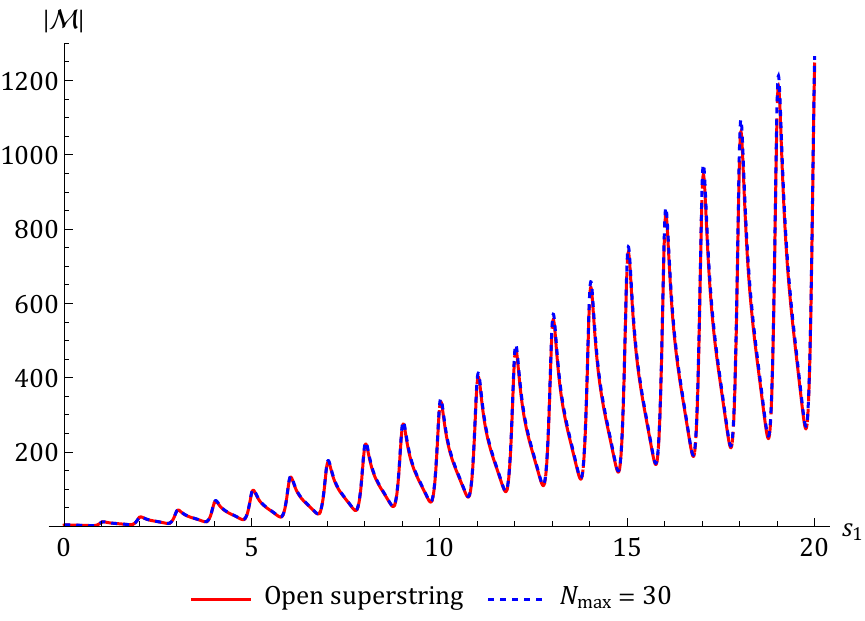} }}%
    \caption{Bootstrapping tree-level open superstring theory. We plot $\big|\mathcal{M}\left(s_1+\frac{i}{10},s_2\right)\big|$ versus $s_1$ for different values of $s_2$. The agreement with the exact answer is excellent.}%
   \label{OpenString}
\end{figure}

\begin{figure}[H]
    \centering
    \subfloat[\centering Leading Regge trajectory]{{\includegraphics[width=7cm]{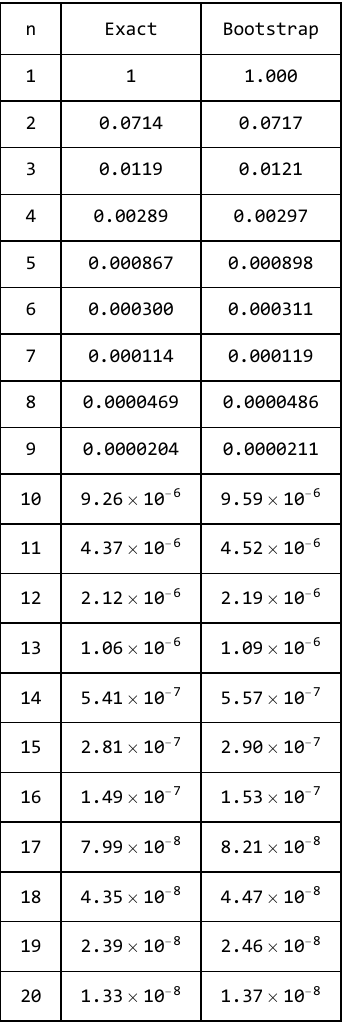} }}%
    \qquad
    \subfloat[\centering Subleading Regge trajectory]{{\includegraphics[width=7cm]{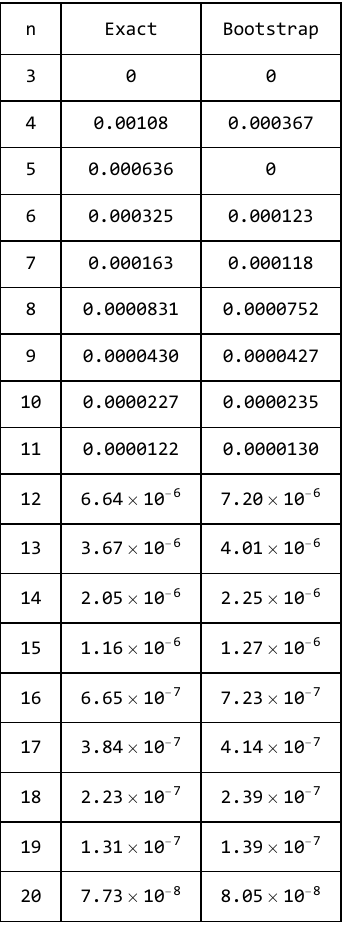} }}%
    \caption{Values of $c_\ell^{(n)}$s on the leading and the subleading Regge trajectories for the open superstring amplitude from the exact expression compared with the corresponding values from the bootstrap.}%
   \label{LeadingSubleading}
\end{figure}

\begin{figure}[H]
    \centering
    \includegraphics[scale=0.8]{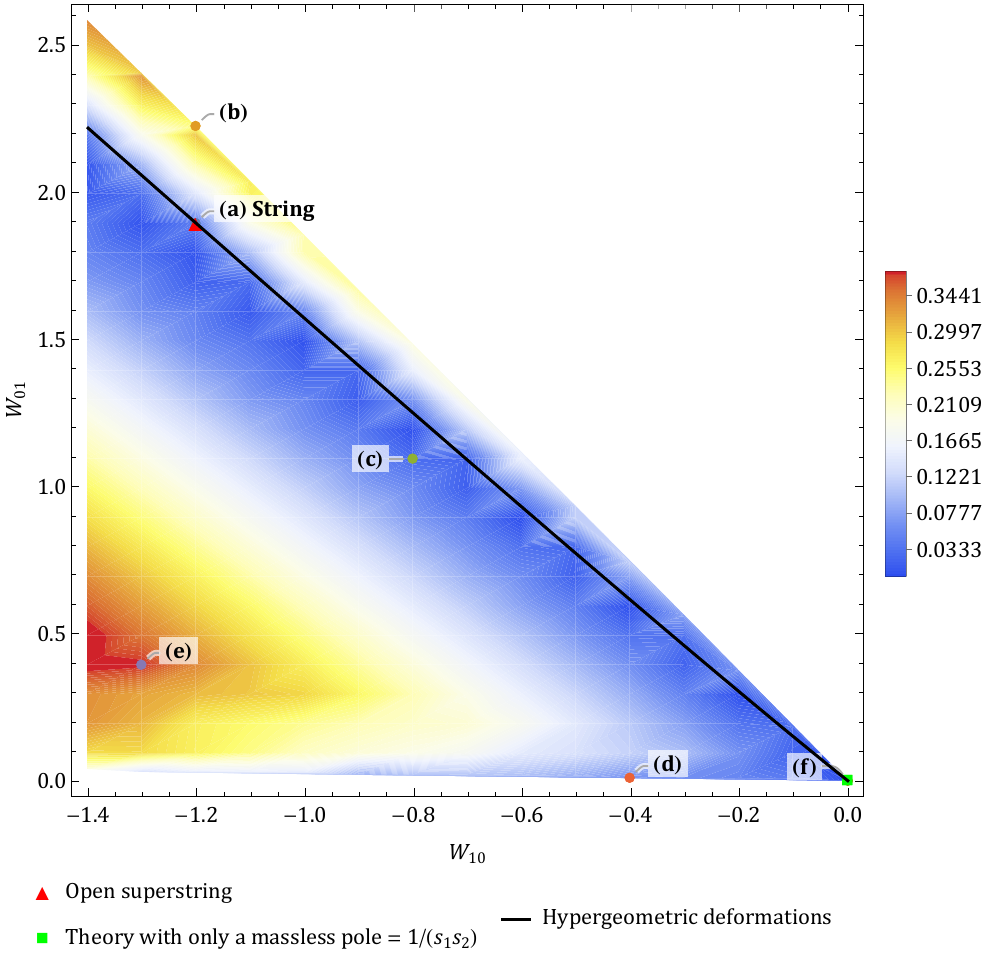}
    \caption{Allowed region of S-matrices for category-I theories in the space of Wilson coefficients $(W_{10},W_{01})$. The different shades of colors in this contour plot denote the variation of $|D_M|$ as we scan the entire space. The red triangle denotes the open superstring. It is located in a region shaded in deep blue, demonstrating that it satisfies the dual resonance model. The green square at the origin $(W_{10},W_{01})=(0,0)$ denotes the S-matrix $\mathcal{M}(s_1,s_2)=1/(s_1 s_2)$. The black line passing through the string point represents a family of hypergeometric deformations of the open superstring \cite{Mansfield:2024wjc}. The S-matrices indicated by $(a)$ to $(f)$ are plotted in fig.(\ref{Cat1}). No solution is found in the white region on the upper right half of this figure and for $W_{01}<0$. So, this region is excluded for category-I theories.}
    \label{fig:dualitycontourplotCat1}
\end{figure}

\begin{figure}[H]
    \centering
    \includegraphics[scale=0.64]{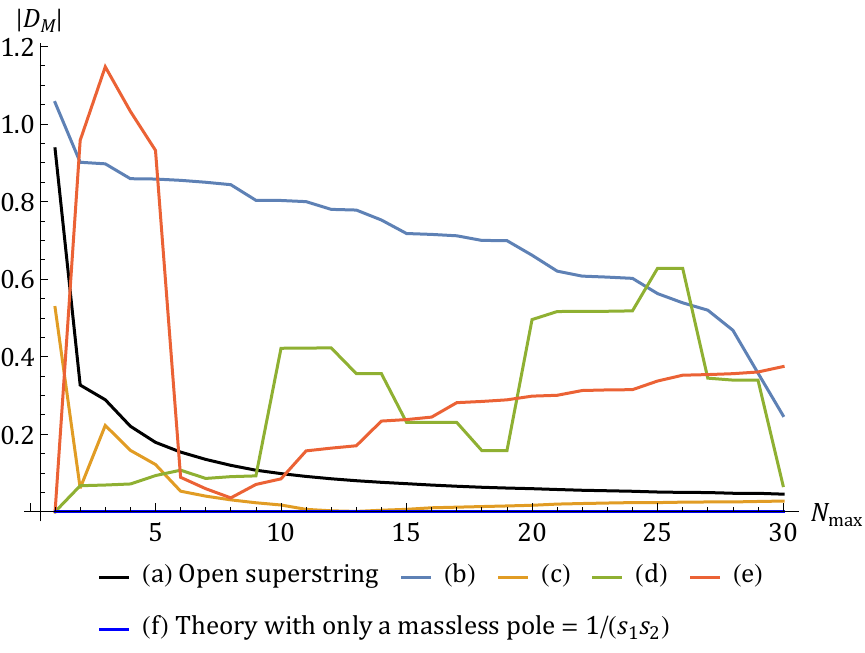}
    \caption{Convergence of the partial sums of $|D_M|$ as a function of $N_{\text{max}}$ for the S-matrices denoted by $(a)$ to $(f)$ in fig.(\ref{fig:dualitycontourplotCat1}).}
    \label{fig:dualitypartialsum}
\end{figure}

\begin{figure}[H]
    \centering
    \subfloat[\centering $W_{10}=-\zeta(3),W_{01}=\frac{7}{4}\zeta(4)$]{{\includegraphics[width=8cm]{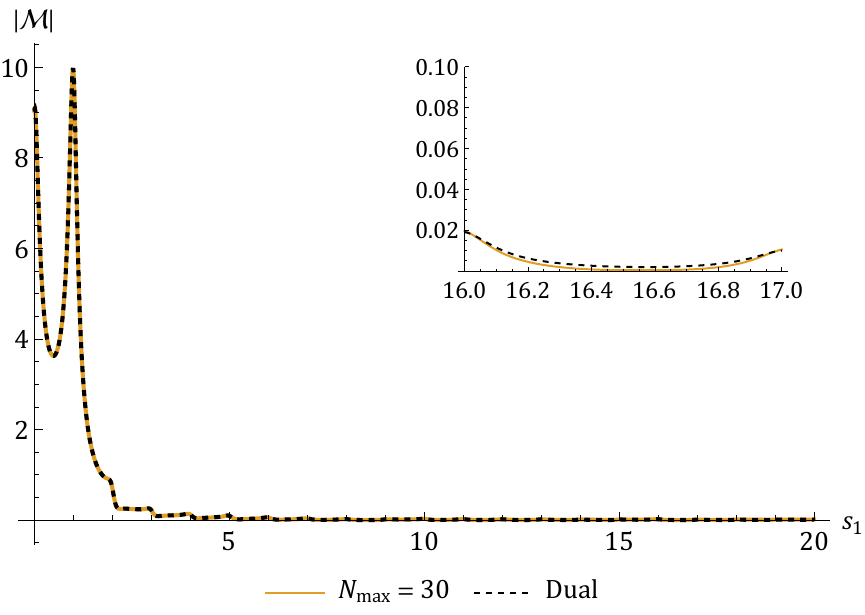} }}%
    \qquad
    \subfloat[\centering $W_{10}=-\zeta(3),W_{01}=\frac{7}{4}\zeta(4)+\frac{33}{100}$]{{\includegraphics[width=8cm]{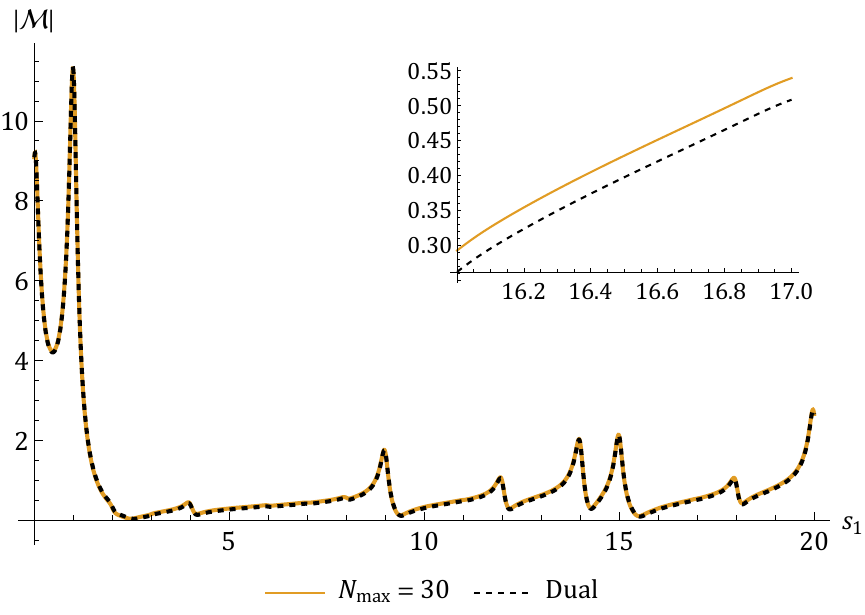} }}%
    \qquad
    \subfloat[\centering $W_{10}=-\zeta(3)+\frac{4}{10},W_{01}=\frac{7}{4}\zeta(4)-\frac{8}{10}$]{{\includegraphics[width=8cm]{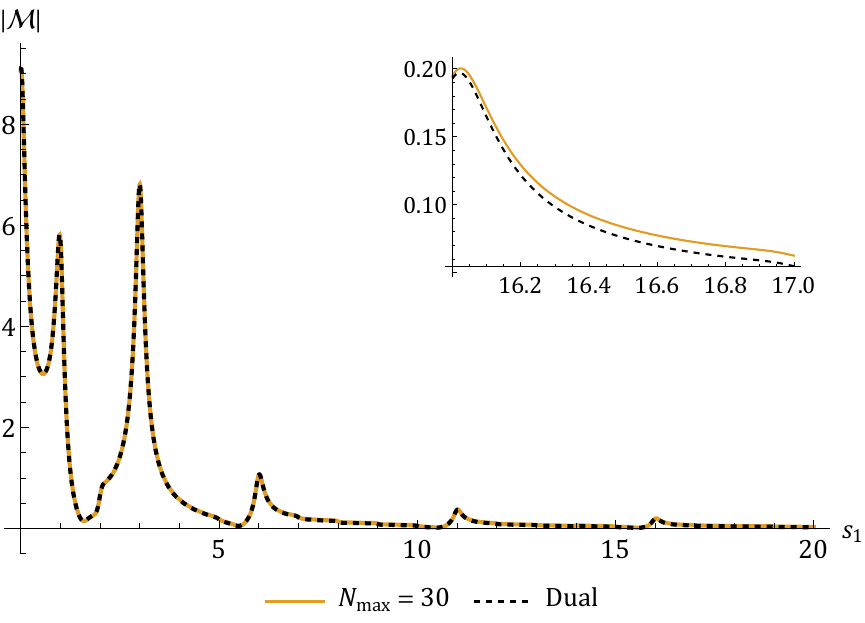} }}%
    \qquad
    \subfloat[\centering $W_{10}=-\zeta(3)+\frac{8}{10},W_{01}=\frac{1}{100}$]{{\includegraphics[width=8cm]{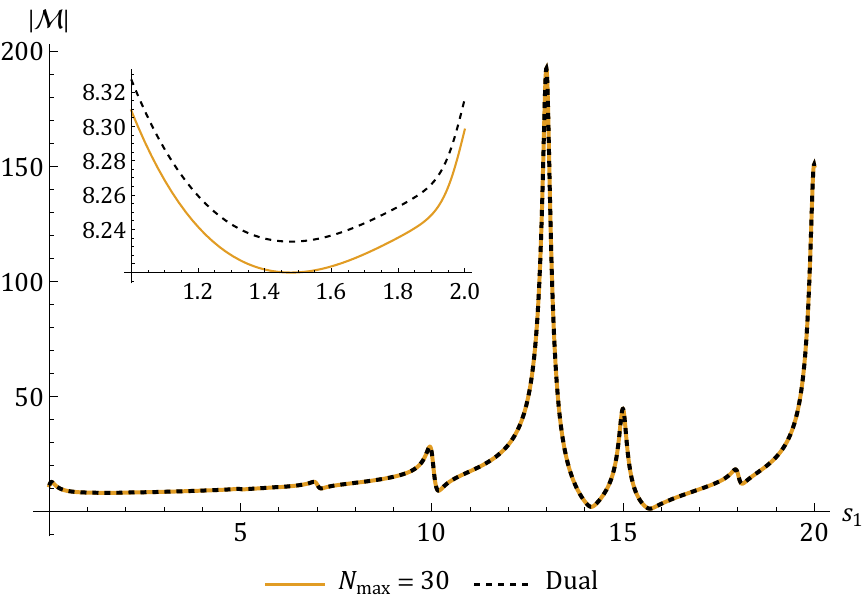} }}%
    \qquad
    \subfloat[\centering $W_{10}=-\zeta(3)-\frac{1}{10},W_{01}=\frac{7}{4}\zeta(4)-\frac{15}{10}$]{{\includegraphics[width=8cm]{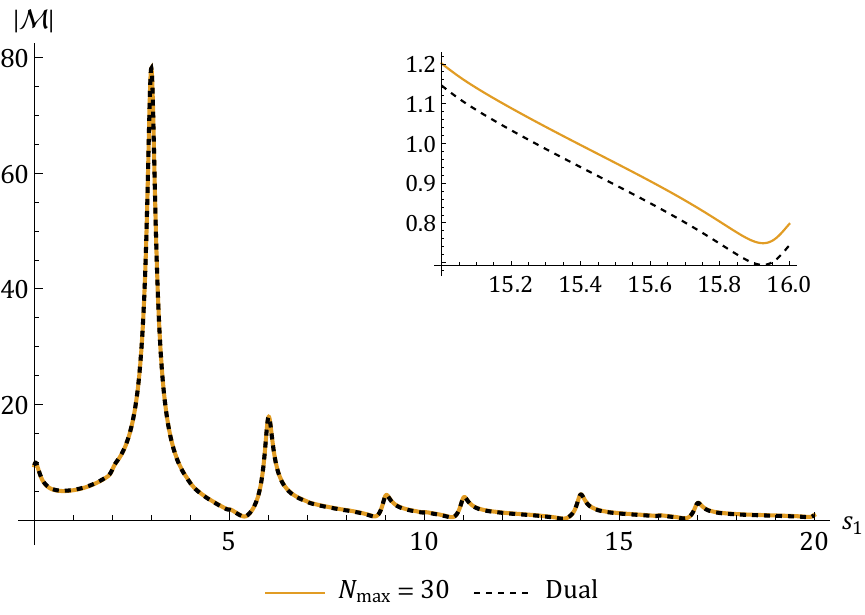} }}%
    \qquad
    \subfloat[\centering $W_{10}=0,W_{01}=0$]{{\includegraphics[width=8cm]{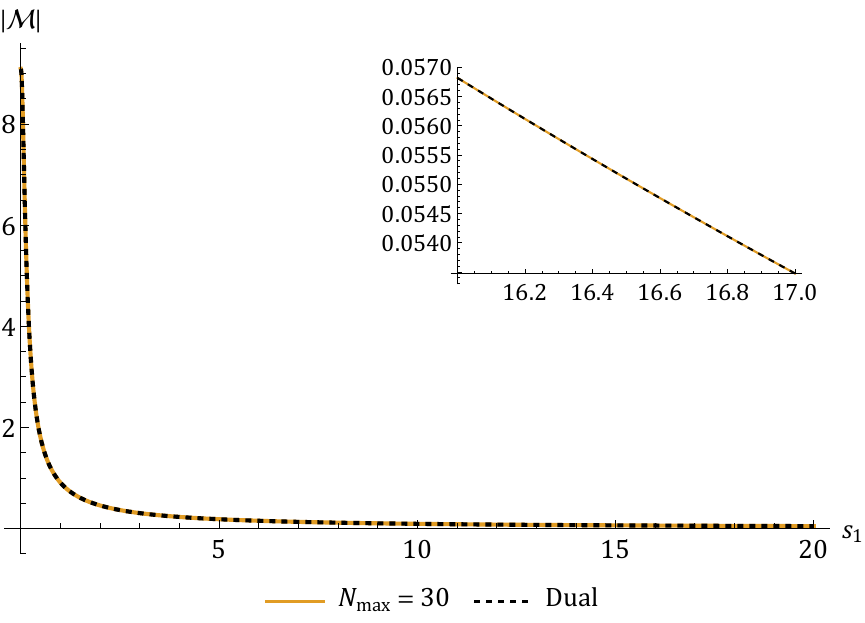} }}%
    \caption{We plot specific S-matrices ($s_1\rightarrow s_1+i/10$) in the ($W_{10}, W_{01}$) plane in fig.(\ref{fig:dualitycontourplotCat1}) along with the corresponding S-matrix in the dual resonance model for different values of $W_{10}$ and $W_{01}$ for $s_2=-1.1$. Fig.\ref{fig:dualitypartialsum} suggests that $(b), (d)$ and $(e)$ do not obey eq.(\ref{dualcheck}) well. The insets in the plots show zoomed in portions where $(b), (d)$ and $(e)$ do not satisfy the dual resonance model well. We have marked the location of these S-matrices in fig.(\ref{fig:dualitycontourplotCat1}).}%
   \label{Cat1}
\end{figure}

First, let us fix $W_{10}, W_{01}$ to the known values for the open superstring case: $W_{10}=-\zeta(3)$, $W_{01}=\frac{7}4\zeta(4)=\frac{7\pi^4}{360}$. Then we run the bootstrap. Remarkably, the S-matrix we find on minimizing the entangling moment leads to an almost exact agreement with the full superstring amplitude, as shown in fig.(\ref{OpenString}). The lower bound obtained on minimization of the entangling moment in this manner is 
\begin{equation}\label{epopenstring}
	min\left(\int_{1}^\Lambda ds\, \frac{\Delta\mathcal{E}}{s^2}\right)_{\text{open-string}}=1.5997\,.
\end{equation}
For a better comparison, we show the values of $c_\ell^{(n)}$s for the leading and subleading Regge trajectories in fig.(\ref{LeadingSubleading}). Amazingly, we find a very close match with the values from the exact expression for both the leading and the subleading Regge trajectories. In fig.(\ref{fig:dualitycontourplotCat1}) we show the allowed region of S-matrices in the space of the low-energy coefficients $W_{10}$ and $W_{01}$, which are now fixed to different values. The color coding is set by $|D_{M}|$ in \eqref{dualcheck}. As the plot shows, there are several models that respect \eqref{dualcheck} to the same tolerance as the string theory. We quantify this further in fig.(\ref{fig:dualitypartialsum}), where we plot the partial sums of $|D_M|$ in \eqref{dualcheck}, to check for convergence. As this plot indicates, the S-matrices labeled by $(b),(e)$ in fig.(\ref{fig:dualitycontourplotCat1}) do not respect \eqref{dualcheck} as well as the others, while the S-matrix $(d)$ has questionable convergence. The S-matrices for various cases are shown in fig.(\ref{Cat1}). In the same plots, we show how they compare with the single-channel representation at $s_2=-1.1$. The S-matrix $(c)$ obeys \eqref{dualcheck} and is similar in nature as what arises in the hypergeometric deformations (see appendix \ref{Hypergeometric deformations}). Note that the strengths of the resonances vary substantially from plot to plot. In appendix \ref{General spectrum} we find that the open superstring amplitude can be bootstrapped with a general integer spaced spectrum with all spins $0\leq \ell\leq n$ at mass level $n$.

\subsection*{Leading and subleading Regge trajectories}

\subsubsection{Spin-0 dominance and existence of infinite higher masses and spins}
We have already explained analytically that $W_{01}\leq -2 W_{10}$---the upper boundary in  fig.\ref{fig:dualitycontourplotCat1} respects this. Let us now comment on the lower boundary. It is curious to note that here $W_{01}>0$---the question arises, why? These theories obey \eqref{dualcheck} so let us examine this equation more carefully. 
\eqref{dualcheck} can be used to check for the presence of spin-0 dominance in category-I theories. Taking derivative of $D_M$ w.r.t $\lambda$ and then putting $\lambda = 0$ we get
\begin{equation}\label{der}
	\frac{\partial D_M}{\partial \lambda}\bigg|_{\lambda=0}=\sum_{n=1}^{\infty}\sum_{\ell}\frac{c_\ell^{(n)}}{n^2}\left\{2(D-3)\mathcal{G}_{\ell-1}^{\left(\frac{D-1}{2}\right)}\left(1\right)-\mathcal{G}_{\ell}^{\left(\frac{D-3}{2}\right)}\left(1\right)\right\} = 0\,.
\end{equation}
The $\mathcal{G}_{\ell-1}$ terms contribute from $\ell=1$ onward. The above equation can be separated into two parts: one corresponding to spin-0 and the other corresponding to higher spins,
\begin{equation}
	\frac{\partial D_M}{\partial \lambda}\bigg|_{\lambda=0}=\underbrace{-\sum_{n=1}^{\infty}\frac{c_0^{(n)}}{n^2}}_{\text{spin-0}}+\underbrace{
	\sum_{n=1}^{\infty}\sum_{\ell>0}\frac{c_\ell^{(n)}}{n^2}\left\{2(D-3)\mathcal{G}_{\ell-1}^{\left(\frac{D-1}{2}\right)}\left(1\right)-\mathcal{G}_{\ell}^{\left(\frac{D-3}{2}\right)}\left(1\right)\right\}}_{\text{higher spins}} = 0\,.
\end{equation} 
The spin-0 contribution is always negative and it can be easily checked that the contribution from higher spins is always positive which proves that spin-0 must exist and furthermore the contribution from spin-0 must be greater in magnitude than the contribution from any particular spin in the higher spin sum. Similarly, calculating the $p_{\text{th}}$ derivative of $D_M$ w.r.t $\lambda$ and checking the contributions from various terms it can be concluded that $n\geq p+1$ and $\ell\geq p$ will be needed to satisfy $ \frac{\partial^p D_M}{\partial \lambda^p}\big|_{\lambda=0}=0$. This means that infinite $n$'s and infinite $\ell$'s have to be turned on. The implication of this spin-0 dominance now feeds into \eqref{W10W01}. By examining the signs of various spins for $W_{01}$ in \eqref{W10W01}, we find that at $\lambda=0$, spin-0 contributes positively and all other spins contribute negatively. By comparing the $n$ dependence in eq.(\ref{der}), it is then suggestive that the spin-0 dominance would lead to $W_{01}>0$, which is what we find in fig.(\ref{fig:dualitycontourplotCat1}). Let us perform a numerical check to verify this. Consider the S-matrix with $W_{10}=-\zeta(3)+\frac{8}{10}$ and $W_{01}=\frac{1}{100}$ labelled $(d)$ in fig.(\ref{fig:dualitycontourplotCat1}). This S-matrix lies exactly on the lower boundary of the allowed region and numerically we find that 
\begin{equation}
    \frac{(W_{01})_{\text{spin-0}}}{(W_{01})_{\text{higher spins}}}=-1.501
\end{equation}
which demonstrates spin-0 dominance.

\subsection{Category-II theories: away from dual resonance}
\begin{figure}[H]
    \centering
    \begin{subfigure}{0.45\textwidth}
        \includegraphics[width=9.5cm]{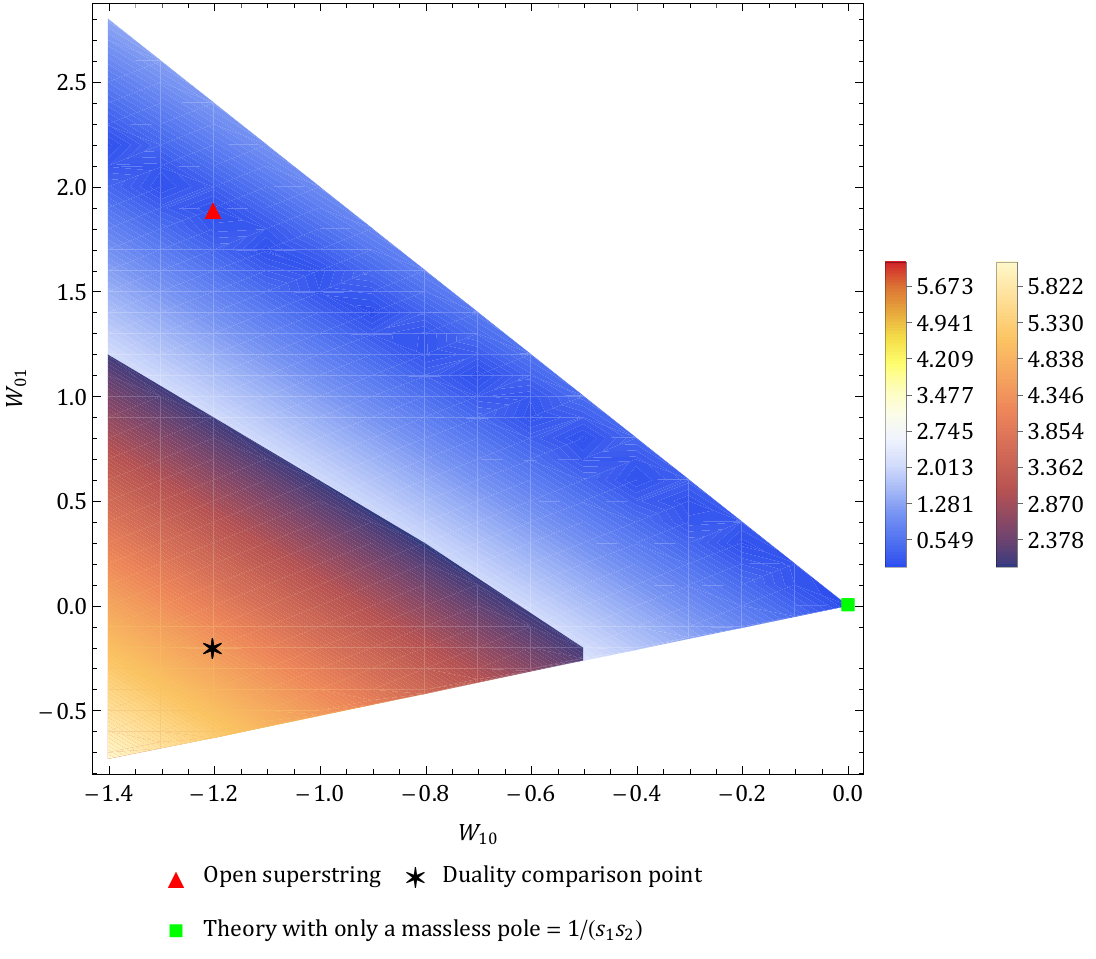}
        \caption{Duality contour plot for category-II.}
        \label{contour}
    \end{subfigure}
    \hfill
    \begin{subfigure}{0.45\textwidth}
        \includegraphics[width=7.5cm]{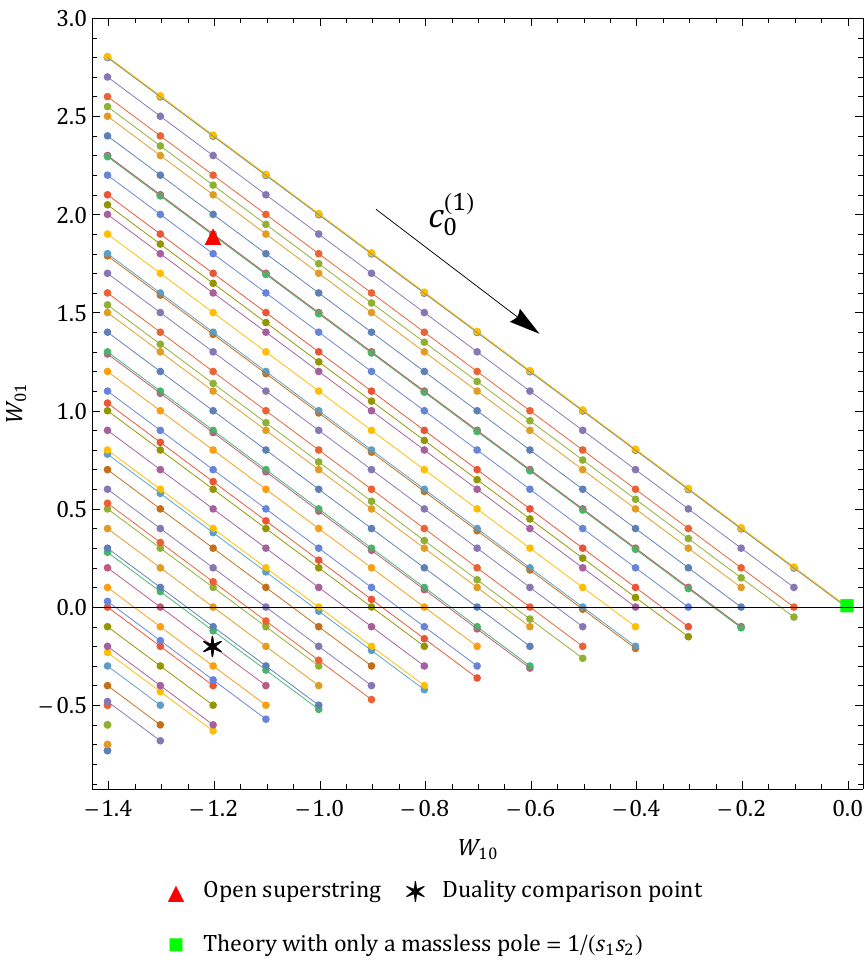}
        \caption{Universality classes for category-II.}
        \label{Reggeclass}
    \end{subfigure}
    \caption{Allowed region in the space of ($W_{10},W_{01}$) for category-II theories.}%
   \label{fig:dualitycontourplotCat2}
\end{figure}

The category-II S-matrices are found in the region shown below. Since we use \eqref{ansatzfull}, we have to specify $W_{00}$. We choose this to be such that the boundary term in \eqref{t-fix} is zero for $s_2=0$. In this manner close to $s_2=0$, we expect to have Regge behavior. The legend in fig.(\ref{contour}) quantifies $|D_M|$ defined in \eqref{dualcheck}, and is a measure of how closely the S-matrices obey the dual resonance condition. In this particular case, since we started with \eqref{ansatzfull}, we find an interesting feature in fig.(\ref{Reggeclass}). The theories along the straight lines are related to each other by a scalar deformation. The minimization picks up only the $n=1$ mode for this deformation. We can say that theories on the same line belong to the same universality class. In the category-I theories, this was not possible--the reason is that a single scalar deformation in the ansatz used in this case is not $\lambda$-independent. For the rest of this section, we will focus our attention on a specific category-II S-matrix, indicated by the black star in fig.(\ref{contour}). This S-matrix will turn out to obey duality only in the forward limit and will deviate elsewhere. It is also observed that the upper boundary of the allowed region for category-II theories has only $c_0^{(1)}$ turned on with $c_0^{(1)}=-W_{10}$. Therefore, we can obtain the S-matrices on the upper boundary using the bootstrap by simply restricting the $\ell$ spectrum to the leading Regge trajectory \textit{i.e.} $\ell=n-1$. Note that the upper boundary in fig.(\ref{fig:dualitycontourplotCat2}) is exactly the upper bound on $W_{01}\leq -2W_{10}$ in \eqref{W01ub}.

\subsubsection{High-energy behavior}
As mentioned in the introduction, Polchinski and Strassler in \cite{polchinskistrassler}, used insights from the AdS/CFT correspondence to suggest how the power-law behavior in glueball scattering amplitudes could arise, despite the string amplitude in 10 dimensions having Regge behavior. They found that there could be a transition from Regge behavior for low $t$ to the expected power law behavior at higher $t$. The condition on $t$ depends on $s$ and it was found that for fixed $t$, for sufficiently large $s$, the amplitude exhibited power-law dependence on $s$. We will refer to this as a Polchinski-Strassler type transition and ask if the bootstrap gives amplitudes showing similar transitions. The hypergeometric deformations considered in \cite{Cheung:2023adk, Haring:2023zwu} are of this nature, arising from amplitudes that obey the dual resonance condition. Here, we will demonstrate that similar transitions also happen in theories that do not obey the duality hypothesis except in the low-energy regime. 

We consider the black star point in fig.(\ref{contour}) using the fit function
\begin{equation}
    \mathcal{M}_{\text{fit}}(s_1,s_2)=c+s_1^{a_0+a_1s_2},\hspace{1cm} s_1\gg 1, s_2<0.
\end{equation}
Here, $c$, $a_0$ and $a_1$ are the fit parameters. We use \eqref{ansatzfull} for category-II S-matrices. We choose the boundary term in \eqref{t-fix} to vanish in the forward limit, that is, $\mathcal{B}(s_2=0)=0$. Making this choice for the S-matrices in category-II, $W_{00}$ in \eqref{ansatzfull} is given by \eqref{W00}. We use \eqref{ansatzfull} and \eqref{W00} to make a data table of $\mathcal{M}(s_1,s_2)$ for the black star S-matrix obtained from the bootstrap for fixed $s_1=19.5$ and varying $s_2$ such that $-10.1\leq s_2\leq -0.1$ with a step size of $0.01$\,. We use the \texttt{FindFit} function in Mathematica with \texttt{MaxIterations -> 10000} to obtain $\mathcal{M}_{\text{fit}}(s_1,s_2)$ using this data table. The results obtained for the fit parameters are 
\begin{eqnarray}
   c  =  -4.3939\,,\quad
 a_0  =  0.4804 \,,\quad
           a_1  =  0.2019\,.
\end{eqnarray}
Since $a_1>0$, for any negative $s_2\lesssim -2.4$ and sufficiently large $s_1$, $c$
dominates, similar to the Polchinski-Strassler case reviewed above.

\begin{figure}[H]
    \centering
    \begin{subfigure}{0.45\textwidth}
        \includegraphics[width=8cm]{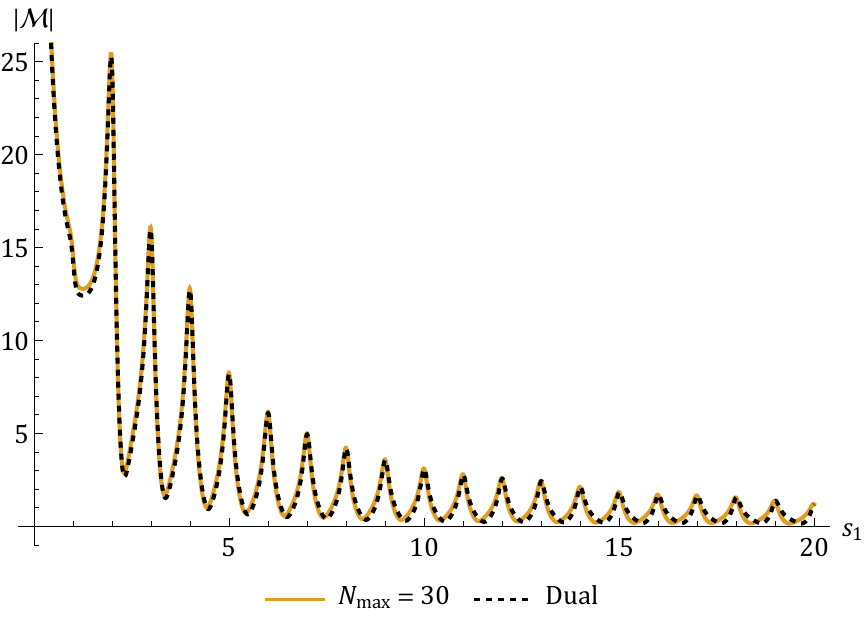}
        \caption{$s_2=-0.1$}
        \label{forward}
    \end{subfigure}
    \qquad
    \begin{subfigure}{0.45\textwidth}
        \includegraphics[width=8cm]{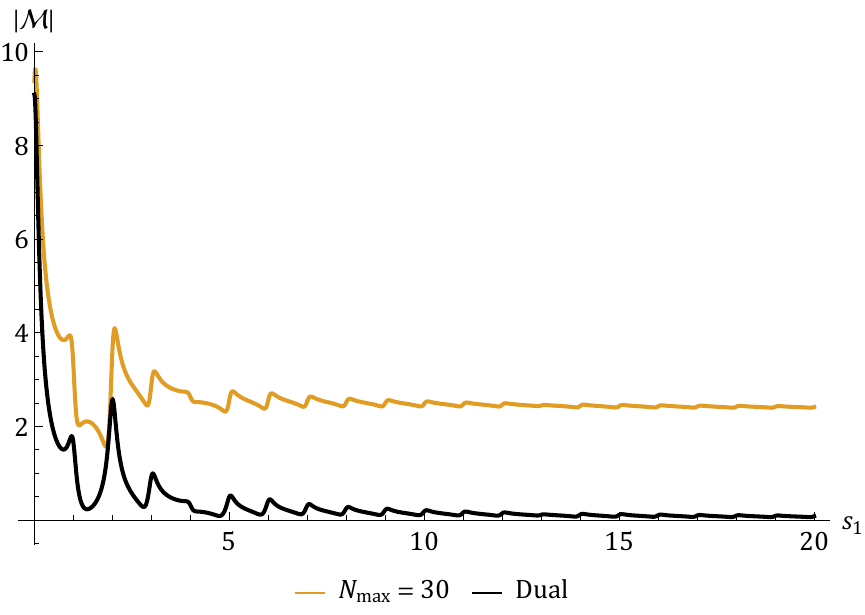}
        \caption{$s_2=-1.1$}
        \label{nonforward}
    \end{subfigure}
    \caption{We plot $\big|\mathcal{M}\left(s_1+\frac{i}{10},s_2\right)\big|$ for the black star along with the corresponding dual S-matrix versus $s_1$ for fixed $s_2=-0.1$ (close to the forward limit) and $s_2=-1.1$ (away from the forward limit). We observe that this amplitude goes to a constant at large $s_1$, away from the forward limit, hence violating the dual resonance model.}%
   \label{fig:polstras}
\end{figure}

To obtain $W_{00}$ for the black star S-matrix using this fit we notice from \eqref{ansatzfull} that
\begin{equation}
    W_{00} = c- \sum_{n=1}^{N_{\text{max}}}\sum_{\ell}c_{\ell}^{(n)} \left(\frac{2}{n}-\frac{1}{\lambda+n}\right)\mathcal{G}_{\ell}^{\left(\frac{D-3}{2}\right)}\left(1-\frac{2\lambda}{\lambda+n}\right)=-3.2108\,.
\end{equation}
This perfectly matches with $W_{00}$ from \eqref{W00} which is
\begin{equation}
    W_{00} = -\sum_{n=1}^{N_{\text{max}}}\sum_{\ell}\frac{1}{n}c_{\ell}^{(n)}\mathcal{G}_{\ell}^{\left(\frac{D-3}{2}\right)}\left(1\right)=-3.2084\,.
\end{equation}
Of course, this is equal to the bound on minimization of the entangling moment \eqref{epowerexp}. We plot our results in fig.(\ref{fig:polstras}). Fig.(\ref{forward}) clearly demonstrates the amplitude obeying the duality condition for low values of $s_2$. As fig.(\ref{nonforward}) illustrates, at higher negative values of $s_2$, the amplitude does not obey the duality hypothesis. A final point we wish to remind the reader here is that we made an assumption about what spins contribute at each mass level (see assumption 3 in the introduction). We relax this assumption in the appendix \ref{General spectrum} to see what happens. The conclusion is that the leading Regge trajectories are similar while differences appear for subleading ones. At the level of the plots, the differences are negligible.
\subsection{Upper critical dimension}
 From the bootstrap we can also calculate the upper critical dimension $D_{\text{uc}}$ for the category-I and category-II S-matrices that we have investigated so far. This means that if we consider an S-matrix with the $10D$ $c_\ell^{(n)}$s from the bootstrap solution and then compute the $c_\ell^{(n)}$s in $D=11$ using $11D$ Gegenbauer polynomials then some specific $11D$ $c_\ell^{(n)}$s (\textit{e.g.} $c_0^{(3)},c_1^{(4)},c_2^{(5)}$) become negative indicating violation of unitarity. For example, consider $c_0^{(3)}$. For the open superstring S-matrix in fig.(\ref{fig:dualitycontourplotCat1}), the $10D$ $c_0^{(3)}=0$, but the $11D$ $c_0^{(3)}=-0.00425$. Similarly, for the black star S-matrix in fig.(\ref{fig:dualitycontourplotCat2}) we find the $10D$ $c_0^{(3)}=0$, but the $11D$ $c_0^{(3)}=-0.022$. We summarize our results in table $\ref{ucd}$.

 This follows from the property that higher $D$ Gegenbauers can be written as positive linear combinations of lower $D$ Gegenbauers. So a unitary solution in $D=10$ must be unitary in all $D\leq 10$, but unitarity for $D>10$ is not guaranteed. The upper critical dimension of all the S-matrices in category-I in fig.(\ref{fig:dualitycontourplotCat1}) is $D_\text{uc}=10$. In category-II all the S-matrices in fig.(\ref{fig:dualitycontourplotCat2}) have $D_\text{uc}=10$ except those living on the upper boundary for which only $c_{0}^{(1)}$ is non-zero and rest all other coeeficients are zero. This is because for the S-matrices where only spin-0 coefficients are non-zero, for example, the theories living on the upper boundary in fig.(\ref{fig:dualitycontourplotCat2}) where only $c_{0}^{(1)}$ is turned on, there is no need for higher spins and the sole spin-0 coefficient remains positive when calculated in all $D$.
\begin{center}
\begin{table}[H]
    \centering
    \begin{tabular}{|c|c|c|c|c|c|c|}
    \hline
   \multirow{2}{*}{$c_\ell^{(n)}$}& \multicolumn{2}{c|}{Open superstring} & \multicolumn{2}{c|}{Black star}\\ \cline{2-5}
    & $D=10$ & $D=11$ & $D=10$ & $D=11$\\
    \hline
    $c_0^{(3)}$ & $0$ & $-0.00425$ & $0$ & $-0.022$ \\
    \hline 
    $c_1^{(4)}$ & $0.000367$ & $-0.000655$ & $0.0232$ & $0.0152$ \\
    \hline 
    $c_2^{(5)}$ & $0$ & $-0.000278$ & $0$ & $-0.00147$\\
    \hline 
    \end{tabular}
   \caption{Table showing that the upper critical dimension for the open superstring and the black star is $D_{\text{uc}}=10$.}
   \label{ucd}
\end{table}
\end{center}

\section{Closed-string}
\label{sec:clst}
Here, we will briefly examine whether entanglement minimization in 2-2 scattering of gravitons (assuming type II supersymmetry) can pick up the closed-string amplitude. The answer is yes, but the convergence is slower than that of an open-string.
\subsection{Parametric representation}

In \cite{Saha:2024qpt}, parametric representations for the closed-string tree amplitude were derived by recycling the 2-channel symmetric representation. 
A parametric series representation resembling the closed-string tree amplitude, maintaining 3-channel symmetry, for a general case where $s_{1}+s_{2}+s_{3}=\kappa$, has been obtained in \cite{rosengren} using the method of partial fractions; here we will use an approach motivated from dispersion relation (see \cite{bhatClStr} for details). From the equation $y=-s_{1}s_{2}s_{3}$, we find
\begin{equation}\label{shifted-s2}
	s_{2}^{\pm}\left(s_{1}\right)=-\frac{s_{1}-\kappa}{2}\Biggl\{1\pm\sqrt{1+\frac{4y}{s_{1}\left(s_{1}-\kappa\right)^{2}}}\Biggr\}.
\end{equation}
Let us consider a general version of closed-string amplitude, given by 
\begin{equation}\label{genclstring}
	\mathcal{M}^{(\alpha,\beta)}_{\text{cl}}\left(s_{1},s_{2},s_{3}\right) = \frac{\Gamma\left(\alpha-s_{1}\right)\Gamma\left(\alpha-s_{2}\right)\Gamma\left(\alpha-s_{3}\right)}{\Gamma\left(\beta+s_{1}\right)\Gamma\left(\beta+s_{2}\right)\Gamma\left(\beta+s_{3}\right)}.
\end{equation}
 Here $\alpha$ and $\beta$ are any non-zero parameters. We can work out a series representation of the above amplitude by using local crossing symmetric dispersion relation \cite{Song:2023quv},
\begin{equation}\label{cldisp}
	\mathcal{M}_{\text{cl}}\left(s_{1},s_{2},s_{3}\right) = \frac{1}{\pi}\int_{s_{0}}^{\infty}\mathrm{d}\sigma\left[\frac{1}{s_{1}-\sigma}+\frac{1}{s_{2}-\sigma}+\frac{1}{s_{3}-\sigma}+\frac{1}{\sigma}\right]\mathcal{A}^{\left(s_{1}\right)}\left(\sigma,s_{2}^{\pm}\left(\sigma\right)\right).
\end{equation} 
Here, $\mathcal{A}^{(s_{1})}\left(s_{1},s_{2}^{\pm}\right)$ is the discontinuity of the amplitude in $s_{1}$ channel and $s_{2}^{\pm}$ is given by \eqref{shifted-s2}. $p_{0}$ denotes the location of the lowest pole in $s_{1}$.
Note that \eqref{genclstring} is invariant under $s_{i}\rightarrow s_{i}+\lambda$, $\alpha\rightarrow\alpha+\lambda$, $\beta\rightarrow\beta-\lambda$ and $\kappa\rightarrow\kappa+3\lambda$. If we perform these shifts, followed by setting $\alpha=0$, $\beta=1$ and $\kappa=0$, we then obtain a one-parameter family of representations for the closed-string amplitude,
\begin{eqnarray}\label{clst-1para}
	 \frac{\Gamma\left(-s_{1}\right)\Gamma\left(-s_{2}\right)\Gamma\left(-s_{3}\right)}{\Gamma\left(1+s_{1}\right)\Gamma\left(1+s_{2}\right)\Gamma\left(1+s_{3}\right)} 
	& = & -\frac{1}{s_{1}s_{2}s_{3}} + \sum_{n=1}^{\infty}\frac{1}{\left(n!\right)^{2}}\left[\frac{1}{s_{1}-n} + \frac{1}{s_{2}-n}+\frac{1}{s_{3}-n} + \frac{1}{\lambda+n}\right]	\nonumber\\
	&& \left(1-\frac{n}{2}+\frac{n-2\lambda}{2}\sqrt{1-\frac{4\left(s_{1}+\lambda\right)\left(s_{2}+\lambda\right)\left(s_{3}+\lambda\right)}{\left(n+\lambda\right)\left(n-2\lambda\right)^{2}}}\right)_{n-1}^2
\end{eqnarray}
The series converges absolutely for $\text{Re}(\lambda)>-1$. The case $\lambda=0$ agrees with \cite{Saha:2024qpt} and the general case with \cite{rosengren}\footnote{One has to massage the expressions in this paper into the form quoted above.}.

\subsection{Bootstrap analysis}
Following the same argument as in Sec.(\ref{sec:lambda-amp-deriv}), we denote discontinuity of the amplitude as
\begin{equation}
	\mathcal{A}^{(s_{1})}\left(s_{1},s_{2}^{\pm}\right) = -\pi\sum_{n=1}^{\infty}\sum_{\ell\in\text{spectrum}}c_{\ell}^{(n)}\mathcal{G}_{\ell}^{\left(\frac{D-3}{2}\right)}\left(1+\frac{2\left(s_{2}^{\pm}-\alpha\right)}{s_{1}-\alpha}\right)\delta\left(s_{1}-\alpha-n\right).
\end{equation}
We use the above equation in \eqref{cldisp}. It is implicitly assumed that the amplitude is invariant under $s_{i}\rightarrow s_{i}+\lambda$, $\alpha\rightarrow \alpha+\lambda$ and $\kappa\rightarrow\kappa+\lambda$. Since we are interested in massless external states, we set $\alpha=0$ and $\kappa=0$. We then obtain
\begin{eqnarray}\label{clansatz}
 \mathcal{M}\left(s_{1},s_{2},s_{3}\right) & = & -\frac{1}{s_{1}s_{2}s_{3}} + \sum_{n=1}^{\infty}\sum_{\ell\in\text{spectrum}}\left[\frac{1}{s_{1}-n}+\frac{1}{s_{2}-n}+\frac{1}{s_{3}-n}+\frac{1}{\lambda+n}\right]\nonumber\\
	&& c_{\ell}^{(n)}\mathcal{G}_{\ell}^{\left(\frac{D-3}{2}\right)}\left[\pm\left(1-\frac{2\lambda}{n}\right)\sqrt{1-\frac{4\left(s_{1}+\lambda\right)\left(s_{2}+\lambda\right)\left(s_{3}+\lambda\right)}{\left(\lambda+n\right)\left(n-2\lambda\right)^{2}}}\right].\nonumber\\
\end{eqnarray}
In this case, at mass level $n$ the spectrum contains, $0\leq\ell\leq 2n-2$ in steps of $2$. Therefore, the spectrum contains even spins only, and we can choose either sign before the square root. For even spins, we only have even powers of the argument, and hence the representation maintains locality. 
\subsubsection{Negativity from crossing}
\begin{figure}[H]
    \centering
    \subfloat[Negativity domain ($\lambda=0$). ``Grumpy Scissorhands"]{{\includegraphics[width=7.5cm]{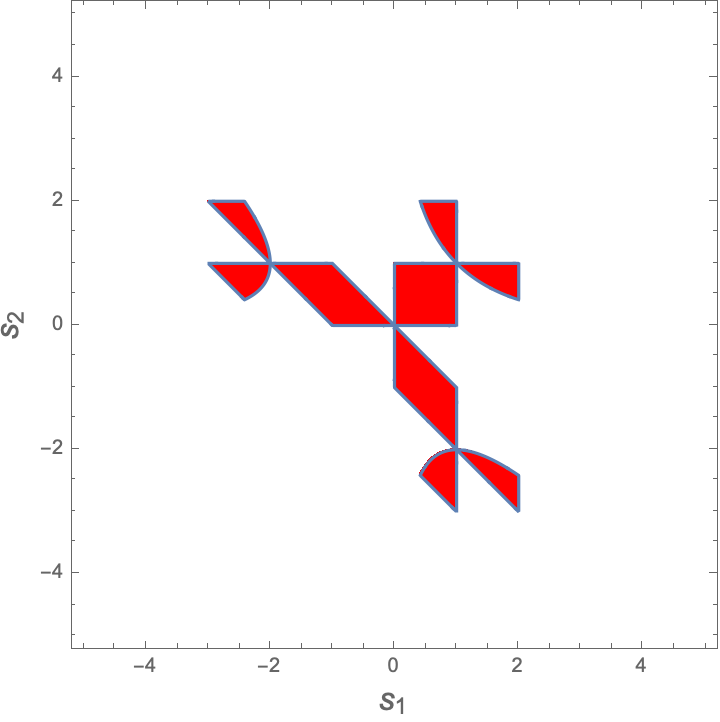} }}%
    \qquad
    \subfloat[ Extended domain obtained by dialing $-1<\lambda<0$.]{{\includegraphics[width=7.5cm]{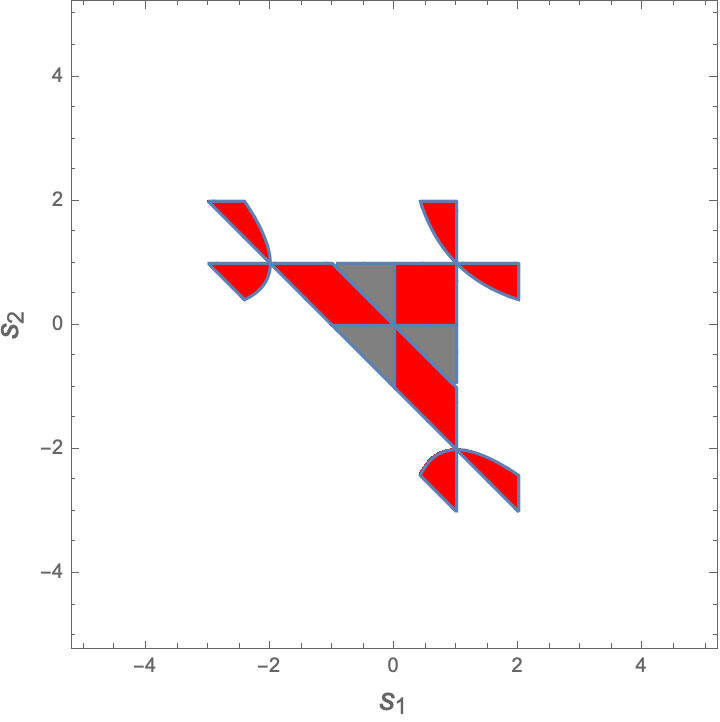} }}%
    \caption{Negativity domains for closed-string.}%
   \label{negdomcl}
\end{figure}

A consequence of \eqref{clansatz} is that it implies that $\mathcal{M}$ is negative in the $s_1, s_2$-region showed in fig.\ref{negdomcl}(a) below. Here, we set $\lambda=0$. This is the region where $1/(s_1-n)+1/(s_2-n)+1/(s_3-n)+1/n$ is negative and $s_1 s_2 s_3<0$ so that the argument of the Gegenbauers is greater than 1. This would imply that each term in the non-zero mode sum in \eqref{ansatz} is negative. Hence, the full amplitude is negative in this region, which is given by
\begin{eqnarray}
    &&\bigl\{s_{1}s_{2}\left(s_{1}+s_{2}\right)>s_{1}^{2}+s_{2}^{2}+s_{1}s_{2}-1\;\cap \; 2>s_{1}s_{2}\left(s_{1}+s_{2}\right)>0\bigr\}\\
    &\cup &\bigl\{2<s_{1}s_{2}\left(s_{1}+s_{2}\right)<s_{1}^{2}+s_{2}^{2}+s_{1}s_{2}-1\; \cap\; 16>s_{1}s_{2}\left(s_{1}+s_{2}\right)>2\left(s_{1}^{2}+s_{2}^{2}+s_{1}s_{2}\right)-8\bigr\}.\nonumber
\end{eqnarray}

The extended domains by dialing $-1<\lambda<0$ are shown in fig.\ref{negdomcl}(b). The amplitude now is negative in the combined red-gray region\footnote{It will be interesting to see if the positivity in the context of Celestial Holography, observed in \cite{celestial} is related to this.}.

\subsubsection{Numerics}
\begin{figure}[H]
    \centering
    \subfloat[\centering $s_2=-0.1$]{{\includegraphics[width=8cm]{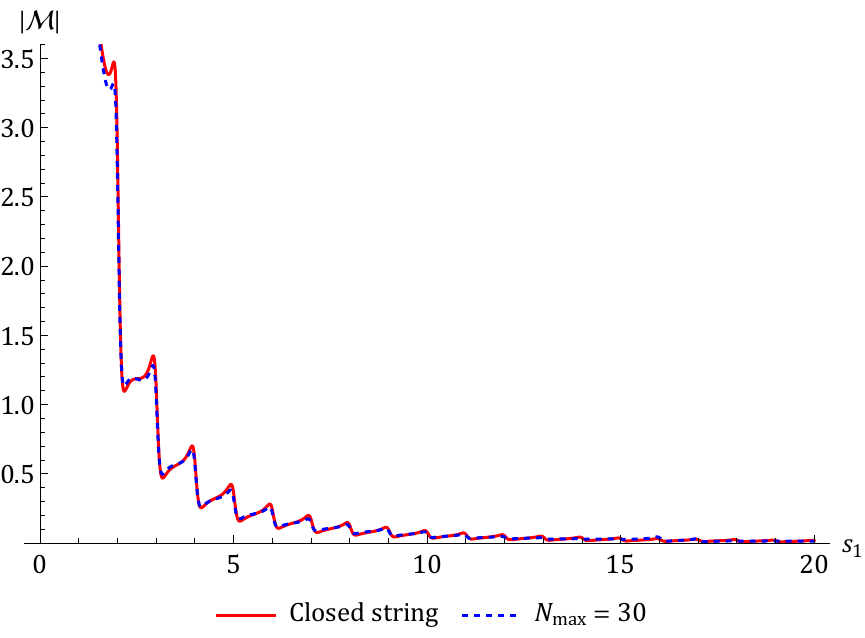} }}%
    \qquad
    \subfloat[\centering $s_2=1.1$]{{\includegraphics[width=8cm]{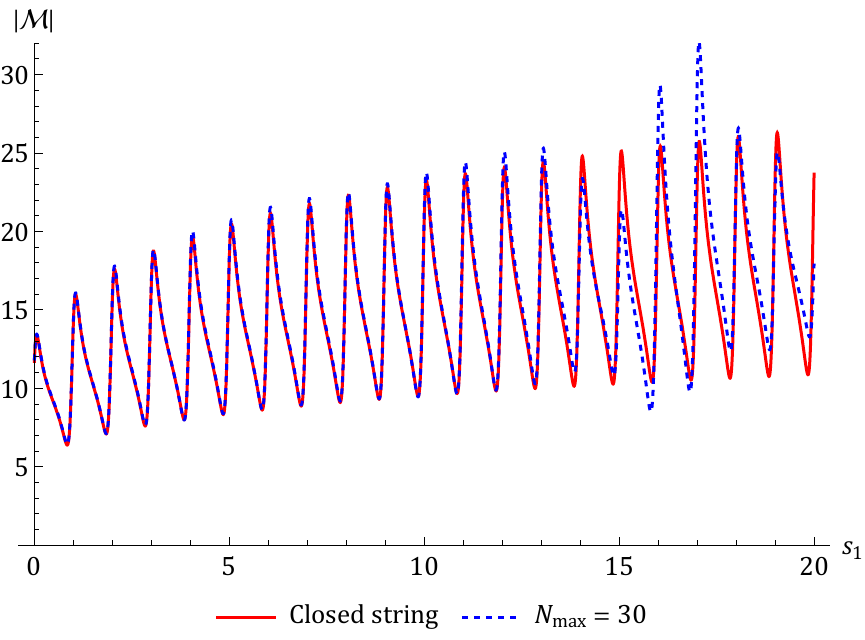} }}%
    \qquad
    \subfloat[\centering $s_2=-0.1$]{{\includegraphics[width=8cm]{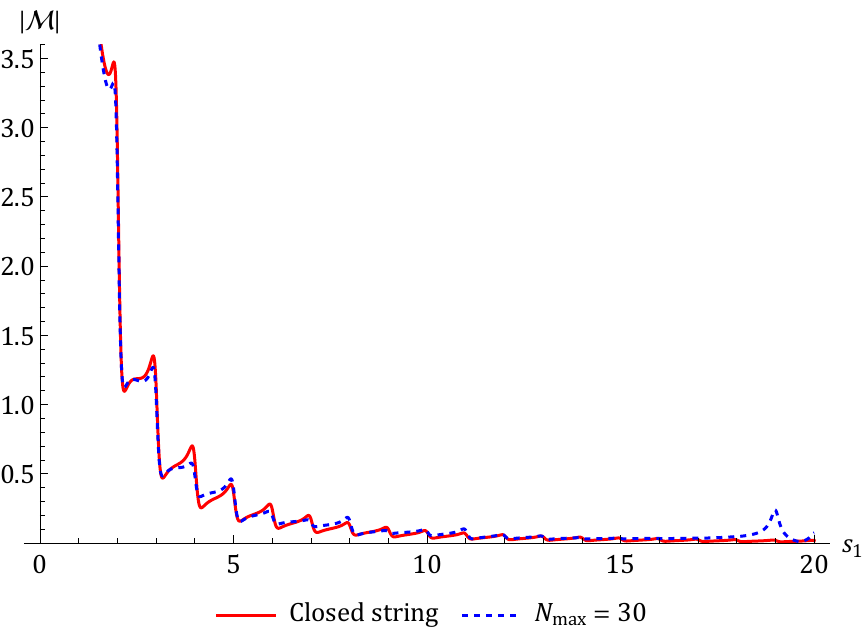} }}%
    \qquad
    \subfloat[\centering $s_2=1.1$]{{\includegraphics[width=8cm]{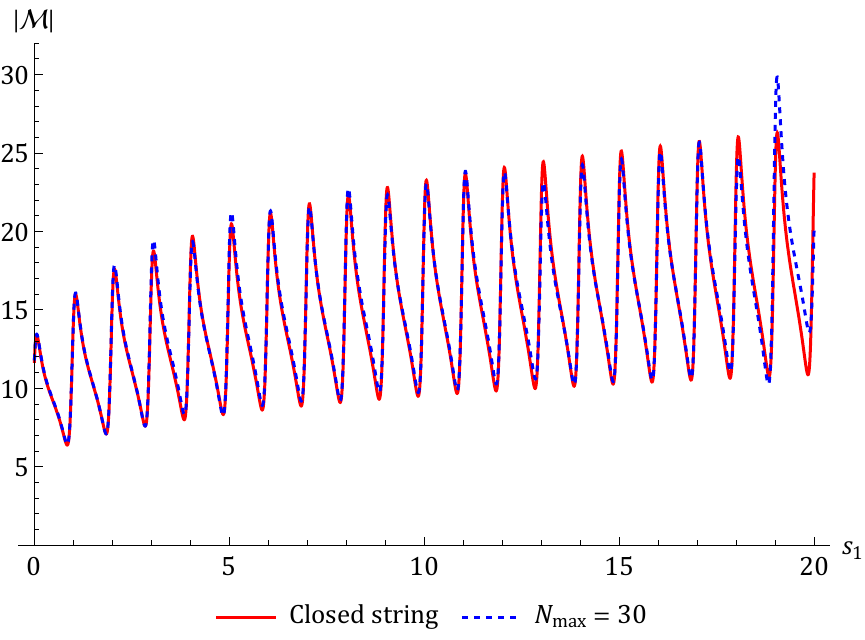} }}%
    \caption{Bootstrapping tree-level closed-string theory. We plot $\big|\mathcal{M}\left(s_1+\frac{i}{10},s_2\right)\big|$ versus $s_1$ for different values of $s_2$ for $N_{\text{max}}=30$ in $D=10$ with $\lambda=1.6$. The tolerance we used was $T = 10^{-6}$. In (a) and (b), we plot the amplitude resulting from the minimization of $\int_1^\Lambda ds\, \frac{\Delta\mathcal{E}}{s^3}$ which yields the bound $1.533$. In (c) and (d), we plot the amplitude resulting from the minimization of $\int_1^\Lambda ds\,\frac{\Delta\mathcal{E}}{s^4}$ which yields the bound $1.185$. }%
   \label{ClosedString}
\end{figure}

\begin{figure}[H]
    \centering
    \subfloat[\centering Minimization of $\int_1^\Lambda ds\, \frac{\Delta\mathcal{E}}{s^3}$.]{{\includegraphics[width=8cm]{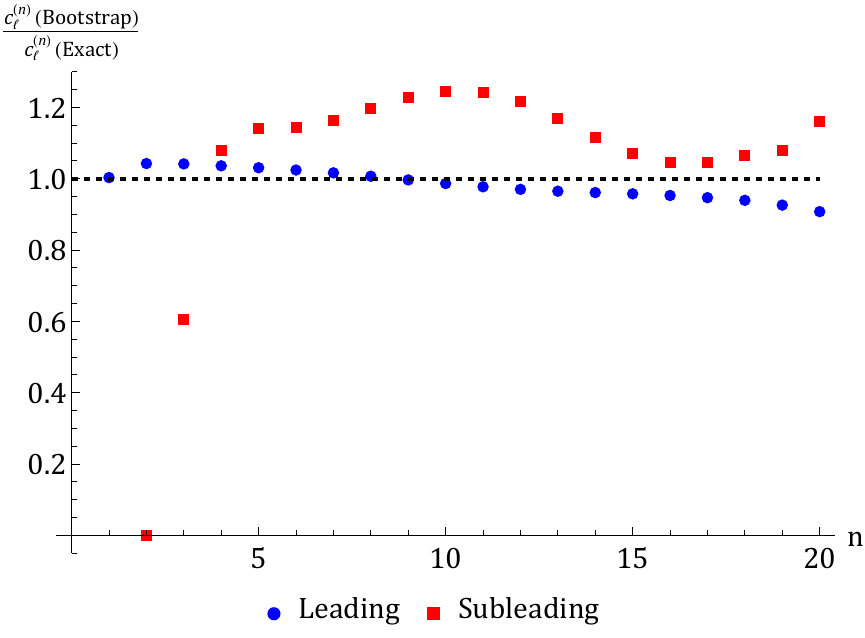} }}%
    \qquad
    \subfloat[\centering Minimization of $\int_1^\Lambda ds\, \frac{\Delta\mathcal{E}}{s^4}$.]{{\includegraphics[width=8cm]{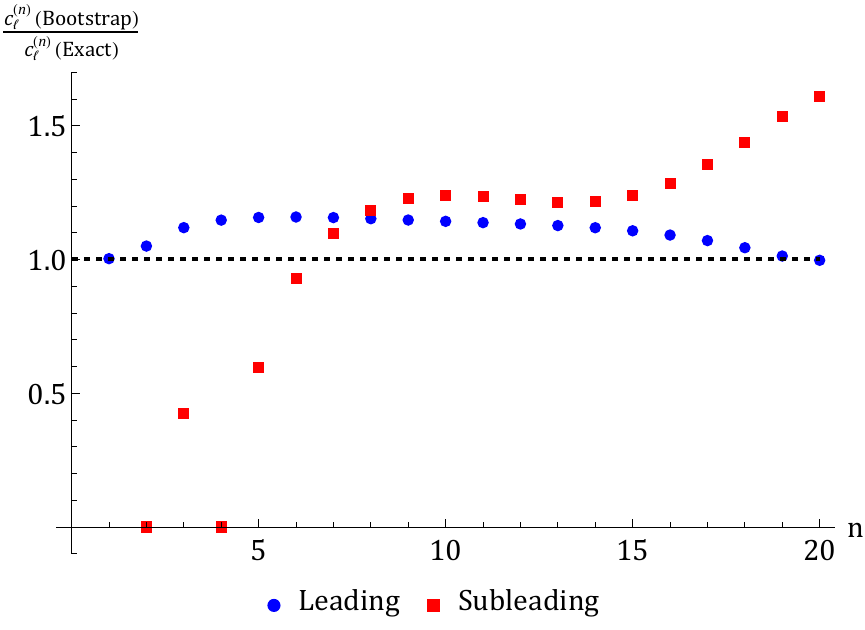} }}%
    \caption{Ratio of $c_\ell^{(n)}$ from bootstrap to the $c_\ell^{(n)}$ from the exact solution for closed-string.}%
   \label{ClosedStringcells}
\end{figure}

In fig.(\ref{ClosedString}) we show our results for the closed-string amplitude. We fixed $W_{10}$ and $W_{01}$ to the closed-string values: $W_{10}=-2\zeta(5), W_{01}=2\zeta(3)^2$ and used the same grid as the open-string. We used $N_{\text{max}}=30$ in $D=10$ with $\lambda=1.6$ as our seed. The tolerance we used was $T=10^{-6}$. As the plots indicate, the closed-string is obtained from minimization. Compared to the open-string case, here we expect the entangling power to be proportional to $s^4 \mathcal{N} Im {\mathcal M}$. The $s^4$ follows from the expectation that in supersymmetric closed-string theories, the leading correction is proportional to $Riemann^4$. The residue goes like $1/n^2$ for $n\gg 1$. Thus, the $1/s^3$ moment of the entangling power will be finite. However, for this class of amplitudes, the convergence of \eqref{clansatz} suggests that the $1/s^4$ moment is generally finite. We have considered both in the plots below. Both give comparable results. However, the convergence of the numerics is slower than what we found in the open-string case. In fig.(\ref{ClosedStringcells}) we show the our results for the ratio of $c_\ell^{(n)}$ from bootstrap to the $c_\ell^{(n)}$ from the exact solution for closed-string for the leading and subleading Regge trajectories. Ideally this ratio should be 1 and indeed we find nice agreement with 1 especially for the leading Regge trajectory.

\section{Non-Linear constraints via Machine Learning}
\label{sec:ML}
A thorny issue in the numerics that we have avoided addressing so far is how do we decide what tolerance to use for the magnitude of the $\lambda$-independence constraints.  One could take the approach that since we know at least one solution to our bootstrap problem - the string theory solution, we can compute how well the string solution (truncated to some $N_{max}$) satisfies the $\lambda$-independence constraints and use that to guide our estimate for the tolerance. Certainly, we want the string theory amplitude to be allowed, so the tolerance cannot be smaller than that dictated by the string solution. This approach however requires prior knowledge of the full string amplitude which somewhat defeats the purpose of bootstrap. In addition, since the magnitude of the amplitude may differ widely for various values of $s_1, s_2$, the derivatives (truncated to some $N_{max}$) themselves may differ in magnitude significantly, making a naive approach based on setting a small tolerance infeasible. 

One way to circumvent these difficulties is to consider ratios of the derivatives with respect to the amplitude. In this case, using a small tolerance is reasonable and one can impose ($0 < \epsilon << 1)$
\begin{equation}
\label{DerRatioCons}
   \left| \frac{1}{\mathcal{M}_{\lambda}(s_1,s_2)} \frac{\partial^k \mathcal{M}_{\lambda}(s_1,s_2)}{\partial\lambda^k} \right| \leq \epsilon, \quad \text{for } k =1,2,..., k_{max}
\end{equation}
Another approach could be to directly impose 
\begin{equation}
\label{AmpRatioCons}
\left | \frac{\mathcal{M}_{\lambda_1}(s_1,s_2)}{\mathcal{M}_{\lambda_2}(s_1,s_2)} -1 \right | \leq \epsilon \,,
\end{equation}
However, both the above constraints are non-linear in the parameters $c_{\ell}^{(n)}$ and not imposable via the usual linear (or semi-definite) optimization methods. One could try to recast them in a linear fashion but that would require knowledge of positivity properties of the amplitude. For example, if the amplitude were known to be positive at some $s_1,s_2$, we could impose $\left | \frac{\partial^k \mathcal{M}(s_1,s_2)}{\partial\lambda^k} \right| \le \epsilon ~ \mathcal{M}(s_1,s_2)$ or $\left | \mathcal{M}_{\lambda_1}(s_1,s_2) - \mathcal{M}_{\lambda_2}(s_1,s_2)\right| \le \epsilon ~ \mathcal{M}_{\lambda_2}(s_1,s_2)$. Since such positivity properties are not known in general, imposing such linear constraints would leave out a subspace of allowed amplitudes. 

In this section, we explore the use of Machine Learning methods to address this issue. Neural networks are well-known to be universal function approximators \cite{HORNIK1989359}. In particular, Physics-Informed Neural Networks (PINNs)\cite{RAISSI2019686} can be used to learn functions that satisfy a wide variety of constraints and have found several physics applications such as numerical learning of Calabi-Yau metrics \cite{CYML1,CYML2}, the conformal bootstrap, \cite{CBootML1,CBootML2,CBootML3,CBootML4,Trenta:2024nni,CBootML5}, and scattering in quantum mechanics \cite{QMML} among others. Recently, PINNs have also been implemented for the S-matrix bootstrap to constraint the phase of the scattering amplitude when the modulus is given \cite{SbootML1,SbootML2}. PINNs are characterized by the fact that the constraints equations are directly incorporated into the loss function and as the network trains to minimize the loss function, it approximates the true solution to the constraint equations better and better. The use of PINNs to do bootstrap could be a very promising direction. Most notably, this approach removes the need to make an ansatz to set up the numerical S-matrix bootstrap, which can prove to be very helpful in cases where a suitable ansatz is hard to write down. In our case, we will employ PINNs to learn the scattering amplitudes ($c_{\ell}^{(n)}$s) that satisfy the non-linear constraints \eqref{DerRatioCons} and \eqref{AmpRatioCons} which are otherwise not possible to impose via linear optimization methods. 

\subsection{Machine Learning Implementation}
\begin{figure}[H]
    \centering
    \includegraphics[width=12cm]{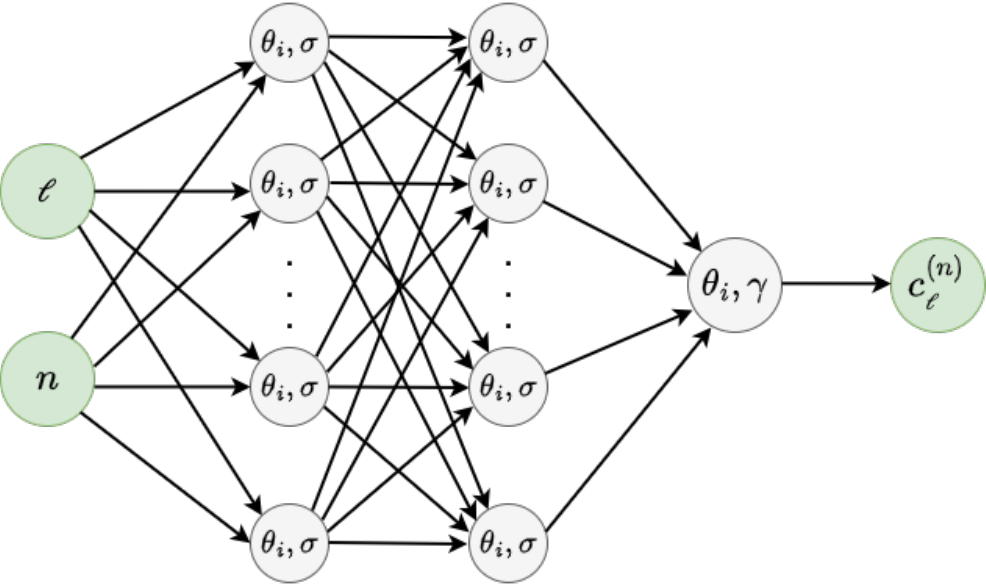} 
    \caption{The input layer has 2 neurons, one each for $\ell$ and $n$ values. We use 2 and 3 hidden layers of size 64 and an output layer of size 1. Each neuron in the layers has its own set of weights and biases, $\theta_i$ and a ReLU activation function denoted by $\sigma$. Final layer has the SoftPlus activation function $\gamma$ that ensures positivity of the output $c_{\ell}^{(n)}$ values.}
   \label{fig:OpenStringPINN}
\end{figure}

A neural network is basically an input-to-output map with several tunable parameters. In our case, we use a fully-connected, feed-forward  neural network that we implement using \textit{PyTorch} \cite{PyTorch}. Since we want to learn a two-variable function $c_{\ell}^{(n)}$, the input layer of our neural network has size 2 (see fig. \ref{fig:OpenStringPINN}). Let us pause here and elaborate a bit more as to why this is a useful thing to do. One may argue that instead of learning this two-variable function $c_{\ell}^{(n)}$, why don't we directly optimize using an optimizer? There are two points to make here: a) We want to examine Machine-Learning methods for bootstrap and more importantly for the future b) Finding this two-variable function will enable us to go into the complex $\ell$ plane treating $t=n$ as a continuous variable and examine Regge theory. A preliminary attempt towards this end is given below.

Each input corresponds to a specific neuron in the input layer; so we have a neuron for each of the $\ell$ and $n$ inputs. The input layers are followed by hidden layers of neurons. The number of hidden layers required and their sizes depend on the complexity of the function that the neural network is trying to learn. We use 2 hidden layers, both of size 64.  Each connection between neurons has associated tunable parameters called weights and biases and outputs a weighted sum of its inputs plus a bias. The output is then acted upon by an activation function before being passed on to the next layer. The activation functions add non-linearity to the neural network and are crucial for it to learn complex relationships between inputs and outputs. We use the very commonly used ReLU (Rectified Linear Unit) activation function defined as $\sigma(x)=max(0,x)$. ReLU is computationally simple, leading to a faster training process. The final layer is the output layer. In our case, it just has size 1 and gives the learned $c_{\ell}^{(n)}$ values for given input $\ell,n$ values. To ensure that the $c_{\ell}^{(n)}$ values our neural network outputs are always positive, as required by unitarity, we pass the output through the Softplus activation function defined as $\gamma(x) = \log(1+ e^x)$.  We give the full details of the various hyperparameters of our neural network in Table \ref{tab:HyperTable}.

PINNs are a sub-class of neural networks where the loss function includes some physical constraint equations. The true solution to the constraint equations is learned iteratively as the PINN updates its parameters (weights and biases) every iteration (called an epoch) via the Gradient descent method to approach the minimum of the loss function. In our case, the positivity constraint on the $c_{\ell}^{(n)}$s is already imposed as the neural network output is passed via the Softplus activation function. The constraints that remain left to impose are fixing $W_{10}$ and $W_{01}$ to the open string values, and the non-linear $\lambda$-independence constraints. The task of PINN is to maximize $W_{00}$ given these constraints. This can be achieved via the following loss function:
\begin{equation}
    \label{Loss_Func}
    \mathcal{L} = - W_{00} + \beta_1 \bigg{(}W_{10} - (-\zeta(3))\bigg{)}^2 + \beta_2 \left(W_{01} - \frac{7\zeta(4)}{4}\right)^2 + \beta_3 \frac{1}{C} \sum_{s_1,s_2 \in \mathcal{C}} \tilde{\mathcal{L}}(s_1,s_2)
\end{equation}
Here, $\tilde{\mathcal{L}}(s_1,s_2)$ are the $\lambda$-independence constraints imposed on a set $\mathcal{C}$ of $s_1,s_2$ points of length $C$. $\beta_i$s are positive weights. As different terms compete with each other during the minimization of the loss function, we can tune the magnitude of $\beta_i$s according to the tolerance required to satisfy the respective constraints they multiply. By construction, $max(W_{00}) = min(\mathcal{L})$; thus, as the PINN trains to minimize loss, it also learns the $c_{\ell}^{(n)}$s that correspond to $max(W_{00})$ and satisfy all constraints.  

\begin{table}[hbt!]
    \centering
        \begin{tabular}{|c|c|} 
            \hline
             Input \textbar{} Hidden \textbar{} Output & 2 \textbar{} 64, ..., 64 \textbar{} 1  \\
            \hline
            Activation Function & ReLU \\
            \hline
            Final Layer & Softplus \\
            \hline
            Optimizer & Adam \\
            \hline
            Learning Rate &  $10^{-4}$ \\
            \hline
        \end{tabular}
    \caption{Hyperparameters for the network.}
    \label{tab:HyperTable}
\end{table}

\subsection{Bootstrap results for Open String}
\begin{figure}[H]
    \centering
    \begin{subfigure}{0.45\textwidth}
        \includegraphics[width=8cm]{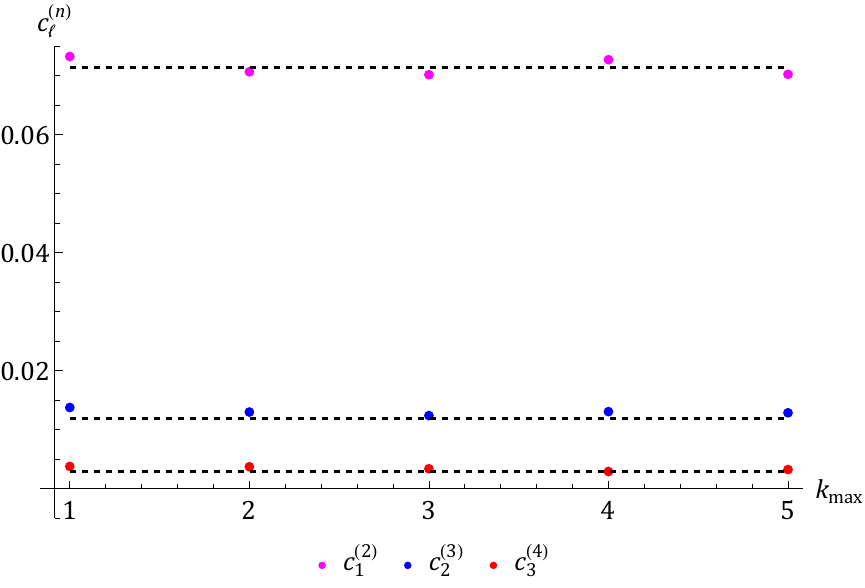}
        \caption{Convergence with $k_{max}$.}
        \label{derivconvML}
    \end{subfigure}
    \qquad
    \begin{subfigure}{0.45\textwidth}
        \includegraphics[width=8cm]{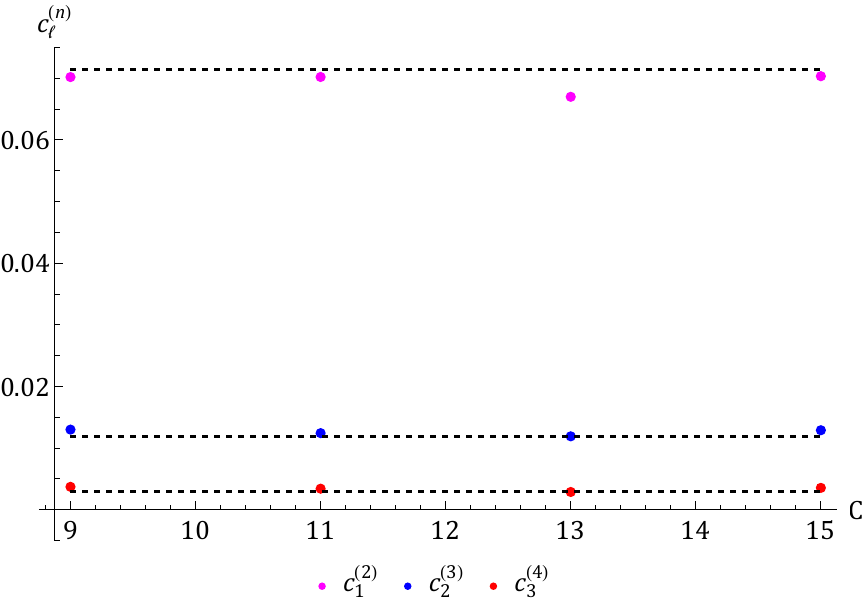}
        \caption{Convergence with $C$.}
        \label{consconvML}
    \end{subfigure}
    \caption{Convergence of the $c_\ell^{(n)}$s on the leading Regge trajectory with $k_{max}$ and $C$. The black, dashed lines denote the exact values of the corresponding $c_\ell^{(n)}$ from tree-level open superstring theory.}%
   \label{fig:convML}
\end{figure}

We now present the results from imposing the non-linear constraints in \eqref{AmpRatioCons} and \eqref{DerRatioCons} via the PINN implementation discussed above. 
\begin{itemize}
    \item \textbf{Case 1:} When $\tilde{\mathcal{L}}(s_1,s_2) =\left(1 - \frac{\mathcal{M}_{\lambda_1}(s_1,s_2)}{\mathcal{M}_{\lambda_2}(s_1,s_2)}\right)^2$. \\
    We choose $\lambda = 5.6, N_{max} = 13, \lambda_1 = 5.1, \lambda_2 = 6.1$. We also pick $\beta_1 = \beta_2 = \beta_3 = 10^4$. We impose $\lambda$-independence on the following set of points
    \begin{equation}
        \mathcal{C} = \{ (s_1, s_2)\, |\, -5.5 \leq s_1 \leq 5.5 \, , -0.2\leq s_2 \leq 0.2  \, , \D_{s_1} = 1, \D_{s_2} = 0.4\}
    \end{equation}   
    \\
     With these parameters, we train the neural network for $2 \times 10^5$ epochs. In practice, we do this multiple times and then pick the solution for which the product of all the constraints is the smallest. This leads us to the following solution.
    \begin{equation}
    \begin{split}
        max(W_{00}) = -1.506 \, &, \quad  \,  W_{10}-(-\zeta(3)) = 2.8 \times 10^{-6}  \\
        W_{01}-\frac{7\zeta(4)}{4} = - 5.2 \times 10^{-5} \, &, \quad  \, \tilde{\mathcal{L}}_{mean}  = 8.5 \times 10^{-4}. 
    \end{split}
    \end{equation}
    where we have defined $\tilde{\mathcal{L}}_{mean}=\left(\frac{1}{C} \sum_{s_1,s_2 \in \mathcal{C}} \tilde{\mathcal{L}}(s_1,s_2)\right)^{\frac{1}{2}}$. For the leading Regge trajectory, the solution compares to the open string values as follows:
\begin{table}[H]
    \centering
    \begin{tabular}{|c|c|c|c|c|c|c|} 
        \hline
        & $c_{0}^{(1)}$ & $c_{1}^{(2)}$ & $c_{2}^{(3)}$ & $c_{3}^{(4)}$ & $c_{4}^{(5)}$ & $c_{5}^{(6)}$ \\ 
        \hline
        Open String  & 1 & 0.0714 & 0.0119  & 0.00289 & 0.000867 & 0.000300\\ 
        \hline
        PINN & 0.999 & 0.0715 & 0.0121  & 0.00300  & 0.000922 & 0.000332 \\ 
        \hline
    \end{tabular}
    \label{tab:sample1}
\end{table}

    \item \textbf{Case 2:} When $\tilde{\mathcal{L}}(s_1,s_2) = \sum_{k = 1}^{k_{max}} \left( \frac{1}{\mathcal{M}_{\lambda}(s_1,s_2)} \frac{\partial^k \mathcal{M}_{\lambda}(s_1,s_2)}{\partial\lambda^k} \right)^2$. \\
    We will focus on the three derivatives case ($k_{max}=3$) and pick $\lambda = 14.6$, $N_{max} = 20$, $\beta_1 = \beta_2= 10^6$. We choose $\beta_3 = 10^8$ for $k=1$ whereas $\beta_3 = 10^6$ for $k=2,3$.  We use 3 hidden layers of size 64 each. We choose the following set of points for the $\lambda$-independence constraints:
    \begin{equation}
        \mathcal{C} = \{ (s_1, s_2)\, |\, 0.4 \leq s_1 \leq 10.4 \, ,  s_2 = 10.1  \, , \D_{s_1} = 1\}
    \end{equation}   
   Close to $s_1,s_2 \approx 10$, for $N_{max} =20$, the open string amplitude is very large and its first derivative is $O(1)$. For example, at $(s_1,s_2) = \left(10.4,10.1\right)$, the amplitude and its first derivative are $1.34 \times 10^5$ and $-2.51$ respectively, making an approach based on setting a small tolerance for the first derivative itself infeasible. Since the ratio is $O(10^{-5})$, we can impose the small ratio condition safely using the PINN implementation. \\
   Training for $3 \times 10^5$ epochs gives us the following solution (again selected by looking for the solution with the smallest product of the constraint terms)
   \begin{equation}
    \begin{split}
        max(W_{00}) = -1.601 \, &, \quad  \,  W_{10}-(-\zeta(3)) = 9.3 \times 10^{-8}  \\
        W_{01}-\frac{7\zeta(4)}{4} = 5.2 \times 10^{-8} \, &, \quad  \, \tilde{\mathcal{L}}_{mean}  = 9.8\times10^{-6}. 
    \end{split}
    \end{equation}
Here we have defined $\tilde{\mathcal{L}}_{mean}=\frac{1}{k_{max}}\sum_{k = 1}^{k_{max}}\left(\frac{1}{C} \sum_{s_1,s_2 \in \mathcal{C}} \left( \frac{1}{\mathcal{M}_{\lambda}(s_1,s_2)} \frac{\partial^k \mathcal{M}_{\lambda}(s_1,s_2)}{\partial\lambda^k} \right)^2\right)^{\frac{1}{2}}$. 

The solution for the leading Regge trajectory is as follows:
\begin{table}[H]
    \centering
    \begin{tabular}{|c|c|c|c|c|c|c|} 
        \hline
        & $c_{0}^{(1)}$ & $c_{1}^{(2)}$ & $c_{2}^{(3)}$ & $c_{3}^{(4)}$ & $c_{4}^{(5)}$ & $c_{5}^{(6)}$ \\ 
        \hline
        Open String  & 1 & 0.0714 & 0.0119  & 0.00289 & 0.000867 & 0.000300\\ 
        \hline
        PINN & 0.998 & 0.0702 & 0.0124  & 0.00341  & 0.000990 & 0.000339 \\ 
        \hline
    \end{tabular}
    \label{tab:sample2}
\end{table}
\end{itemize}
In fig.(\ref{fig:convML}) we show the convergence of values of the coefficients on the leading Regge trajectory, specifically $c_{1}^{(2)},c_{2}^{(3)}$ and $c_{3}^{(4)}$ with the number of derivatives $k_{max}$ and the number of $\lambda$-independence constraints at each derivative order $C$. In fig.(\ref{derivconvML}) we set $C=11$ and vary $k_{max}$ from $1$ to $5$. We find that the coefficients on the leading Regge trajectory have nicely converged to the value obtained from tree-level open superstring theory. In fig.(\ref{consconvML}) we set $k_{max}=3$ and vary $C$ from $9$ to $15$. Again we find convergence of the values of the coefficients on the leading Regge trajectory to the value obtained from tree-level open superstring theory.
\subsection{Regge analysis from Machine Learning}
\begin{figure}[H]
    \centering
    \includegraphics[width= 0.6\linewidth]{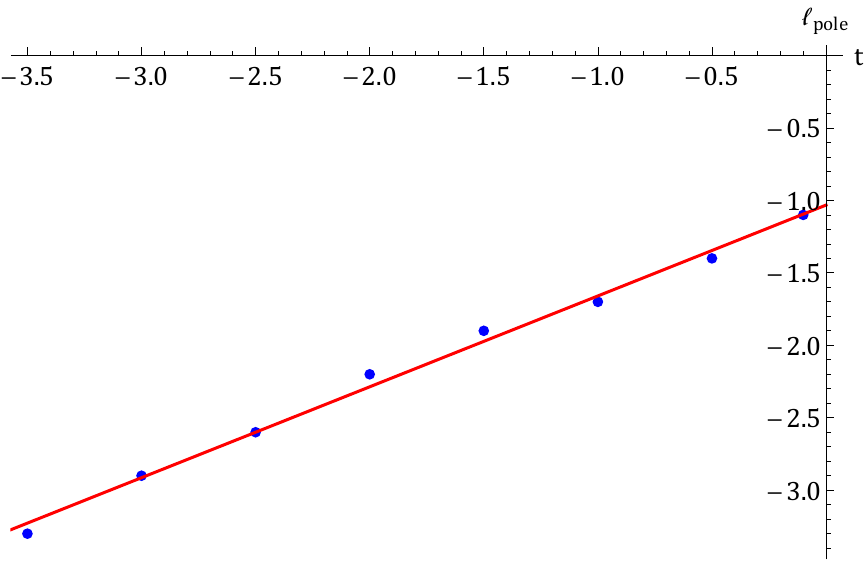}
    \caption{$\ell_\text{pole}$ versus $t$ from machine learning. The blue dots denote the location of the Regge pole $\ell_\text{pole}$ of $c(\ell,t)$ as a function $t$. We find that these blue dots lie on the red straight line obtained by making a fit with our data. The intercept of this straight line $\approx -1$ hence matching the value of the Regge intercept for the open superstring amplitude.}
    \label{fig:ReggePoleML}
\end{figure}
An advantage of implementing the neural network in the way described above is that we are learning solutions - $c_{\ell}^{(n)}$s that are continuous functions of $\ell$ and $n$. This allows us to examine whether the neural network also learns the leading Regge pole. To highlight the continuous nature of the solutions, we will use the notation $c(\ell, t) \equiv c_{\ell}^{(n)}$.  Our working definition for the Regge pole is the following.  We make a data table of $\frac{1}{|c(\ell,t)|}$ as a function of $\ell$ for fixed $t < 0$ and interpret the global minimum as the Regge pole of $c(\ell,t)$. We repeat this for several $t<0$ and plot $\ell_{\text{pole}}$ versus $t$. The results are shown in fig.(\ref{fig:ReggePoleML}). Remarkably, we find that this curve is a straight line. We interpret this as the linear Regge trajectory
\begin{equation}
    \ell_\text{pole}=\alpha_0+\alpha't
\end{equation} 
where $\alpha_0$ is the Regge intercept and $\alpha'$ is the Regge pole. Our fit gives the values
\begin{equation}
    \alpha_0=-1.03\:, \qquad\alpha'=0.63\:.
\end{equation}
Note that Regge intercept $\alpha_0$ matches the leading Regge intercept of the open superstring amplitude for which we have 
\begin{equation}
    \left(\ell_\text{pole}\right)_{\text{open-string}}=-1+t\:.
\end{equation}
We leave a more systematic examination of Regge theory using these ideas for future work. Our findings suggest that this is an interesting direction to pursue.

\section{A unified dispersion relation}
\label{sec:parametricLCSDR}
\begin{figure}[H]
    \centering
    \includegraphics[width= 0.4\linewidth]{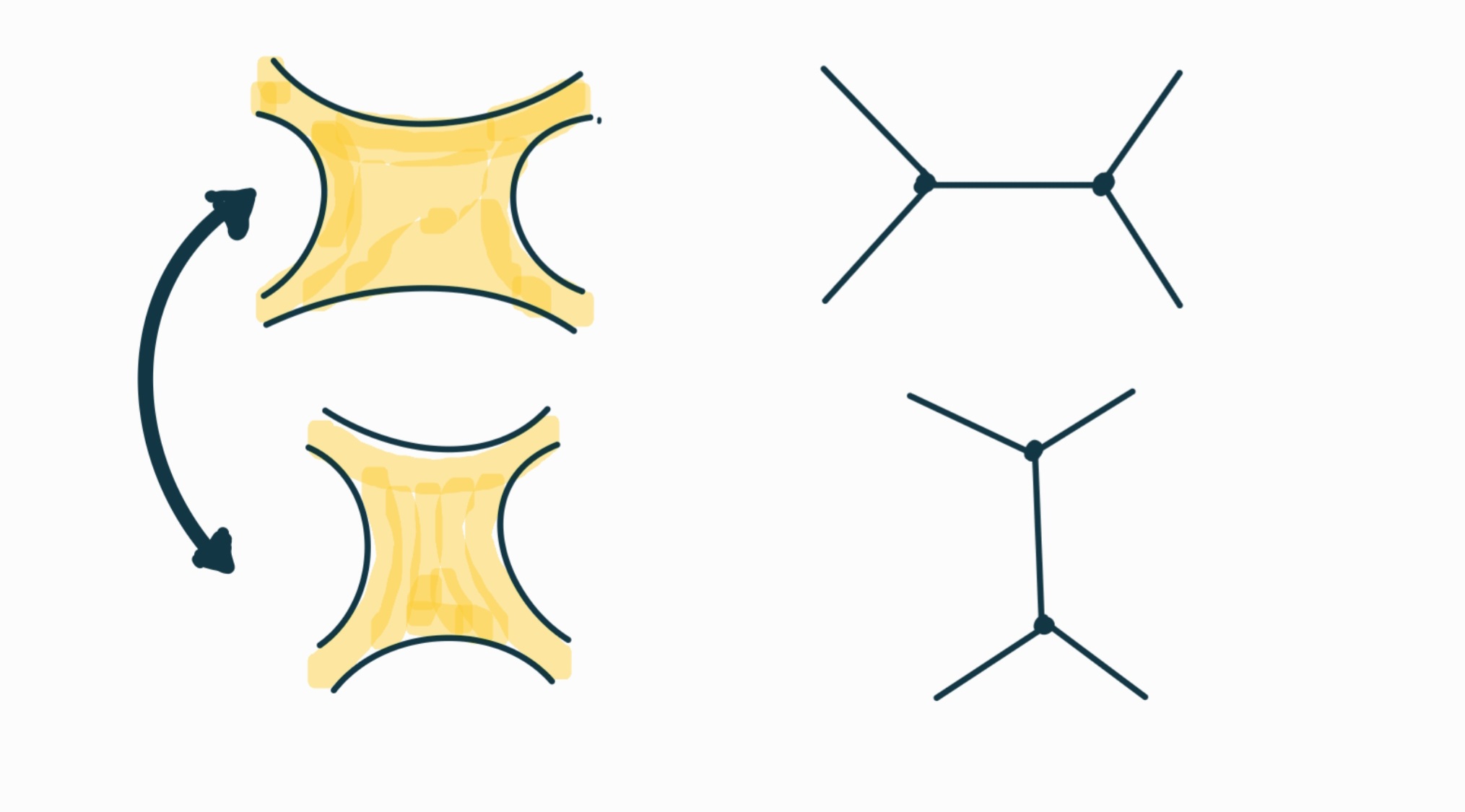}
    \caption{The worldsheet  duality between channels.}
    \label{fig:wsdual}
\end{figure}

Before we conclude, we would like to give a unified dispersion relation involving the parameter $\lambda$. We will focus on the 2-channel symmetric case; a similar analysis can be made for the 3-channel case and will be presented elsewhere. The motivational question we ask is the following. In fig.\ref{fig:wsdual}, we exhibit pictorially the famous dual resonance hypothesis, where in the string picture, the two channels are equivalent descriptions, getting deformed into each other. Is there a way that the fixed-$s$ dispersive representation can be deformed into the fixed-$t$ one?  We will show that different limits of $\lambda$ give various known dispersive representations giving the cartoon in the figure below. 
\begin{figure}[H]
    \centering
    \includegraphics[width= 0.4\linewidth]{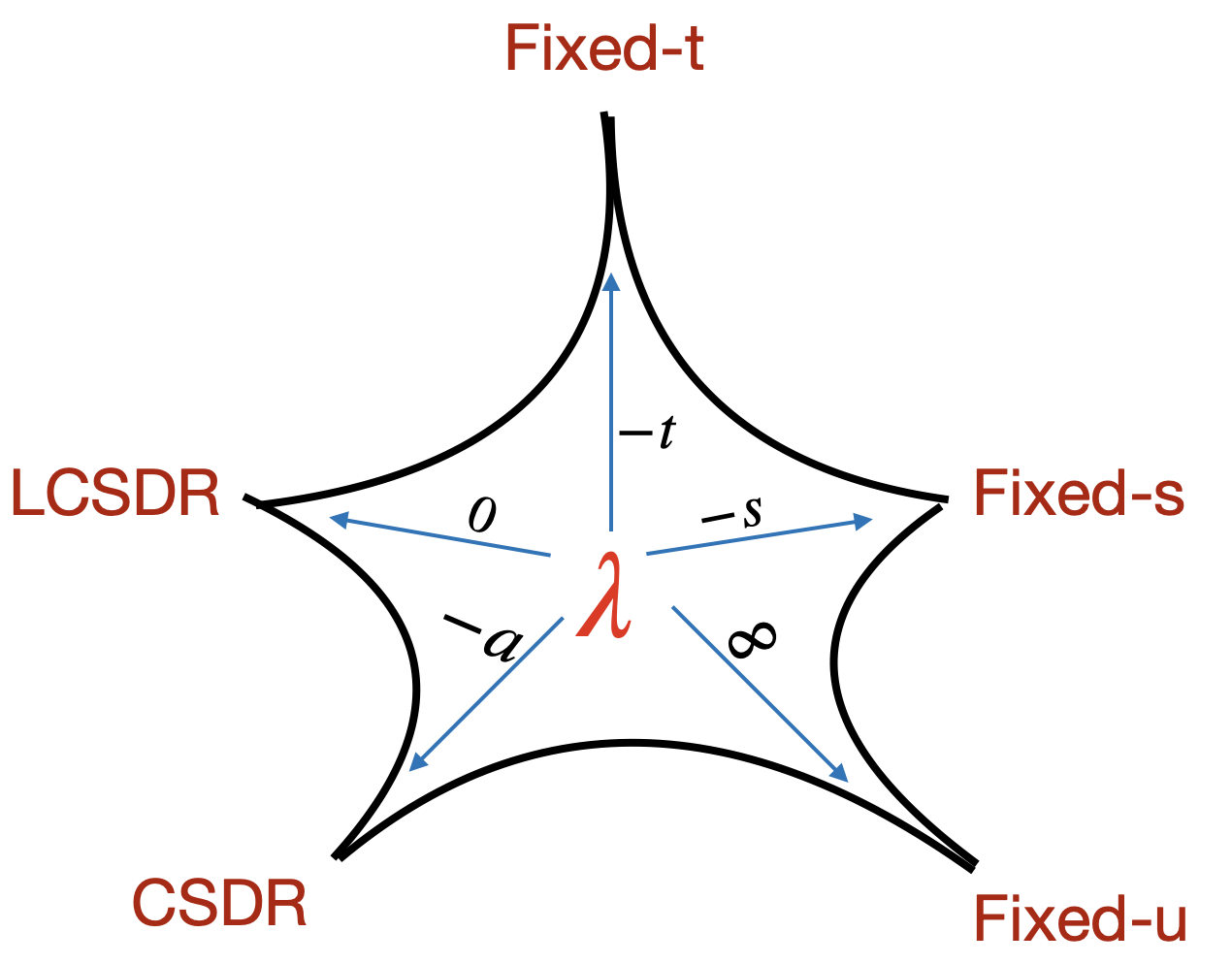}
    \caption{The ``web" of dispersion relations related by limits of $\lambda$.}
    \label{fig:mdisp}
\end{figure}

Given a function $\mathcal{M}(s_1,s_2)$ which is $s_1,s_2$ symmetric, satisfies $\displaystyle \lim_{s_1 \rightarrow \infty} \dfrac{\mathcal{M}(s_1,s_2)}{s_1} = 0$ for some fixed $s_2$ and has branch cuts (or poles) along $s_1 \in (s_0, \infty)$ and no other singularities, we can represent it via the following dispersion relation
\begin{equation}
	\mathcal{M}\left(s_{1}, s_{2}\right) = \mathcal{M}\left(0,0\right)+ \frac{1}{\pi}\int_{s_{0}}^{\infty}\mathrm{d}\sigma\left[\frac{1}{\sigma-s_{1}}+\frac{1}{\sigma-s_{2}}-\frac{1}{\sigma}\right]\mathcal{A}^{(s_{1})}\left(\sigma,\frac{y}{\sigma}\right) - \frac{1}{\pi}\int_{s_{0}}^{\infty}\frac{\mathrm{d}\sigma}{\sigma}\mathcal{A}^{(s_{1})}\left(\sigma,0\right)\:.
\end{equation}
where $\mathcal{A}^{(s_1)}\left(s_1,s_2\right)$ is the $s_1$-channel discontinuity of $\mathcal{M}\left(s_{1}, s_{2}\right)$ along the branch cut. Lets apply the above dispersion relation to the function $\mathcal{M}\left(s_{1}-\alpha, s_{2}-\alpha\right)$ thinking of it as a function of $s_1$ and $s_2$. We can write,
\begin{equation}
\begin{split}
	\mathcal{M}\left(s_{1}-\alpha, s_{2}-\alpha\right) = &\:\mathcal{M}\left(-\alpha,-\alpha \right) - \frac{1}{\pi}\int_{s_{0}+\alpha}^{\infty}\frac{\mathrm{d}\sigma}{\sigma}\mathcal{A}^{(s_{1})}\left(\sigma-\alpha,-\alpha\right)\\&+ \frac{1}{\pi}\int_{s_{0}+\alpha}^{\infty}\mathrm{d}\sigma\left[\frac{1}{\sigma-s_{1}}+\frac{1}{\sigma-s_{2}}-\frac{1}{\sigma}\right]\mathcal{A}^{(s_{1})}\left(\sigma-\alpha,\frac{s_1s_2}{\sigma}-\alpha\right)\:.
\end{split}
\end{equation}
Making the change of variable of integration $\sigma\rightarrow\sigma+\alpha$ we obtain
\begin{equation}
\label{varchange}
\begin{split}
	\mathcal{M}\left(s_{1}-\alpha, s_{2}-\alpha\right) = &\:\mathcal{M}\left(-\alpha,-\alpha \right) - \frac{1}{\pi}\int_{s_{0}}^{\infty}\frac{\mathrm{d}\sigma}{\sigma+\alpha}\mathcal{A}^{(s_{1})}\left(\sigma,-\alpha\right)\\&+ \frac{1}{\pi}\int_{s_{0}}^{\infty}\mathrm{d}\sigma\left[\frac{1}{\sigma+\alpha-s_{1}}+\frac{1}{\sigma+\alpha-s_{2}}-\frac{1}{\sigma+\alpha}\right]\mathcal{A}^{(s_{1})}\left(\sigma,\frac{s_1s_2}{\sigma+\alpha}-\alpha\right)\:.
\end{split}
\end{equation}
The \textit{lhs} of \eqref{varchange} is invariant under the shifts $s_1\rightarrow s_1+\lambda$, $s_2\rightarrow s_2+\lambda$ and $\alpha\rightarrow \alpha+\lambda$ which means that
\begin{equation}
\begin{split}
	\mathcal{M}\left(s_{1}-\alpha, s_{2}-\alpha\right) = &\:\mathcal{M}\left(-\alpha-\lambda,-\alpha-\lambda\right) - \frac{1}{\pi}\int_{s_{0}}^{\infty}\frac{\mathrm{d}\sigma}{\sigma+\alpha+\lambda}\mathcal{A}^{(s_{1})}\left(\sigma,-\alpha-\lambda\right)\\&+ \frac{1}{\pi}\int_{s_{0}}^{\infty}\mathrm{d}\sigma\left[\frac{1}{\sigma+\alpha-s_{1}}+\frac{1}{\sigma+\alpha-s_{2}}-\frac{1}{\sigma+\alpha+\lambda}\right]\\&\hspace{1.5cm}\times\mathcal{A}^{(s_{1})}\left(\sigma,\frac{(s_1+\lambda)(s_2+\lambda)}{\sigma+\alpha+\lambda}-\alpha-\lambda\right)\:.
\end{split}
\end{equation}
Putting $\alpha=0$ in the above equation we obtain \eqref{Gen_Sing_LCSDR}
\begin{equation}\label{gendisp}
\begin{split}
    \mathcal{M}(s_1,s_2)=&\:\mathcal{M}(-\lambda,-\lambda) - \frac{1}{\pi} \int_{s_0}^{\infty} \mathrm{d}\sigma\frac{\mathcal{A}^{(s_{1})}\left(\sigma,- \lambda \right)}{\sigma + \lambda} \\
    &+\frac{1}{\pi} \int_{s_0}^{\infty} \mathrm{d}\sigma  \left[\frac{1}{\sigma -s_1}+\frac{1}{\sigma - s_2}- \frac{1}{\sigma + \lambda}\right] \mathcal{A}^{(s_{1})}\left(\sigma,\frac{(s_1+\lambda)(s_2 + \lambda)}{\sigma + \lambda} - \lambda \right)\:.
\end{split}
\end{equation}
We perform a check of this dispersion relation using the massless box diagram in the appendix \ref{masslessbox}. 
Despite appearances, the above representation is $\lambda$ independent; the role of $\lambda$ is to enable a truncation of the upper limit of the integral, which may be desirable in numerical bootstrap. In case, the function satisfies $\displaystyle \lim_{s_1 \rightarrow \infty} \mathcal{M}(s_1,s_2) = 0$, the first line above vanishes and we get a representation without any additional constant pieces. Note that choosing $\lambda=-s_2$ leads to the usual fixed-$s_2$ dispersion relation, which is another use of the $\lambda$-dependence. Since the fixed-$s_2$ dispersion relation and crossing symmetry together lead to the so-called null constraints or locality constraints, we expect that statement of $\lambda$-independence is equivalent to that of null or locality constraints. Similarly, taking $\lambda=-s_1$ leads to the fixed-$s_1$ dispersion relation, while taking $\lambda\rightarrow \infty$ gives the fixed-$u$ dispersion relation. Curiously, setting $\lambda=-a\equiv -s_1 s_2/(s_1+s_2)$ gives us back the (nonlocal) crossing-symmetric dispersion relation (up to the locality constraints) while $\lambda=0$ gives the local crossing-symmetric dispersion relation \cite{Saha:2024qpt}. These limits give the web in fig.\ref{fig:mdisp}. It also makes it tempting to conjecture a relation to a potential worldsheet description, which we now briefly (and hesitatingly!) elaborate.

The worldsheet description enables us to visualize various individual channels as limits of different ways of pinching the worldsheet as in fig.\ref{fig:wsdual}. However, we also know from string field theory \cite{Polchinski:1998rq} that we can also have a Feynman diagram decomposition if we put in the appropriate contact terms. The $\lambda$-dependent dispersion relation has precisely this flavour as is suggested by fig.\ref{fig:mdisp}. However, we do not have a first principles proof that various corners of the web are accessible by dialing $\lambda$ appropriately. For instance, in the string theory tree-level example, the convergence of the mass-level sum features $\lambda$ (which for the superstring case gives the condition $\text{Re}(\lambda)>-1$). Could convergence issues rule out the interpolation between various corners? This is related to establishing a rigorous analyticity domain which enables eq.(\ref{gendisp}) to converge. Examining how and when an underlying string worldsheet description emerges from the bootstrap is a very interesting open problem for the future.

\section{Discussion}
\label{sec:conclusion}

We will briefly discuss the connection between the present approach and several previous papers. 
In the case of category I amplitudes, we noted that $W_{00} = \frac{1}{\pi}\int_{s_{0}}^{\infty}\frac{\mathrm{d}\sigma}{\sigma}\mathcal{A}^{(s_{1})}\left(\sigma,0\right)$. This condition can be contrasted with the superpolynomial softness of the amplitude mentioned in \cite{Haring:2023zwu}, which states that there always exists $t<0$ such that $j\left(t\right)<-N$, for any positive integer $N$. Here, at high energy with fixed momentum transfer, the amplitude is assumed to grow as $s^{j\left(t\right)}$. In our case, we will require $j\left(0\right)<0$, \textit{i.e.} even in the forward limit amplitude should decay with increasing energy. The open-string amplitude which behaves as $s^{-1+t}$ at large $s$ and fixed $t$ is a very good example of this. In addition, our analysis is complementary to that followed in \cite{Berman:2024wyt, Albert:2024yap}. In our case, the spectrum is given as an input and we fix two Wilson coefficients to determine the partial-wave coefficients. The open-string S-matrix lies at $\left(W_{10}=-\zeta\left(3\right), W_{01}=\frac{7}{4}\zeta\left(4\right)\right)$ in fig.(\ref{fig:dualitycontourplotCat1}). In \cite{Berman:2024wyt, Albert:2024yap} low-energy expansion of the amplitude is used and the Wilson coefficients are constrained by certain assumptions---for {\it e.g.,} mass gap is present between any two states in the spectrum and the lowest massive state is a scalar. By tuning the scalar coupling to the string value, the authors could carve out narrow regions near the boundary of the allowed region in the space of Wilson coefficients, which corresponds to the open-string S-matrix. \eqref{gen-disp-1} renders a broader framework of S-matrices which do not admit duality or unsubtracted fixed-$t$ dispersion relation in general. We have classified such amplitudes as category II. In this category there exist some amplitudes which admit single-channel expansion only in certain limited domains, one class being scalar deformations to string amplitudes. These amplitudes also have infinite tower of Regge trajectories, which is an essential condition for having Regge behavior \cite{Eckner:2024pqt}.

The main point of view that we have advocated in this paper is the following.

{\it It is worth considering the dual resonance hypothesis in an approximate sense, which may enable us to make better contact with experiments.}

See \cite{yellin} for an old critical review of the duality hypothesis and phenomenological arguments why duality could be ``badly broken" in nature.

Of course, this opens up several questions, some of which we list as promising future directions:

\begin{itemize}
\item It will be interesting to re-examine the hadron scattering data using a basis of functions, motivated by the considerations in this paper. As we have summarized in the introduction, if one supplements the so-called interference models of the 1960s with contact terms, one would get a basis that would manifest crossing symmetry but still maintain the desirable features of the string model. The single channel analysis of \cite{DHS} could only capture the asymptotic features of the amplitude; perhaps with our new basis, the Breit-Wigner type features could also be captured, leading to a more faithful contact with experiments.

\item It is extremely unlikely that the entire space of tree-level S-matrices that we have found using the techniques here, will meet all the requirements to come from consistent theories. For instance, in \cite{nimacons}, five-particle consistency was used to rule out certain analogs of string models which satisfied the dual resonance condition \cite{Cheung:2023uwn}. It is quite likely that similar considerations will rule out swathes of theory space. It is a challenge however, how to go about considering five particle versions of the 2-2 S-matrices found in this paper. 

\item Perhaps a more do-able question is to set up a study of  perturbative unitarity. In this case, we would need to consider a coupling parameter, which would give an imaginary part to the resonances. One could conceive of an analysis similar to \cite{gu1,gu2} but starting with the S-matrices found in this paper. Perhaps demanding the emerging non-analyticity due to one-loop corrections to be of a specific kind will enable us to rule out theory space.

\item If entanglement minimisation is supposed to pick out the open string amplitude exactly, it will be interesting to find a good analytical explanation for the same. One can also look at the corresponding dual optimization problem to the primal problem we set up in this study. We did this and found that the dual problem converges much faster and matches the primal bounds to high numerical accuracy. Since the duality gap is zero in this case, one can also reconstruct the corresponding extremal primal solution from the dual solution using the complementary slackness conditions. 

\item We demonstrated that Machine Learning methods can prove very useful while imposing non-linear constraints that traditional linear or semi-definite optimization methods fail at. In particular, the approach via PINNs may be indispensable in cases when its hard to write down an ansatz for the amplitude to set up the numerical bootstrap. Since the tree-level string amplitude has an analytic structure which is just a series of poles, there was a natural choice of ansatz for the residue at the poles in terms of Gegenbauers. Beyond tree-level however, branch cuts appear. In such cases, one can use a PINN as an ansatz for the absorptive part of the amplitude. It will be interesting to explore this direction.

\item While PINNs have the advantage of being very general and therefore allow one to impose a wide variety of physical constraints, its very hard to achieve the kind of numerical accuracy that more conventional numerical methods can achieve. This could prove to be a drawback in some high-precision bootstrap studies. It will be very helpful to find ways to make PINNs more accurate without prohibitively slowing down the convergence rate. 

\item{Double copy:}
Finally, let us comment on the prospect of having the double-copy KLT type relation \cite{KAWAI19861, Polchinski:1998rq} level-wise.
Let us consider the closed string amplitude in \eqref{clst-1para}. Near the poles, $s_{1}=n$ and $s_{2}=n$, for $n\ge 1$, the amplitude takes the form
\begin{equation}
	\mathcal{M}_{\text{cl}}\left(s_{1}\rightarrow n, s_{2}\rightarrow n\right) \approx \frac{1}{\left(n!\right)^{2}}\left(\frac{1}{s_{1}-n}+\frac{1}{s_{2}-n}\right)\left(1+n\right)_{n-1}^{2}.
\end{equation}
The above expression suggests that near the massive poles, closed and open string amplitudes are related as
\begin{equation}\label{doublecopy1}
	\mathcal{M}_{\text{cl}}\left(s_{1}\rightarrow n, s_{2}\rightarrow n\right) = \mathcal{M}_{\text{op}}\left(s_{1}\rightarrow n, s_{2}\rightarrow n\right)\frac{1}{\frac{1}{s_{1}-n}+\frac{1}{s_{2}-n}}\mathcal{M}_{\text{op}}\left(s_{1}\rightarrow n, s_{2}\rightarrow n\right).
\end{equation}
$\mathcal{M}_{op}$ denotes the open string amplitude given in \eqref{op-st}. We can also derive \eqref{doublecopy1} by expanding the double copy relation \cite{KAWAI19861,Polchinski:1998rq},
\begin{equation}
	\mathcal{M}_{\text{cl}}\left(s_{1}, s_{2}, s_{3}\right) = \mathcal{M}_{\text{op}}\left(s_{1}, s_{2}\right)\frac{\sin\left(\pi s_{1}\right)\sin\left(\pi s_{2}\right)}{\pi\sin\left(\pi\left(s_{1}+s_{2}\right)\right)}\mathcal{M}_{\text{op}}\left(s_{1}, s_{2}\right)
\end{equation}
around $s_{1}=n$, $s_{2}=n$. We leave a detailed study of the double copy relations from the parametric series representations of the open and closed string amplitudes, reported in this paper, for future work.

\end{itemize}

\section*{Acknowledgments} We thank David Gross, Shiraz Minwalla, Hjalmar Rosengren, Ashoke Sen,  Piotr Tourkine, and Spenta Wadia for useful discussions. AS thanks the participants of ``What is string theory: Weaving perspectives together," workshop at KITP, Santa Barbara and the S-matrix bootstrap 2024 conference at Reykjavik for useful discussions and encouragement. APS would like to thank the participants of ``Future Perspectives on QFT and Strings" conference held at IISER Pune and Eurostrings 2024 conference held at University of Southampton, where parts of this work were presented, for insightful comments. APS is grateful to Queen Mary University of London and Swansea University, and especially to Prem Kumar, Ricardo Monteiro, Carlos Nunez and Congkao Wen for warm hospitality and helpful discussions during the final stage of this work. FB thanks Sumanth Kumar for very helpful discussions about PINNs. AS acknowledges support from the SERB core grant CRG/2021/000873 and a Quantum Horizons Alberta chair professorship. APS is supported by the DST INSPIRE Faculty Fellowship (IFA22-PH 282). FB is supported by the Prime Minister's Research Fellowship (PMRF). This research was supported in part by grant NSF PHY-1748958 to the Kavli Institute for Theoretical Physics (KITP).

\appendix
\section{Two-channel symmetric dispersion relation}
We briefly review the dispersion relation \cite{Raman:2021pkf} for amplitudes that are symmetric in two channels, say $s$ and $t$. Let us first define the following variables,
\begin{equation}
	s_{1} = s - \frac{\mu}{3}, \qquad s_{2}=t-\frac{\mu}{3}, \qquad s_{3}=u-\frac{\mu}{3},
\end{equation}
where, $s+t+u=\mu$. We consider the amplitude to be a function of $s_{1}$ and $s_{2}$, so in this case $u$ and analogously $s_{3}$ is held fixed. 

Using the following parametrization we can map $s_{1}$ and $s_{2}$ to complex plane,
\begin{equation}\label{2ch-parametric}
	s_{k} = a\left[1-\frac{\left(z+z_{k}\right)^{2}}{\left(z-z_{k}\right)^{2}}\right] = \frac{4 a z e^{i \pi  k} }{\left(z+e^{i \pi  k}\right)^2}, \qquad k= 1,2.
\end{equation} 
Using the above equation we can express the amplitude $\mathcal{M}\left(s_{1},s_{2}\right)$ as $\widetilde{\mathcal{M}}\left(z,a\right)$, where $a$ is held constant and is given by  
\begin{equation}
	a = \frac{s_{1}s_{2}}{s_{1}+s_{2}}.
\end{equation}
The above equation implies
\begin{equation}
	s_{2} = \frac{a s_{1}}{s_{1}-a}.
\end{equation}
Using \eqref{2ch-parametric} with $a$ kept fixed, we find
\begin{equation}\label{z-s1fn}
	z = \frac{s_{1}-2a\pm2i\sqrt{a\left(s_{1}-a\right)}}{s_{1}}.
\end{equation}
We choose the positive sign in the above equation to denote $z$ corresponding to $s_{1}$. Therefore, when $s_{1}\ge a$, $z$ lies on the unit circle. Thus, \eqref{2ch-parametric} maps the singularities of the amplitude to a unit circle in the complex plane. From \eqref{z-s1fn} we see that when $s_{1}=a$ implies $z=-1$ and when $s_{1}\rightarrow \infty$ we get $z\rightarrow 1$. For $a\le s_{1}<\infty$, $z$ can be expressed as 
\begin{equation}
	z = e^{i\phi}, \qquad \phi=\cos^{-1}\left(\frac{s_{1}-2a}{s_{1}}\right).
\end{equation}
Our goal is to obtain the amplitude from its discontinuity in a particular channel. For this purpose, the above parametrization will be useful. 

For later reference we define $x$ and $y$ variables as
\begin{equation}
	x= s_{1}+s_{2}, \qquad y=s_{1}s_{2}.
\end{equation}

\subsection{Dispersion relation}
We make the following assumptions about $\widetilde{\mathcal{M}}\left(z,a\right)$.
\begin{itemize}
    \item The singularities of the amplitude are located on the unit circle $|z|=1$.
    \item The amplitude satisfies the complex conjugation, $\left[\widetilde{\mathcal{M}}\left(z,a\right)\right]^{\ast}=\widetilde{\mathcal{M}}\left(\frac{1}{z^{\ast}},a\right)$.
    \item Since the amplitude is symmetric in $s_{1}$ and $s_{2}$, it must be a function of even powers of $z$.
\end{itemize}
Let us consider the following contour integration 
\begin{eqnarray}
	&& \frac{1}{2\pi i}\oint_{|z'|>1}\frac{\mathrm{d}z'}{z'-z}\frac{z^{'2}-1}{z^{'2}}\widetilde{\mathcal{M}}\left(z',a\right) - \frac{1}{2\pi i}\oint_{|z'|<1}\frac{\mathrm{d}z'}{z'-z}\frac{z^{'2}-1}{z^{'2}}\widetilde{\mathcal{M}}\left(z',a\right) \nonumber\\
 &=& \frac{1}{\pi}\int_{|z'|=1}\frac{\mathrm{d}z'}{z'-z}\frac{z^{'2}-1}{z^{'2}}\widetilde{\mathcal{A}}\left(z',a\right).
\end{eqnarray}
Here the $\widetilde{\mathcal{A}}$ is the discontinuity of the amplitude at $|z|=1$ and is expressed as
\begin{equation}
	\widetilde{\mathcal{A}}\left(z,a\right) = \lim_{\epsilon\rightarrow 0}\frac{1}{2i}\left[\widetilde{\mathcal{M}}\left(\left(1+\epsilon\right)e^{i\phi},a\right)-\widetilde{\mathcal{M}}\left(\left(1-\epsilon\right)e^{i\phi},a\right)\right].
\end{equation}
This leads us to the dispersion relation
\begin{eqnarray}
	\widetilde{\mathcal{M}}\left(z,a\right) & = & \frac{z^{2}}{1-z^{2}}\biggl\{-\widetilde{\mathcal{M}}\left(\infty,a\right)+ \frac{1}{z^{2}}\left(\widetilde{\mathcal{M}}\left(0,a\right)+z\frac{\partial}{\partial z}\widetilde{\mathcal{M}}\left(z=0,a\right)\right)\biggr\} \nonumber\\
	&&+\frac{1}{\pi}\frac{z^{2}}{1-z^{2}}\int_{|z'|=1}\frac{\mathrm{d}z'}{z'-z}\frac{z^{'2}-1}{z^{'2}}\widetilde{\mathcal{A}}\left(z',a\right) \\
	& = & \widetilde{\mathcal{M}}\left(0,a\right) + \frac{z}{1-z^{2}}\frac{\partial}{\partial z}\widetilde{\mathcal{M}}\left(z=0,a\right) +\frac{1}{\pi}\frac{z^{2}}{1-z^{2}}\int_{|z'|=1}\frac{\mathrm{d}z'}{z'-z}\frac{z^{'2}-1}{z^{'2}}\widetilde{\mathcal{A}}\left(z',a\right), \nonumber
\end{eqnarray}
where we have used $\mathcal{M}\left(\infty,a\right)=\mathcal{M}\left(0,a\right)$. Assuming 
\begin{equation}\label{M-zexpansion}
	\widetilde{\mathcal{M}}\left(z,a\right) = \sum_{n=0}^{\infty}c_{n}z^{2n},
\end{equation}
we can discard all the terms other than powers of $z^{2}$ in the above dispersion relation. Then we obtain,
\begin{equation}\label{disp-2ch-z}
	\widetilde{\mathcal{M}}\left(z,a\right) = c_{0} + \frac{1}{\pi}\frac{z^{2}}{1-z^{2}}\int_{|z'|=1}\frac{\mathrm{d}z'}{z'}\frac{z^{'2}-1}{z^{'2}-z^{2}}\widetilde{\mathcal{A}}\left(z',a\right).
\end{equation}
In kinematic variables we get a dispersion relation which is symmetric in two channels, $s_{1}$ and $s_{2}$, 
\begin{equation}\label{disp-2ch-s}
	\mathcal{M}\left(s_{1},s_{2}\right) = \mathcal{M}\left(0,0\right) + \frac{1}{\pi}\int_{a}^{\infty}\frac{\mathrm{d}\sigma}{\sigma}\mathcal{A}^{(s_{1})}\left(\sigma,\frac{a\sigma}{\sigma-a}\right)\left[\frac{s_{1}}{\sigma-s_{1}}+\frac{s_{2}}{\sigma-s_{2}}\right].
\end{equation} 
Here, $\mathcal{A}^{(s_{1})}$ is the $s_{1}$ channel discontinuity. 

\subsection{Local dispersion relation}
The kernel appearing in \eqref{disp-2ch-s} can be expressed as  
\begin{equation}\label{localkernel}
	H\left(\sigma ;s_{1},s_{2}\right) = \frac{s_{1}}{\sigma-s_{1}}+\frac{s_{2}}{\sigma-s_{2}} = \frac{x\sigma -2y}{\sigma^{2}-x\sigma+y} = \frac{\sigma^{2}-y}{\sigma^{2}-x\sigma+y}-1.
\end{equation}
Now we make a Taylor series expansion of $\mathcal{A}\left(\sigma,a\right)$ around $a=0$. Since $a$ has $s_{1}+s_{2}$ in the denominator, Taylor series expansion gives rise to additional singularities whenever the denominator vanishes. This means that \eqref{disp-2ch-s} has spurious singularities in addition to the physical singularities of the amplitude. Here we follow the analysis of \cite{Song:2023quv} to get rid of the unphysical singularities in the two-channel symmetric dispersion relation.

We can write $\frac{1}{\sigma^{2}-x\sigma+y}=\frac{1}{\sigma^{2}+y}\sum_{p=0}^{\infty}\left(\frac{x\sigma}{\sigma^{2}+y}\right)^{p}$. Note that $y=ax$; therefore, for any positive integer $n$, we have $a^{n}\frac{1}{\sigma^{2}-x\sigma+y}= \frac{y^{n}}{\sigma^{2}+y}\sum_{p=0}^{\infty}\frac{x^{p-n}\sigma^{p}}{\left(\sigma^{2}+y\right)^{p}}$. Spurious singularities appear when $p<n$. If we remove terms for $p=0,1,\ldots n-1$ from the sum, we are left with 
\begin{equation}
    a^{n}\frac{1}{\sigma^{2}-x\sigma+y} \rightarrow \frac{\sigma^{n}y^{n}}{\left(\sigma^{2}+y\right)^{n+1}}\sum_{m=0}^{\infty}\left(\frac{x\sigma}{\sigma^{2}+y}\right)^{m} = \left(\frac{\sigma y}{\sigma^{2}+y}\right)^{n}\frac{1}{\sigma^{2}-x\sigma+y}
\end{equation}
Hence effectively functional dependence of $a$ in $\mathcal{A}$ becomes $\frac{\sigma y}{\sigma^{2}+y}$. For the additional constant piece in \eqref{localkernel} only the $a^{0}$ order term in $\mathcal{A}$ contributes. This leads us to the following local dispersion relation
\begin{equation}\label{LDR-2ch}
	\mathcal{M}^{(\text{local})}\left(s_{1},s_{2}\right) = \mathcal{M}\left(0,0\right)+\frac{1}{\pi}\int_{a}^{\infty}\frac{\mathrm{d}\sigma}{\sigma}\left[\biggl\{\frac{s_{1}}{\sigma-s_{1}}+\frac{s_{2}}{\sigma-s_{2}}+1\biggr\}\mathcal{A}^{(s_{1})}\left(\sigma,\frac{ y}{\sigma}\right)-\mathcal{A}^{(s_{1})}\left(\sigma,0\right)\right].
\end{equation}

\subsection*{Massless pole}
We demonstrate that the massless pole can be handled in \eqref{LDR-2ch} in a limiting way. Let us consider $\mathcal{M}\left(s_{1},s_{2}\right)=\frac{1}{\left(s_{1}-\alpha\right)\left(s_{2}-\alpha\right)}$, which in the limit $\alpha\rightarrow0$ reduces to $\frac{1}{s_{1}s_{2}}$. $s_{1}$-channel discontinuity of the amplitude is given by 
\begin{equation}\label{massless-alpha}
    \mathcal{A}\left(s_{1},s_{2}\right)=-\frac{\pi}{s_{2}-\alpha}\delta\left(s_{1}-\alpha\right).
\end{equation}
It is easy to see that substituting \eqref{massless-alpha} in \eqref{LDR-2ch} yields the required amplitude.

\subsection{Higher spins}

The one-parameter family of representations of the tree-level amplitude given in \eqref{gen-disp-1} works for theories containing an infinite number of higher spin particles. Let us consider a four-point scattering of massless scalars where the exchanges are a massive scalar and a massive spin-one state. We can assume that massless pole and higher massive exchanges are absent for this case. Using \eqref{localDisp} the amplitude that we obtain has the following form,
\begin{equation}\label{ex1}
    \mathcal{M}\left(s_{1}, s_{2}\right) = W_{00} + \left(\frac{1}{s_{1}-1}+\frac{1}{s_{2}-1}+1\right)\left[c^{(1)}_{0}+c^{(1)}_{1} (D-3) (2 s_{1} s_{2}+1)\right]+c^{(1)}_{0}+c^{(1)}_{1}
   (D-3).
\end{equation}
Here, $W_{00}$ is the Wilson coefficient, $c^{(1)}_{0}$ and $c^{(1)}_{1}$ are the partial-wave coefficients for the scalar and spin-one states, respectively. If we try to apply \eqref{gen-disp-1} for the amplitude obtained above, we will get 
\begin{eqnarray}\label{ex1-1}
    &&\mathcal{M}\left(s_{1}, s_{2}\right) \nonumber\\
    & = & W_{00} +\frac{1}{\lambda
   +1}\left[c^{(1)}_{0}+c^{(1)}_{1} (D-3) (1-2 \lambda )\right] + \frac{2 \lambda  }{\lambda +1} \left[c^{(1)}_{0}+c^{(1)}_{1} (D-3) \bigl\{\left(\lambda -1\right) \lambda
   +1\bigr\}\right]\nonumber\\
   && +\left(\frac{1}{s_{1}-1}+\frac{1}{s_{2}-1} + \frac{1}{\lambda
   +1}\right)
   \left[c^{(1)}_{0}+c^{(1)}_{1} (D-3) \biggl\{\frac{2 (\lambda
   +s_{1}) (\lambda +s_{2})}{\lambda +1}+1-2 \lambda\biggr\}\right].
\end{eqnarray}
The difference between \eqref{ex1} and \eqref{ex1-1} is proportional to $c^{(1)}_{1}$, and thus we conclude that the presence of a single massive spin-one state is inconsistent with the shift symmetry of \eqref{gen-disp-1}. This argument can be generalized for a finite number of massive higher-spin exchanges. 

{The above conclusion is related to the Reggeization argument \cite{Caron-Huot:2016icg} which requires an infinite number of subleading Regge trajectories.}

\subsection{Convergence of individual terms}
\label{indiv-ch-conv}
The summation in \eqref{op-st} can be divided differently into three parts, 
 \begin{equation}
     \mathcal{S}_{1} + \mathcal{S}_{2} + \mathcal{C},
 \end{equation}
 where 
 \begin{equation}
     \mathcal{S}_{1}  =  \sum_{n=1}^{\infty}\frac{1}{n!}\frac{\left(1+s_{2}\right)_{n-1}}{s_{1}-n}, \qquad
     \mathcal{S}_{2}  =  \sum_{n=1}^{\infty}\frac{1}{n!}\frac{\left(1+s_{1}\right)_{n-1}}{s_{2}-n}
 \end{equation}
 are the sum of residues over poles at each channel, and 
 \begin{eqnarray}
     	\mathcal{C} & = & \sum_{n=1}^{\infty}\Biggl\{\frac{1}{n!\left(s_{1}-n\right)}\left(1+s_{2}\right)_{n-1}\Biggl\{\prod_{k=1}^{n-1}\left[1+\frac{\left(s_{1}-n\right)\left(s_{2}+\lambda\right)}{\left(k+s_{2}\right)\left(n+\lambda\right)}\right]-1\Biggr\}\nonumber\\
	&& + \frac{1}{n!\left(s_{2}-n\right)}\left(1+s_{1}\right)_{n-1}\Biggl\{\prod_{k=1}^{n-1}\left[1+\frac{\left(s_{2}-n\right)\left(s_{1}+\lambda\right)}{\left(k+s_{1}\right)\left(n+\lambda\right)}\right]-1\Biggr\}\nonumber\\
	&& + \frac{1}{n!\left(\lambda+n\right)}\left(1-\lambda+\frac{\left(s_{1}+\lambda\right)\left(s_{2}+\lambda\right)}{\lambda+n}\right)_{n-1}\Biggr\}
 \end{eqnarray}
 is the contact term. The product inside the curly brackets, in case of the term corresponding to $n=1$, contributes 1.  Now, it is obvious that each individual piece cannot converge everywhere. For example, $\mathcal{S}_1$ converges when $\text{Re}(s_2)<0$. 

 Since the open-string amplitude admits single channel expansion, therefore, when $s_{2}<0$, $\mathcal{S}_{1}$ converges to $\frac{\Gamma\left(-s_{1}\right)\Gamma\left(-s_{2}\right)}{\Gamma\left(1-s_{1}-s_{2}\right)}-\frac{1}{s_{1}s_{2}}$ and $\mathcal{S}_{2} + \mathcal{C}=0$. Similarly for $s_{1}<0$, $\mathcal{S}_{2}$ converges to the same value and $\mathcal{S}_{1}+\mathcal{C}=0$. In the first quadrant, where $s_1>0$ and $s_2>0$, all of $\mathcal{S}_{1}$, $\mathcal{S}_{2}$ and $\mathcal{C}$ individually diverge but their sum $\mathcal{S}_{1}+\mathcal{S}_{2}+\mathcal{C}$ is a finite quantity and equal to the actual value. In the third quadrant, where $s_1<0$ and $s_2<0$, all of $\mathcal{S}_{1}$, $\mathcal{S}_{2}$ and $\mathcal{C}$ individually converge.

\begin{center}
\begin{table}[H]
    \centering
    \begin{tabular}{|c|c|c|c|c|c|c|c|}
    \hline
   $s_{1}$   &  $s_{2}$ & $\mathcal{S}_{1}$ & $\mathcal{S}_{2}$ & $\mathcal{C}$ &$\widehat{\mathcal{M}}$\\
   \hline
    10.7  & -0.5 & 0.14878 & -7.16970 $\times$ $10^{20}$ & 7.16970 $\times 10^{20}$ & 0.14877 \\
    5.6 & 7.8 & -1.46109 $\times 10^{15}$ & -4.09531 $\times 10^{10}$ & 1.46113 $\times 10^{15}$ & -779.43661 \\
    -2.7 & -9.5 & -0.03875 & -0.03875 & 0.03875 & -0.03875\\
    \hline
    \end{tabular}
   \caption{Values of individual sum obtained numerically after keeping 1000 terms in each series. $\lambda$ is set to 6.5\,. $\widehat{\mathcal{M}} = \frac{\Gamma\left(-s_{1}\right)\Gamma\left(-s_{2}\right)}{\Gamma\left(1-s_{1}-s_{2}\right)}-\frac{1}{s_{1}s_{2}}$ is the actual value. The numerical values are in agreement with the actual values in the given decimal places. }
\end{table}
\end{center}
    
\subsection*{When does the LCSDR form hold?}
Let us examine \eqref{localDisp} for convergence. Notice here that in the absorptive part, instead of the conventional $t$ (here $s_2$) dependence of the second argument, we have $s_1 s_2/\sigma$. Now suppose that the absorptive part in the large $s_1$ limit behaved as $s_1^{p(s_2)}$. Here $p\left(x\right)$ is any arbitrary function with the condition that $p\left(0\right)$ and $p'(0)$ are well defined. Then in \eqref{localDisp}, we have the integrand going as $\sigma^{p(s_1 s_2/\sigma)-1}$. In the large $\sigma$ limit we have $\sigma^{p(0)-1}$. In \eqref{localDisp}, together with the last term, we have the integrand behave as 
$$
s_1 s_2 p'(0)\sigma^{p(0)-2} \ln \sigma \,.
$$
Thus for convergence we need $p(0)<1$. Unlike the fixed-$t$ case, we do not have to specify the sign of $s_2$ for this to hold. That is why our representation also captures the unphysical $s_1,s_2>0$ regime (unlike the fixed-$t$ case) of the string amplitude as our plots indicate (see also \cite{Saha:2024qpt}). However, we emphasize that \eqref{localDisp} needed the locality constraints to hold. As such, the partial waves should be such that these constraints hold.
\subsection{A check of the $\lambda$-CSDR}
\label{masslessbox}
\begin{figure}[H]
    \centering
    \includegraphics[width= 0.6\linewidth]{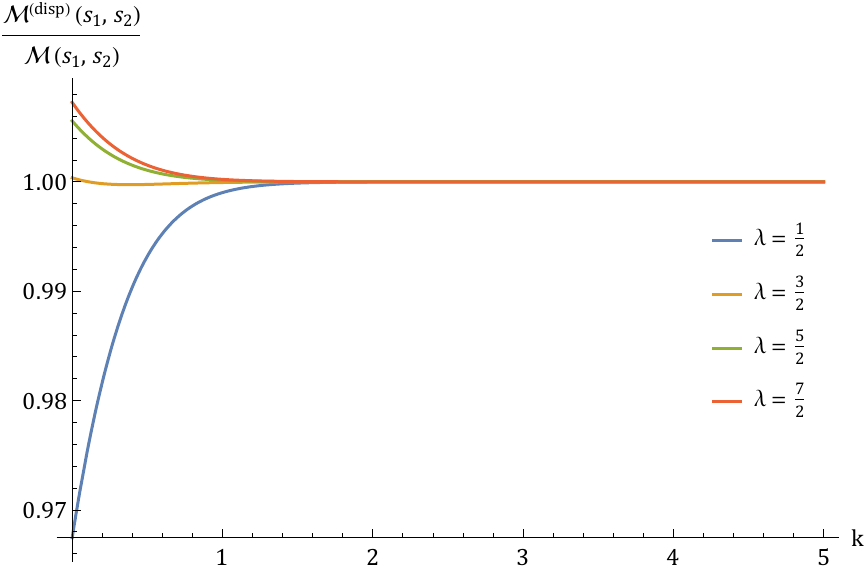}
    \caption{Ratio of the dispersive representation \eqref{Gen_Sing_LCSDR} and the actual function given in \eqref{boxdiag} at $\left(s_1,s_2\right)= \left(-\frac{5}{2},-\frac{1}{3}\right)$ as we increase $\sigma_{max} = 10^k$ for various values of $\lambda$. We have used $\epsilon=\frac{1}{10}$.}
    \label{GenSinglambdaLCSDRCheck}
\end{figure}

We verify the $\lambda$-LCSDR using the following amplitude corresponding to a massless box diagram using dimensional regularization $D=4-2\epsilon$ \cite{Britto:2024mna},
\begin{equation}
\label{boxdiag}
  \mathcal{M}(s_1,s_2) = \frac{1}{s_1s_2} \left[\frac{2}{\epsilon^2}(\left(-s_1\right)^{-\epsilon}+(-s_2)^{-\epsilon})-\left(\log(-s_1)-\log(-s_2)\right)^2-\frac{4\pi^2}{3}\right]
\end{equation}
where the expansion has been performed   $\mathcal{O}(\epsilon^0)$. At $s_1 = 0$, the amplitude blows up both due to the $\log(s_1)$ and the $\frac{1}{s_1}$ factor. Because of this, a direct application of the \eqref{Gen_Sing_LCSDR} leads to ``apparent" divergences. We therefore start with an amplitude where the pole is at $s_1 = q$ and the branch cut starts at $s_1 = s_0$ as follows
\begin{equation}
\label{boxdiaggen}
\begin{split}
  \mathcal{M}(q,s_0,s_1,s_2) = \frac{1}{(s_1-q)(s_2-q)} \bigg[&\frac{2}{\epsilon^2}(\left(s_0-s_1\right)^{-\epsilon}+(s_0-s_2)^{-\epsilon})\\&-\left(\log(s_0-s_1)-\log(s_0-s_2)\right)^2-\frac{4\pi^2}{3}\bigg]
\end{split}
\end{equation}
Applying the \eqref{Gen_Sing_LCSDR} to this amplitude gives
\begin{equation}
\label{boxdiagdisp}
\begin{split}
   \mathcal{M}(q,s_0,s_1,s_2)=\:&\mathcal{M}(q,s_0,-\lambda,-\lambda)+\frac{1}{q+\lambda}\mathop{\mathrm{Res}}_{\sigma = q} \mathcal{M}\left(q,s_0,\sigma,-\lambda\right)-\frac{1}{\pi} \int_{s_0}^{\infty} d\sigma\frac{\mathcal{A}^{(s_1)}\left(q,s_0,\sigma,- \lambda \right)}{\sigma + \lambda}\\&-\left[\left(\frac{1}{q-s_1}+\frac{1}{q-s_2}-\frac{1}{q+\lambda}\right)\mathop{\mathrm{Res}}_{\sigma = q} \mathcal{M}\left(q,s_0,\sigma,\frac{(s_1+\lambda)(s_2 + \lambda)}{\sigma + \lambda} - \lambda\right)\right]\\
    &+\frac{1}{\pi} \int_{s_0}^{\infty} d\sigma  \left(\frac{1}{\sigma -s_1}+\frac{1}{\sigma - s_2}- \frac{1}{\sigma + \lambda}\right) \mathcal{A}^{(s_1)}\left(q, s_0, \sigma,\frac{(s_1+\lambda)(s_2 + \lambda)}{\sigma + \lambda} - \lambda \right)\:.
\end{split}
\end{equation}
Here we have separated out the contribution from the pole at $s_1 = q$. Now we can place the pole at $s_1 = q = 0$ and then bring the branch point at $s_0$ close to 0 in a regulated way to recover the dispersive representation for the massless box diagram \eqref{boxdiag}. In figure \ref{GenSinglambdaLCSDRCheck}, we show for $q=0$ and $s_0=10^{-10}$ that the representation converges and is independent of $\lambda$ as we integrate up to some large $\sigma_{max}$\footnote{We can also derive other interesting relations using \eqref{Gen_Sing_LCSDR}. For example,
\begin{equation}
    \frac{\tan(s_1)}{s_1}\frac{\tan(s_2)}{s_2}=\sum_{n=-\infty}^{\infty}\frac{1}{r_n}\left(\frac{1}{r_n-s_1}+\frac{1}{r_n-s_2}-\frac{1}{r_n+\lambda}\right)\frac{\tan\left(\frac{\left(s_1+\lambda\right)(s_2+\lambda)}{r_n+\lambda}-\lambda\right)}{\left(\frac{\left(s_1+\lambda\right)(s_2+\lambda)}{r_n+\lambda}-\lambda\right)}\, 
\end{equation}
where $r_n=\frac{(2n+1)\pi}{2}$. This series is convergent everywhere except $\left\{s_1,s_2,\lambda=r_n\,|\,n\in\mathbb{Z}\right\}$.}. 

\section{Checks for Convergence}

\subsection{Convergence with $N_\text{{max}}$ and constraints}
\begin{figure}[H]
    \centering
    \subfloat[\centering $s_2=-0.1$]{{\includegraphics[width=8cm]{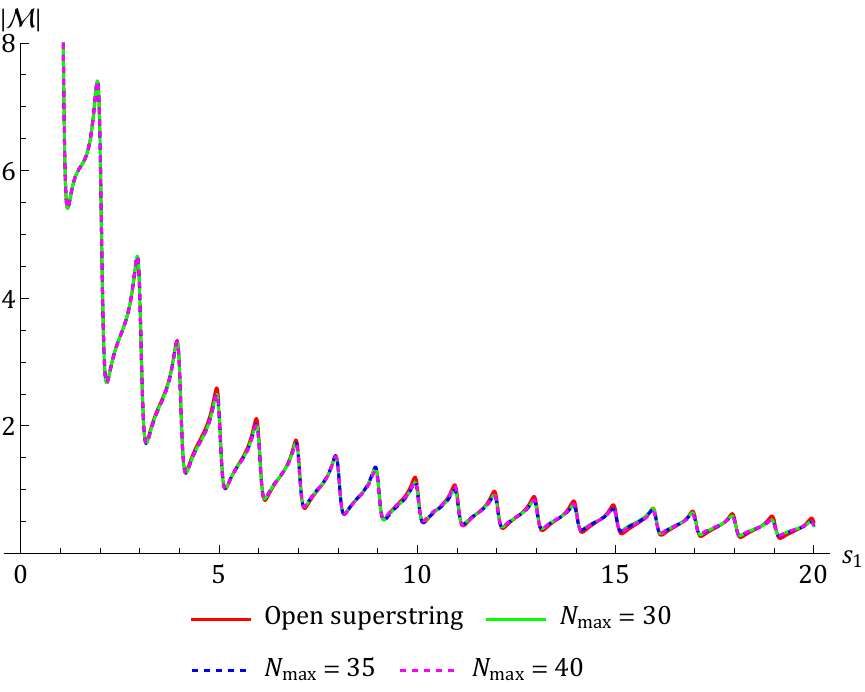} }}%
    \qquad
    \subfloat[\centering $s_2=1.1$]{{\includegraphics[width=8cm]{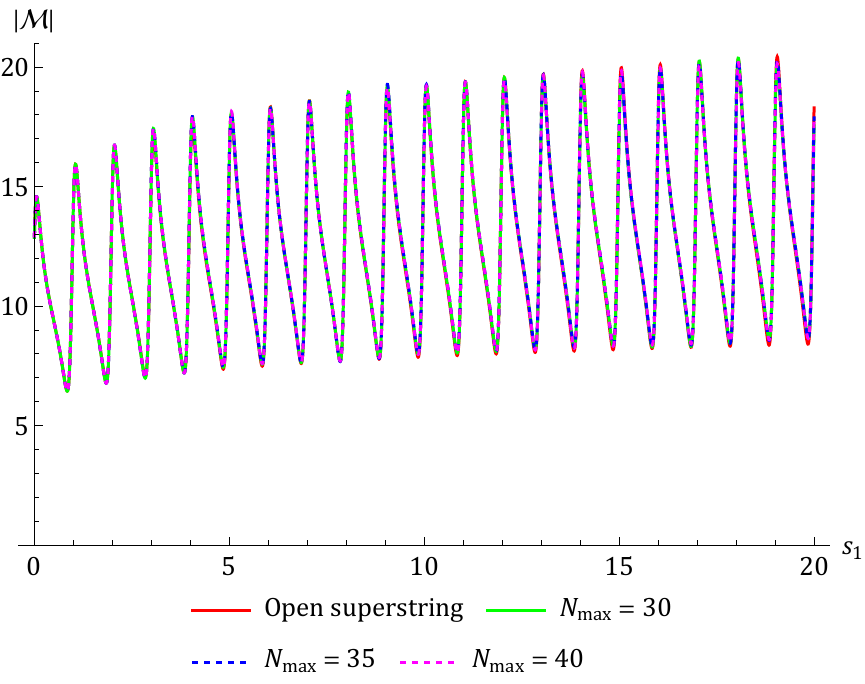} }}%
    \caption{Convergence of the amplitudes with $N_\text{max}$ for the S-matrix corresponding to the open superstring in fig.\ref{fig:dualitycontourplotCat1}.}%
    \label{fig:NmaxConvergenceOpenString}
\end{figure}

\begin{figure}[H]
    \centering
    \subfloat[\centering $s_2=-0.1$]{{\includegraphics[width=8cm]{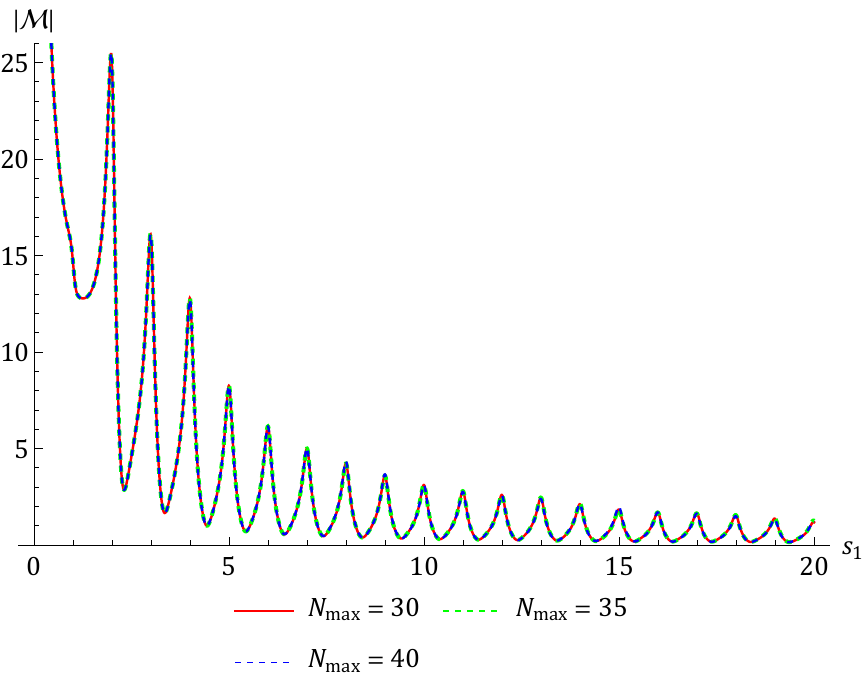} }}%
    \qquad
    \subfloat[\centering $s_2=1.1$]{{\includegraphics[width=8cm]{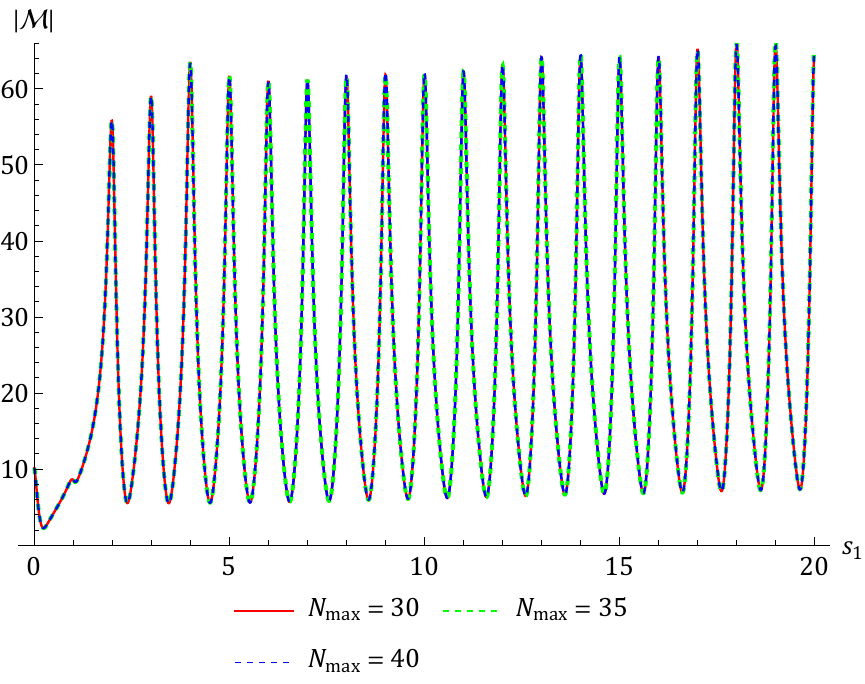} }}%
    \caption{Convergence of the amplitudes with $N_\text{max}$ for the S-matrix corresponding to the black star in fig.\ref{fig:dualitycontourplotCat2}.}%
   \label{fig:NmaxConvergenceNonString}
\end{figure}

Here, we will check the convergence of the amplitudes with $N_{\text{max}}$ and the $\lambda$-constraints \eqref{lambdaconstraints}. For the purpose of demonstration, we will focus on the open superstring amplitude in fig.\ref{fig:dualitycontourplotCat1} and the black star amplitude in fig.\ref{fig:dualitycontourplotCat2}. In fig.\ref{fig:NmaxConvergenceOpenString} and fig.\ref{fig:NmaxConvergenceNonString} we demonstrate the convergence of our numerics for $N_{\text{max}}=30, 35, 40$ in $D=10$ with a grid of 180 $\lambda$-constraints at each derivative order (see fig.\ref{s1s2grid}). In fig.\ref{fig:NmaxConvergenceOpenString} we show our results for the open superstring amplitude. In fig.\ref{fig:NmaxConvergenceNonString} we show our results for the black star amplitude. We find that all of our S-matrices have converged very well with $N_{\text{max}}$.


\begin{figure}[H]
    \centering
    \subfloat[\centering $s_2=-0.1$]{{\includegraphics[width=8cm]{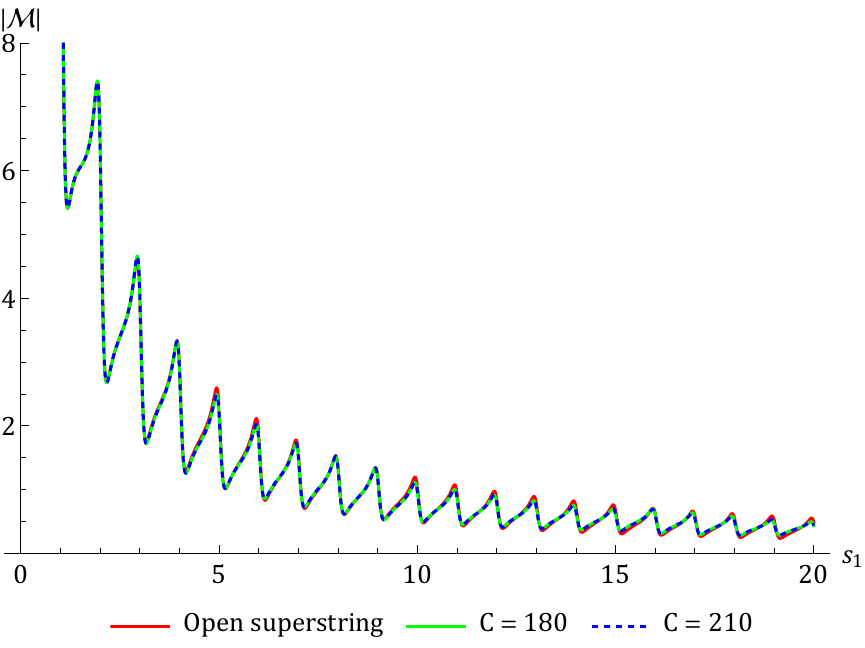} }}%
    \qquad
    \subfloat[\centering $s_2=1.1$]{{\includegraphics[width=8cm]{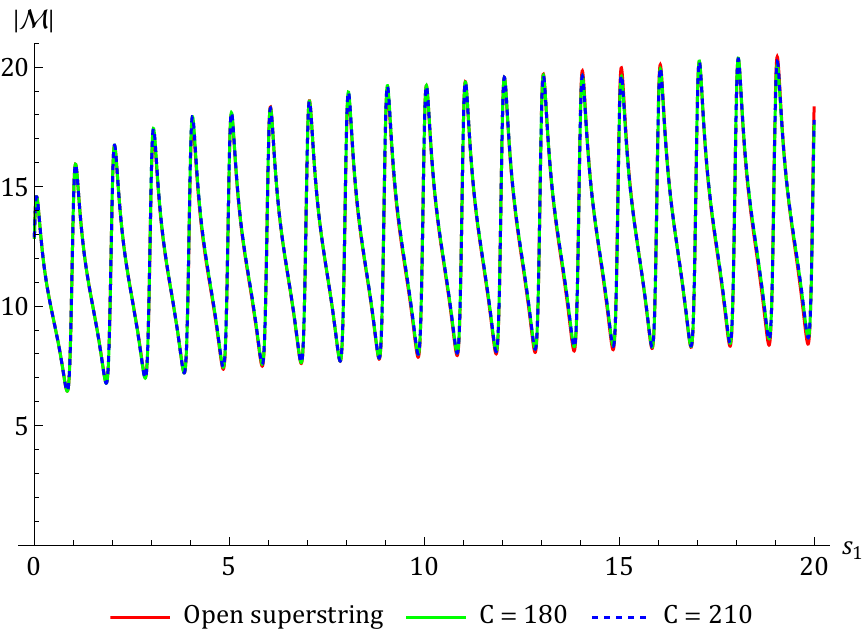} }}%
    \caption{Convergence of the amplitudes with the number of $\lambda$-constraints for the S-matrix corresponding to the open superstring in fig.\ref{fig:dualitycontourplotCat1}.}%
    \label{fig:CConvergenceOpenString}
\end{figure}
\begin{figure}[H]
    \centering
    \subfloat[\centering $s_2=-0.1$]{{\includegraphics[width=8cm]{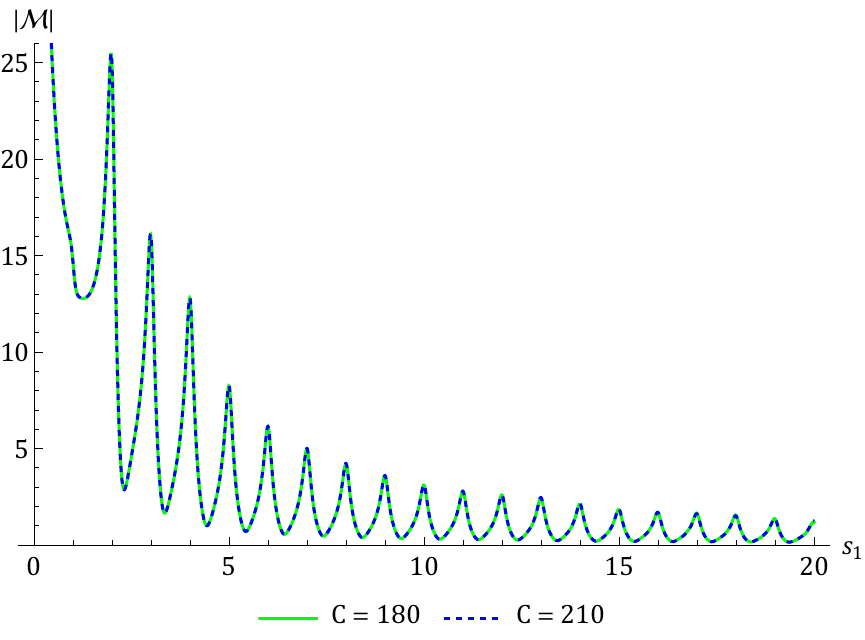} }}%
    \qquad
    \subfloat[\centering $s_2=1.1$]{{\includegraphics[width=8cm]{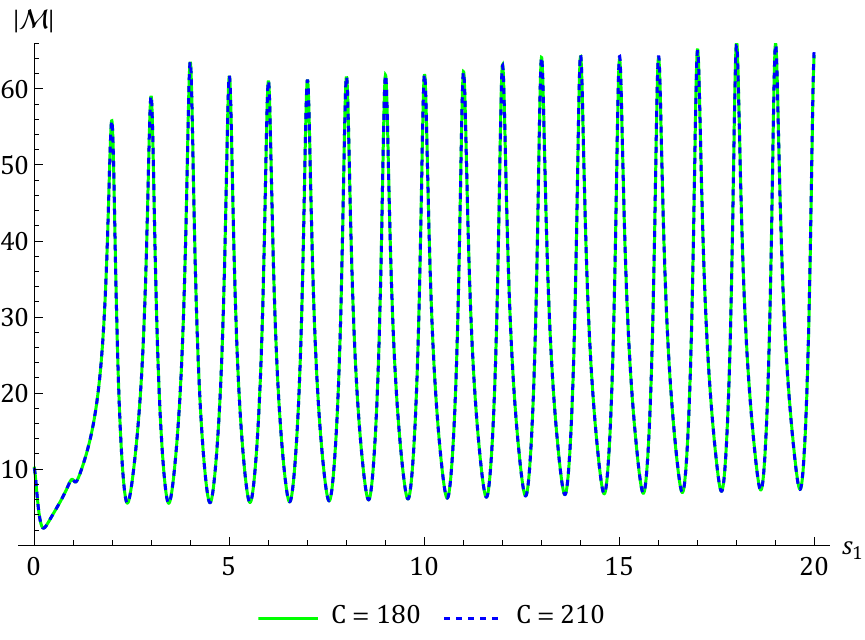} }}%
    \caption{Convergence of the amplitudes with the number of $\lambda$-constraints for the S-matrix corresponding to the black star in fig.\ref{fig:dualitycontourplotCat2}.}%
   \label{fig:CConvergenceNonString}
\end{figure}

Next, we proceed to show convergence of our amplitudes with the number of $\lambda$-constraints at each derivative order. We denote this number by C. The first grid that we choose is the same grid as in fig.\ref{s1s2grid} consisting of 180 $\lambda$-constraints at each derivative order (C = 180). The second grid that we choose consists of 10 points in the 1st quadrant for $|s_i|<10$, 60 points in the 2nd and the 4th quadrants for $|s_i|<25$, 40 points in the 3rd quadrant for $|s_i|<15$ and 40 points in the region $-1<s_1<1$ and $-0.01<s_2<0.01$ adding to a total of 210 $\lambda$-constraints at each derivative order (C = 210) for the second grid. We use $N_\text{max}=30$ in $D=10$ and $\lambda=14.6$. We show our results for the open superstring in fig.\ref{fig:CConvergenceOpenString} and for the black star in fig.\ref{fig:CConvergenceNonString}. We clearly see very good convergence with the number of $\lambda$-constraints.


\subsection{Convergence with $\lambda$}

\begin{figure}[H]
    \centering
    \subfloat[\centering $s_2=-0.1$]{{\includegraphics[width=8cm]{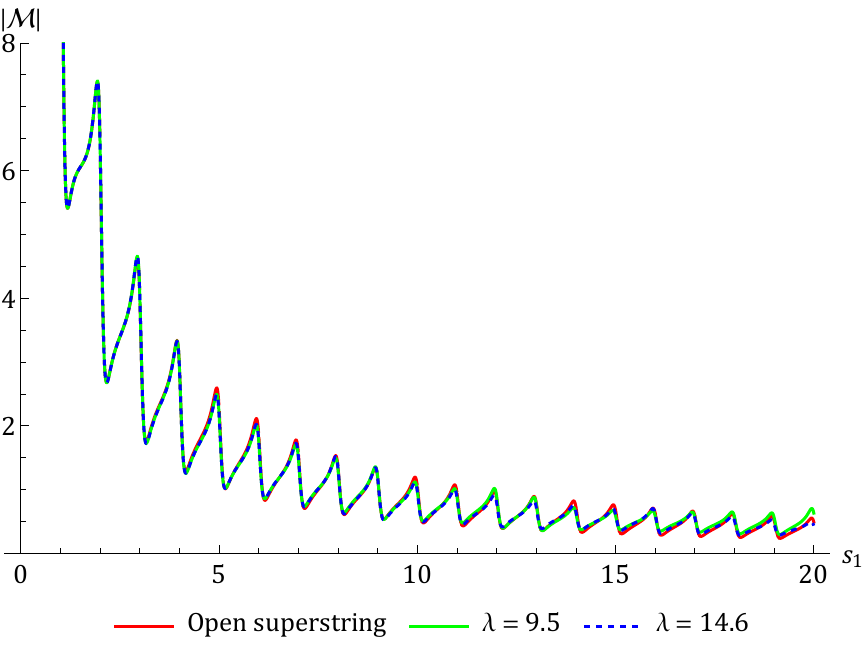} }}%
    \qquad
    \subfloat[\centering $s_2=1.1$]{{\includegraphics[width=8cm]{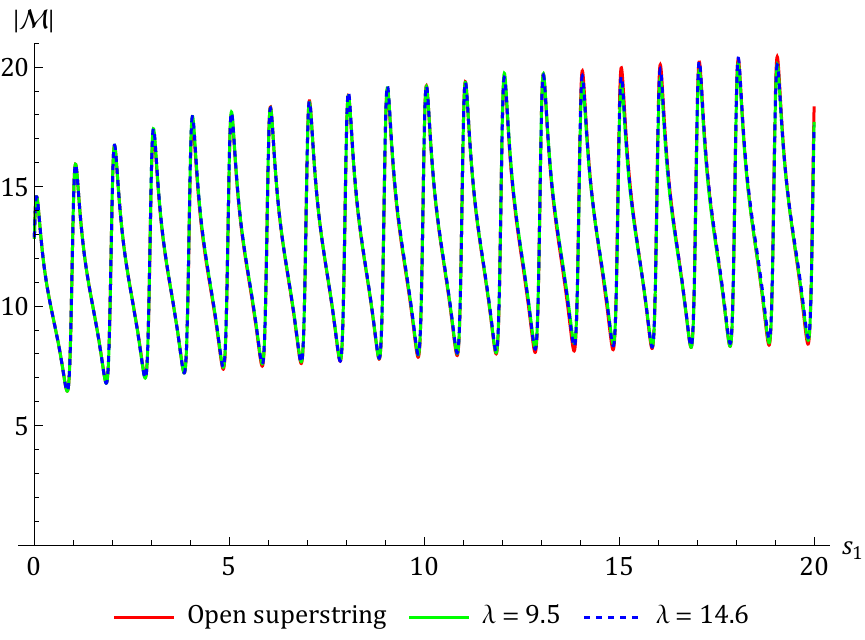} }}%
    \caption{Convergence of the amplitudes with the seed $\lambda$ for the S-matrix corresponding to the open superstring in fig.\ref{fig:dualitycontourplotCat1}.}%
    \label{fig:lConvergenceOpenString}
\end{figure}

\begin{figure}[H]
    \centering
    \subfloat[\centering $s_2=-0.1$]{{\includegraphics[width=8cm]{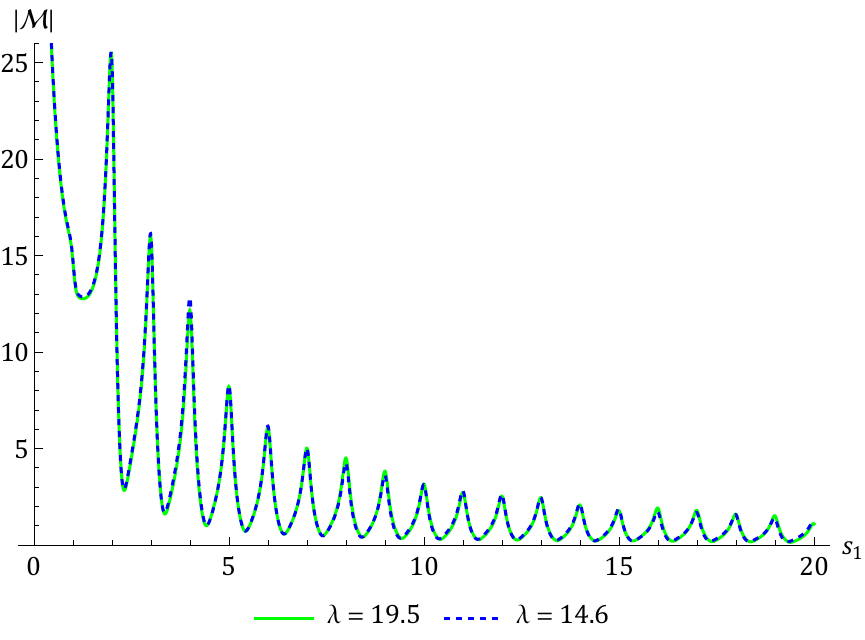} }}%
    \qquad
    \subfloat[\centering $s_2=1.1$]{{\includegraphics[width=8cm]{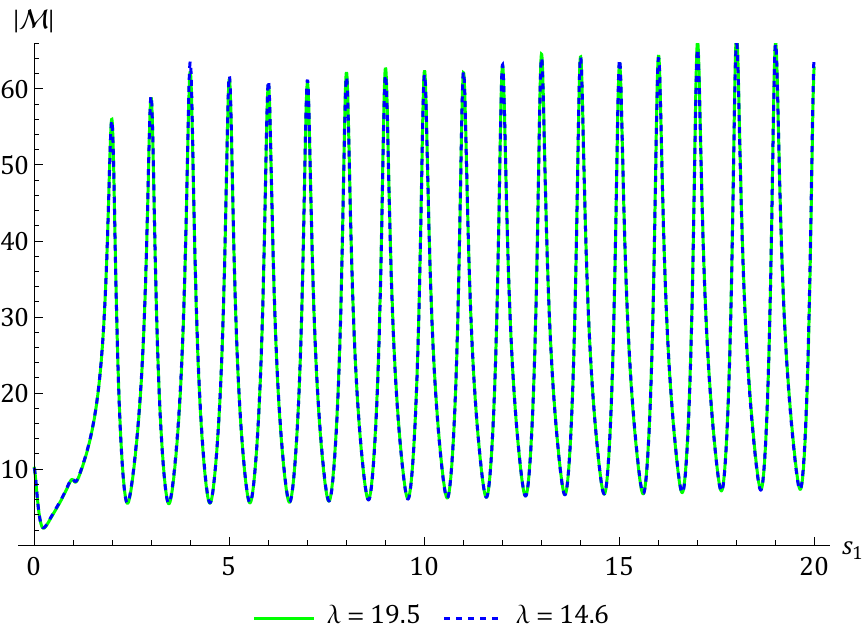} }}%
    \caption{Convergence of the amplitudes with the seed $\lambda$ for the S-matrix corresponding to the black star in fig.\ref{fig:dualitycontourplotCat2}.}%
   \label{fig:lConvergenceNonString}
\end{figure}

\begin{figure}[H]
    \centering
    \subfloat[\centering $s_1=3.1,s_2=-0.1$]{{\includegraphics[width=8cm]{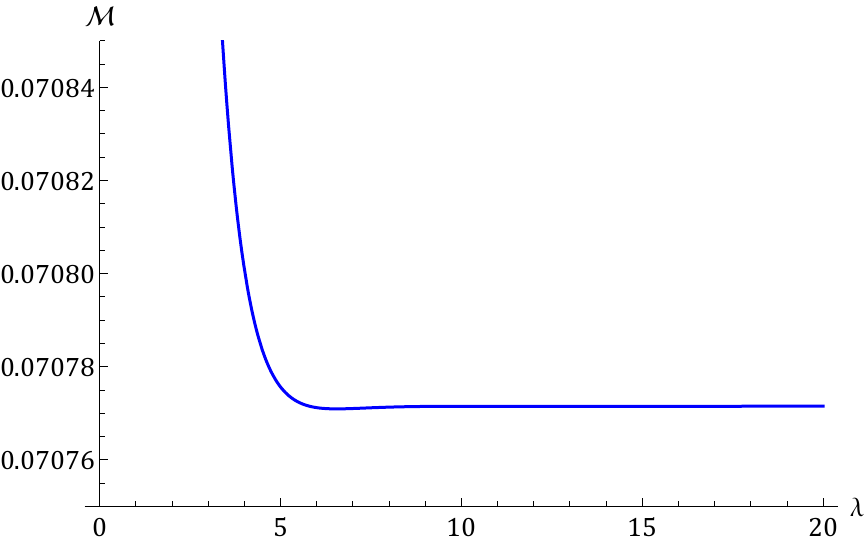} }}%
    \qquad
    \subfloat[\centering $s_1=-5.5,s_2=4.6$]{{\includegraphics[width=8cm]{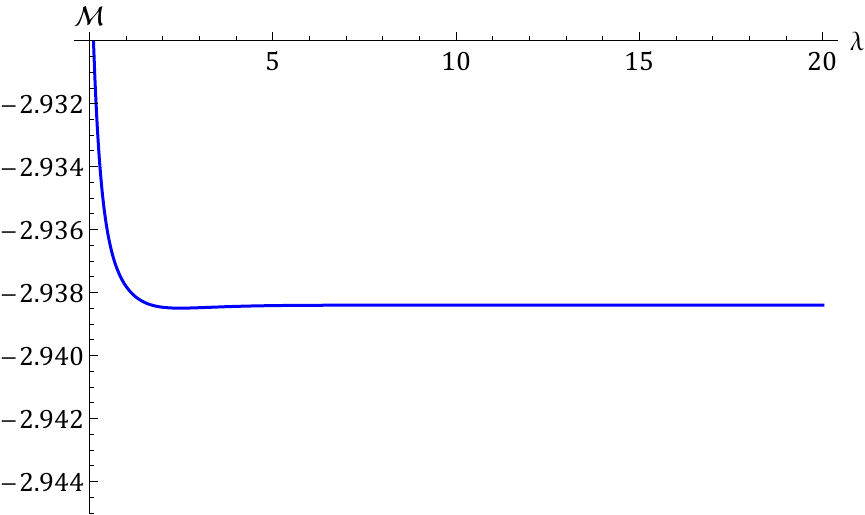} }}%
    \caption{We plot $\mathcal{M}(s_1, s_2)$ versus $\lambda$ for the open superstring in fig.\ref{fig:dualitycontourplotCat1} for different
values of $s_1, s_2$ and find nice plateaus around the seed
value of $\lambda = 14.6$.}
    \label{fig:MvslOpenString}
\end{figure}

\begin{figure}[H]
    \centering
    \subfloat[\centering $s_1=3.1,s_2=-0.1$]{{\includegraphics[width=8cm]{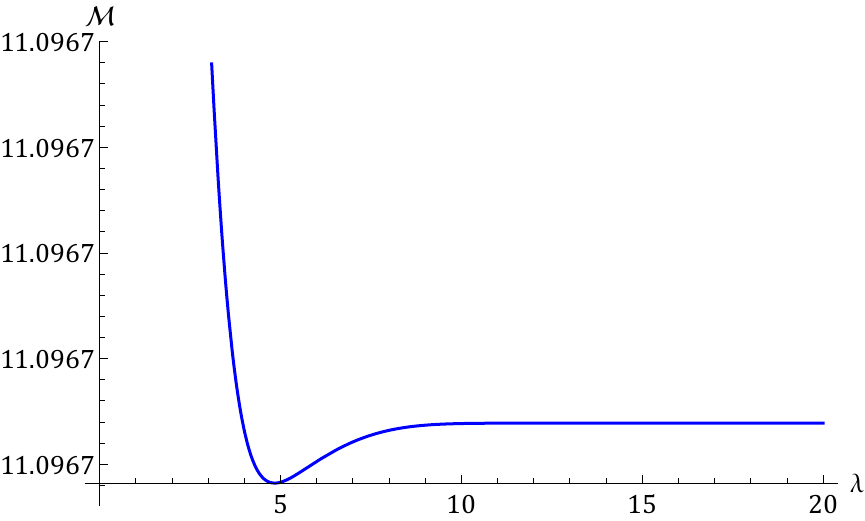} }}%
    \qquad
    \subfloat[\centering $s_1=-5.5,s_2=4.6$]{{\includegraphics[width=8cm]{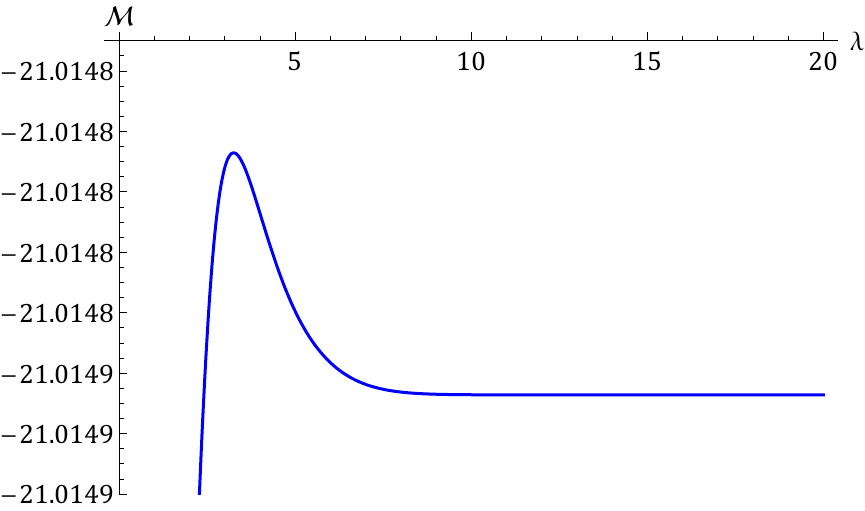} }}%
    \caption{We plot $\mathcal{M}(s_1, s_2)$ versus $\lambda$ for the black star in fig.\ref{fig:dualitycontourplotCat2} for different
values of $s_1, s_2$ and find nice plateaus around the seed
value of $\lambda = 14.6$.}%
   \label{fig:MvslNonString}
\end{figure}

Here, we show that the amplitudes have converged with $\lambda$. In fig.\ref{fig:lConvergenceOpenString}  and fig.\ref{fig:lConvergenceNonString} we plot the amplitudes for the open superstring in fig.\ref{fig:dualitycontourplotCat1} and the black star in fig.\ref{fig:dualitycontourplotCat2} for two different values of seed $\lambda$. For the open superstring we choose $\lambda=9.5$ and $\lambda=14.6$. For the black star we choose $\lambda=14.6$ and $\lambda=19.5$. We use $N_{\text{max}}=30$ in $D=10$. We find great convergence for the respective amplitudes with the seed value of $\lambda$. In fig.\ref{fig:MvslOpenString} and fig.\ref{fig:MvslNonString} we plot $\mathcal{M}(s_1,s_2)$ versus $\lambda$ for the open superstring and the black star with the seed value of $\lambda=14.6$ for different values of $s_1, s_2$. Again we find nice plateaus around this seed value of $\lambda=14.6$. This demonstrates the convergence of our numerics with $\lambda$.

\subsection{Comparison between the Fixed-$s_2$ dispersion relation and the LCSDR}
\begin{figure}[H]
    \centering
    \subfloat[\centering $s_2=-0.1$]{{\includegraphics[width=8cm]{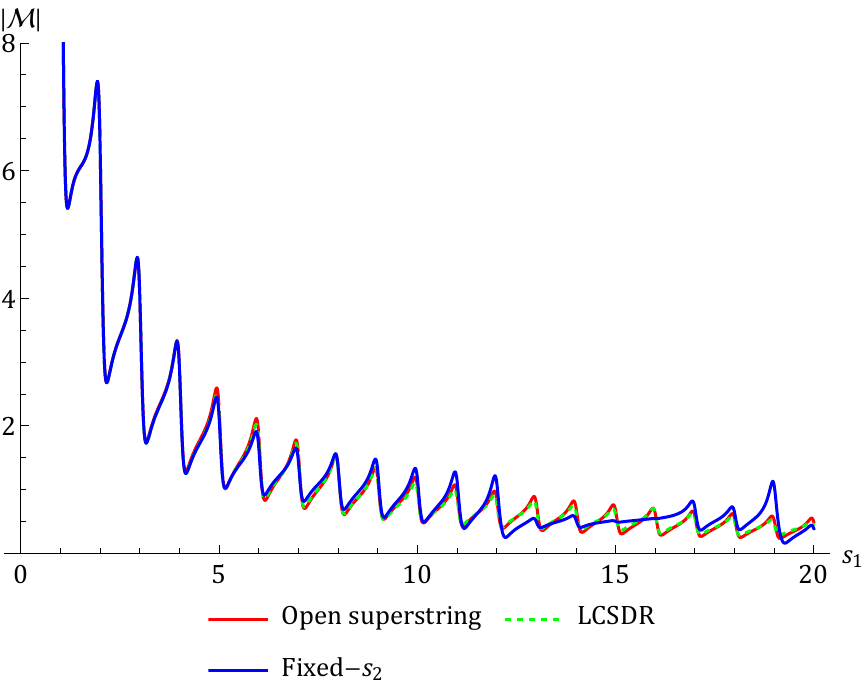} }}%
    \qquad
    \subfloat[\centering $s_2=1.1$]{{\includegraphics[width=8cm]{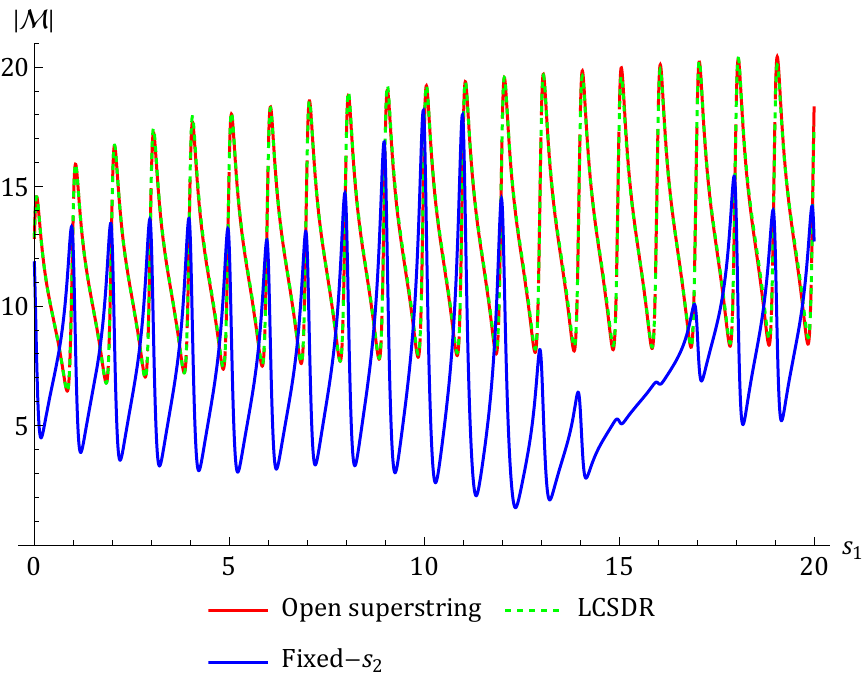} }}%
    \caption{Bootstrapping open superstring theory using the Fixed-$s_2$ dispersion relation. We plot $\big|\mathcal{M}\left(s_1+\frac{i}{10},s_2\right)\big|$ versus $s_1$ for different values of $s_2$.}%
   \label{fig:OpenStringFixedt}
\end{figure}

Here we will try to bootstrap the open superstring amplitude using the Fixed-$s_2$ dispersion relation and compare it with the results obtained from the LCSDR. For this we will use the ansatz obtained directly from the Fixed-$s_2$ dispersion relation for the amplitude \eqref{t-fix}:
\begin{equation}\label{t-fix ansatz}
    \mathcal{M}(s_1,s_2)=P\left(\frac{1}{s_1},\frac{1}{s_2}\right)+\sum_{n=1}^{\infty}\sum_{\ell}c_\ell^{(n)}\frac{1}{s_1-n}\mathcal{G}_\ell^{\left(\frac{D-3}{2}\right)}\left(1+\frac{2s_2}{n}\right).
\end{equation}
Here, we put the boundary term $\mathcal{B}(s_2)\rightarrow 0$ because we are focusing on the open superstring which satisfies the dual resonance model. We use $N_{\text{max}}=30$ in $D=10$ and impose the duality constraints:
\begin{equation}\label{t-fix constraints}
    \left|\sum_{n=1}^{\infty}\sum_{\ell}c_\ell^{(n)}\left\{\frac{1}{s_1-n}\mathcal{G}_\ell^{\left(\frac{D-3}{2}\right)}\left(1+\frac{2s_2}{n}\right)-\frac{1}{s_2-n}\mathcal{G}_\ell^{\left(\frac{D-3}{2}\right)}\left(1+\frac{2s_1}{n}\right)\right\}\right|\leq T\,.
\end{equation}
In the ($s_1,s_2$) plane, the first term in the \textit{lhs} of \eqref{t-fix constraints} is convergent for $s_2\leq 0$ \textit{i.e.} the 3rd and the 4th quadrants. Similarly, the second term in the \textit{lhs} of \eqref{t-fix constraints} is convergent for $s_1\leq 0$ \textit{i.e.} the 2nd and the 3rd quadrants. Therefore, \eqref{t-fix constraints} is valid only in the 3rd quadrant. For the bootstrap, we choose a grid of 40 random points in the 3rd quadrant for $|s_i|<20$. The tolerance we use is $T=10^{-9}$. We show our results in fig.\ref{fig:OpenStringFixedt}. We observe that the Fixed-$s_2$ dispersion relation has slower convergence compared to the LCSDR. Furthermore, we get \texttt{maxComplementarity} in SDPB when we use too many duality constraints, for example if we use a grid of 80 random points instead of 40.
\section{Why $\lambda$-constraints and not the null constraints}
Here we motivate the advantage of using the $\lambda$-constraints over the null constraints. We find that on the exact open superstring solution for the ($s_1,s_2$) grid in fig.(\ref{s1s2grid}) the variation of the null constraints has a much bigger range when compared to the variation of the $\lambda$-contraints. We demonstrate this in the table below. We find that the maximum value of the $\lambda$-constraint is larger than the maximum value of the null constraint in all four quadrants, as well as for the extra grid-points near $s_2=0$. In general, the difference between the maximum and the minimum values is also greater for the null constraints than for the $\lambda$-constraints. Owing to the bigger spread of values, it is expected that with a fixed tolerance, convergence will be poorer for the null constraints, and this is precisely what we find in our numerics.
\begin{center}
\begin{table}[H]
    \centering
    \begin{tabular}{|c|c|c|c|c|c|c|}
    \hline
   \multirow{2}{*}{Constraint region}& \multicolumn{2}{c|}{$\big|\frac{\partial\mathcal{M}(s_1,s_2)}{\partial\lambda}\big|$} & \multicolumn{2}{c|}{$|\text{Null constraint}|$}\\ \cline{2-5}
    & Min & Max & Min & Max\\
    \hline
    1st quadrant & $1.567\times 10^{-11}$ & $7.254\times 10^{-8}$ & $9.883\times 10^{-10}$ & 0.08831 \\
    \hline 
    2nd quadrant & $6.075\times 10^{-14}$ & $2.254\times10^{-9}$ & $1.219\times10^{-14}$ & $1.051\times 10^{-6}$ \\
    \hline 
    3rd quadrant & $6.341\times 10^{-14}$ & $1.576\times 10^{-12}$ & $1.517\times 10^{-14}$ & $2.211\times10^{-11}$\\
    \hline 
    4th quadrant & $1.357\times 10^{-14}$ & $2.011\times 10^{-8}$ & $1.209\times 10^{-16}$ & $2.026\times 10^{-6}$\\
    \hline
    $|s_1|<1, |s_2|<\frac{1}{100}$ & $8.858\times 10^{-13}$ & $5.713\times 10^{-12}$ & $1.046\times 10^{-12}$ & $3.576\times 10^{-10}$\\
    \hline
    \end{tabular}
   \caption{Table for comparison between the $\lambda$-constraints and the null constraints.}
\end{table}
\end{center}
\newpage
\section{Category-III models}
\begin{figure}[H]
    \centering
    \subfloat[\centering $s_2=-0.1$]{{\includegraphics[width=8cm]{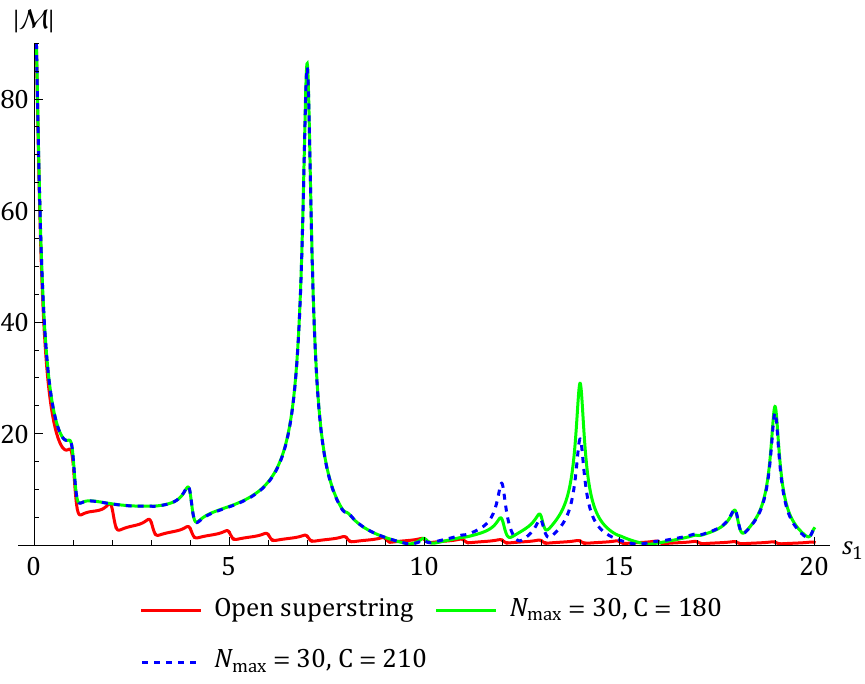} }}%
    \qquad
    \subfloat[\centering $s_2=1.1$]{{\includegraphics[width=8cm]{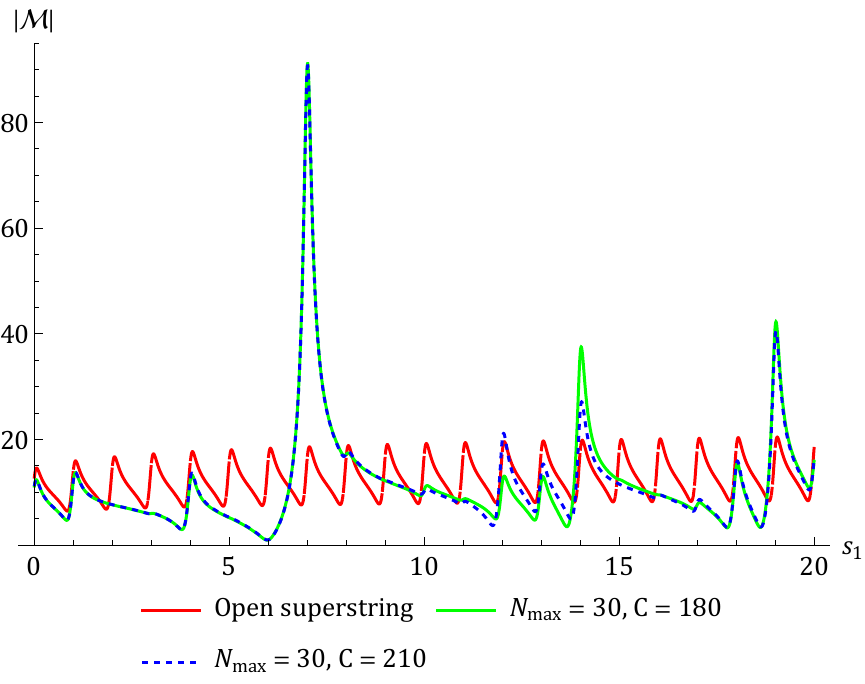} }}%
    \qquad
     \subfloat[\centering $s_2=-0.1$]{{\includegraphics[width=8cm]{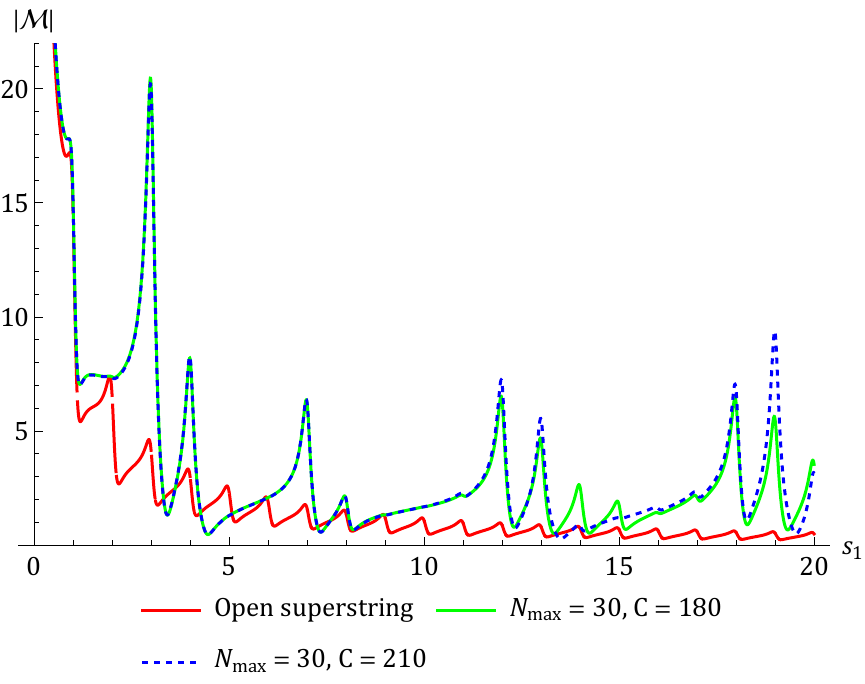} }}%
    \qquad
    \subfloat[\centering $s_2=1.1$]{{\includegraphics[width=8cm]{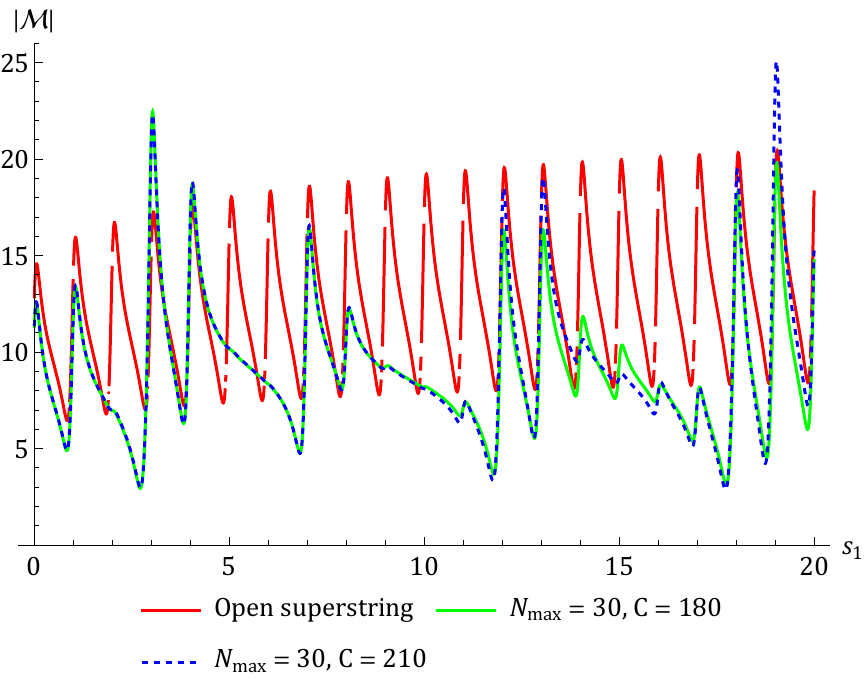} }}%
    \caption{In this figure, (a) and (b) represent results for minimization of $\langle\Delta\mathcal{E}/s^4\rangle$ whereas (c) and (d) represent results for minimization of $\langle\Delta\mathcal{E}/s^5\rangle$. We plot $\big|\mathcal{M}\left(s_1+\frac{i}{10},s_2\right)\big|$ versus $s_1$ for different values of $s_2$.}%
   \label{fig:EPbystp45OpenString}
\end{figure}

A natural question to ask is that what are the quantities that can be minimized using SDPB so that we reproduce string like behavior for the S-matrices. Here we show that the natural quantity for minimization to reproduce the tree-level open superstring amplitude is 
\begin{equation}
\left\langle\frac{\Delta\mathcal{E}}{s^2}\right\rangle=\int_1^\Lambda ds\, \frac{\Delta\mathcal{E}}{s^2}.
\end{equation}
Minimization of all other higher moments for example $\langle\Delta\mathcal{E}/s^4\rangle,\langle\Delta\mathcal{E}/s^5\rangle,\cdots$ does not reproduce the open superstring amplitude. The quantity $\langle\Delta\mathcal{E}/s^3\rangle$ cannot be minimized since it is a linear combination of $W_{10}$ and $W_{01}$ \eqref{W10W01}. Again, we use $N_{\text{max}}=30$ in $D=10$ with $\lambda=14.6$ and impose the $\lambda$-constraints in the same grid as that for the open superstring. In fig.\ref{fig:EPbystp45OpenString} (a) and (b) we show our results for the minimization of $\langle\Delta\mathcal{E}/s^4\rangle$ and in fig.\ref{fig:EPbystp45OpenString} (c) and (d) we show our results for the minimization of $\langle\Delta\mathcal{E}/s^5\rangle$. We clearly see that minimization of these higher moments of the entangling power does not yield the open superstring amplitude as a solution.

\section{Hypergeometric deformations}
\label{Hypergeometric deformations}
Deformations of the open string amplitude have been recently studied in \cite{Mansfield:2024wjc}. Parametric series representation for the hypergeometric deformed amplitudes is given below,
\begin{eqnarray}
	&&\frac{1}{s_{1} s_{2}}-\frac{\Gamma (1-s_{1}) \Gamma (1-s_{2}) \, }{(r+1) \Gamma (-s_{1}-s_{2}+2)}\; _3F_2\left(r+1,1-s_{1},1-s_{2};r+2,-s_{1}-s_{2}+2;1\right)\nonumber\\
	&=&\frac{1}{s_{1} s_{2}}-\sum _{n=1}^{\infty} \frac{(-1)^n \Gamma (r+1)}{\Gamma (n)}  \left(\frac{1}{s_{1}-n}+\frac{1}{s_{2}-n} +\frac{1}{\lambda
		+n}\right) \Gamma \left(1+\lambda -\frac{(s_{1}+\lambda )
		(s_{2}+\lambda )}{n+\lambda }\right) \nonumber\\
		&& _3\tilde{F}_2\left(1-n,r+1,\lambda -\frac{(s_{1}+\lambda )
		(s_{2}+\lambda )}{n+\lambda }+1;r+2,-n+\lambda -\frac{(s_{1}+\lambda ) (s_{2}+\lambda
		)}{n+\lambda }+2;1\right)\nonumber\\
\end{eqnarray}

\begin{figure}[H]
    \centering
    \subfloat[\centering $s_2=-0.1$]{{\includegraphics[width=8cm]{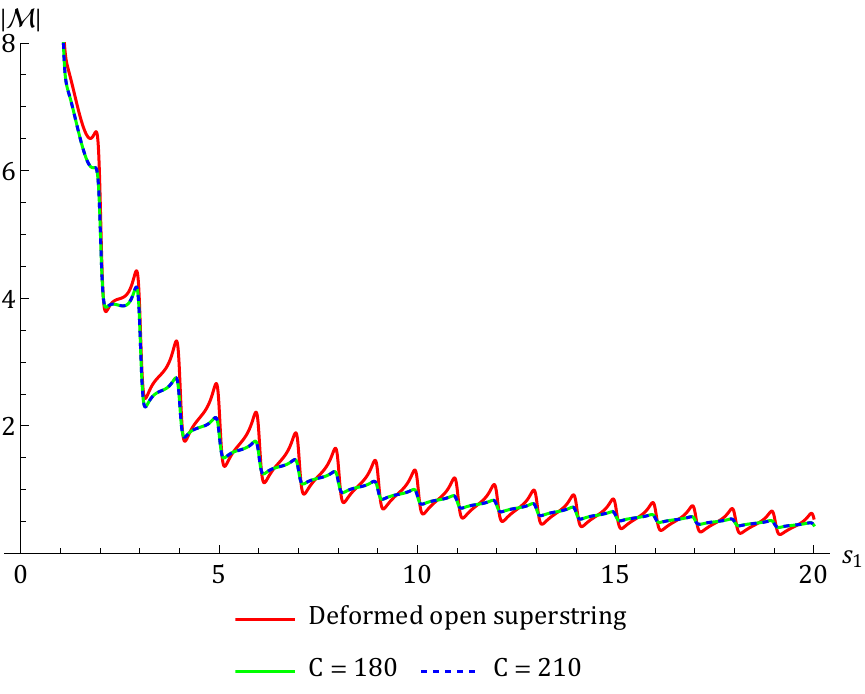} }}%
    \qquad
    \subfloat[\centering $s_2=1.1$]{{\includegraphics[width=8cm]{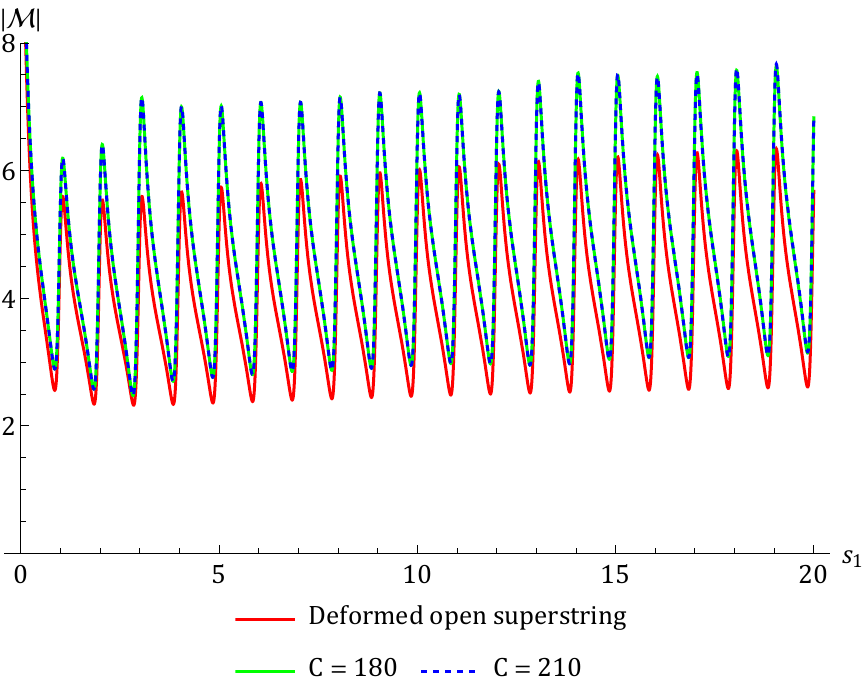} }}%
    \caption{Bootstrapping hypergeometric deformation of open superstring theory. We plot $\big|\mathcal{M}\left(s_1+\frac{i}{10},s_2\right)\big|$ versus $s_1$ for different values of $s_2$.}%
   \label{fig:DeformedOpenString}
\end{figure}

For $r\gg 1$, the amplitude approximates to $1/s_1 s_2$ (which implies $(W_{10},W_{01})=(0,0)$ as indicated by the origin in fig.\ref{fig:dualitycontourplotCat1}), since in this limit the summand goes like $e^{-r}$. As $r\rightarrow -1$, $W_{10}\rightarrow-\frac{1}{1+r}$ and $W_{01}\rightarrow\frac{\pi^2}{6(1+r)}$ at the leading order in their respective series expansions around $r=-1$. Therefore, $r\rightarrow-1$ implies $(W_{10},W_{01})\rightarrow(-\infty,\infty)$. This is also indicated in fig.\ref{fig:dualitycontourplotCat1}.
In fig.\ref{fig:DeformedOpenString} we plot the results for the bootstrap of the hypergeometric deformation of the open superstring amplitude for $r=2$ with $N_{\text{max}}=30$ in $D=10$. We use two different grid of constraints. The first grid is the same as that for the open superstring with 180 $\lambda$-constraints at each derivative order, and the second grid has 210 $\lambda$-constraints at each derivative order in a larger domain in the $(s_1,s_2)$ plane. We find excellent convergence with the number of constraints (C). The theories we find are not exactly the hypergeometric deformations but appear to be close cousins.
\newpage
\section{General integer-spaced spectrum with a higher-spin cutoff}
\label{General spectrum}
\begin{figure}[H]
    \centering
    \subfloat[\centering Open superstring: $s_2=-0.1$]{{\includegraphics[width=8cm]{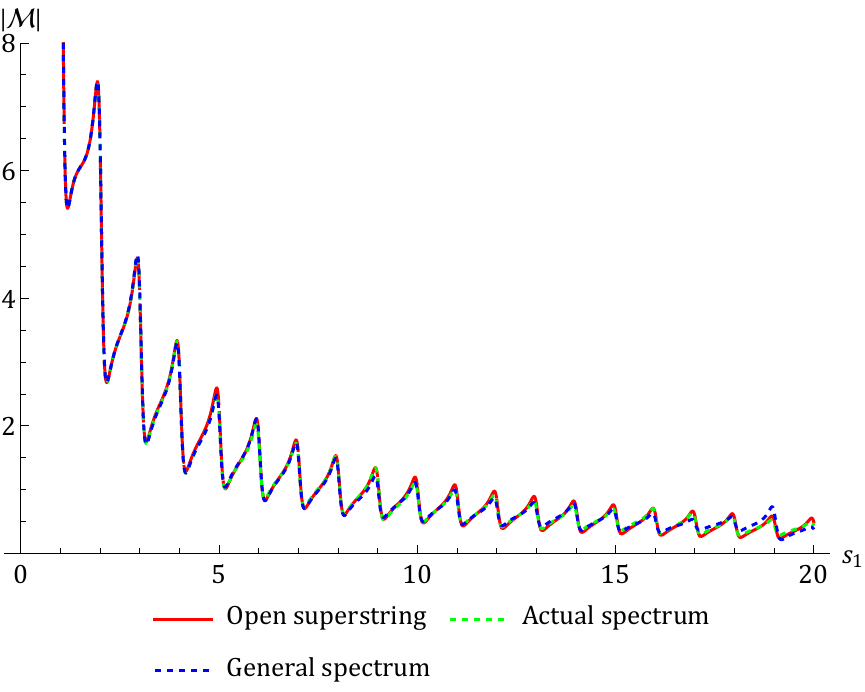} }}%
    \qquad
    \subfloat[\centering Open superstring: $s_2=1.1$]{{\includegraphics[width=8cm]{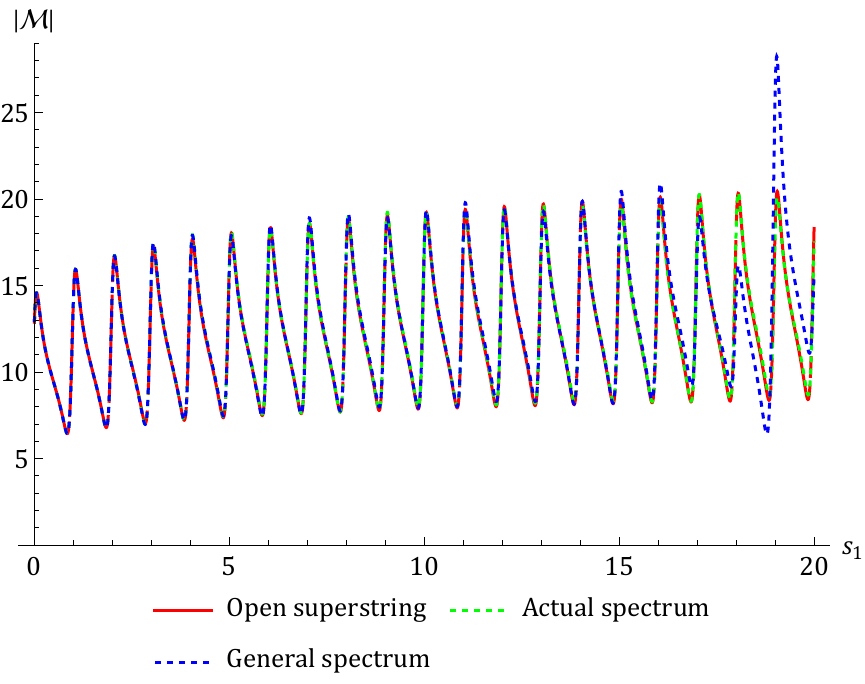} }}%
    \caption{Comparison between the amplitudes with the actual spectrum and the general spectrum for the open superstring in fig.(\ref{fig:dualitycontourplotCat1}).}
   \label{fig:GSPlotopenstring}
\end{figure}

\begin{figure}[H]
    \centering
    \subfloat[\centering Open superstring: Leading Regge trajectory]{{\includegraphics[width=8cm]{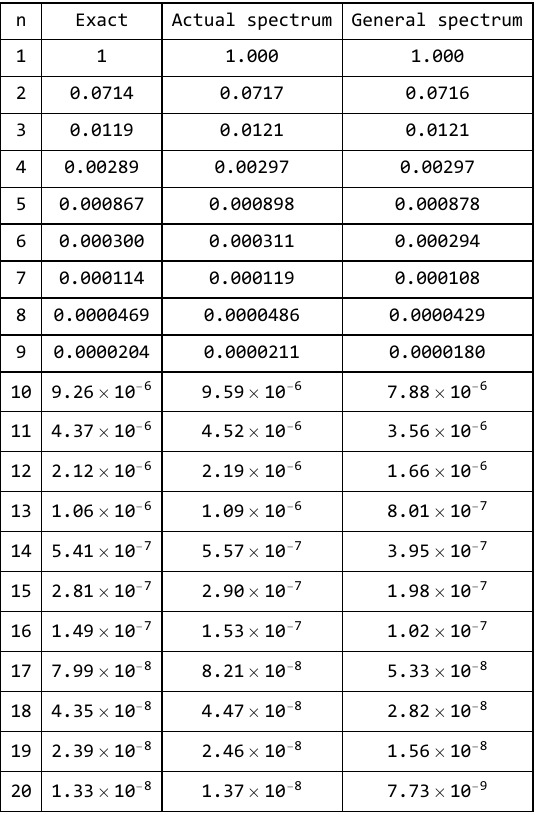} }}%
    \qquad
    \subfloat[\centering Open superstring: Subleading Regge trajectory]{{\includegraphics[width=8cm]{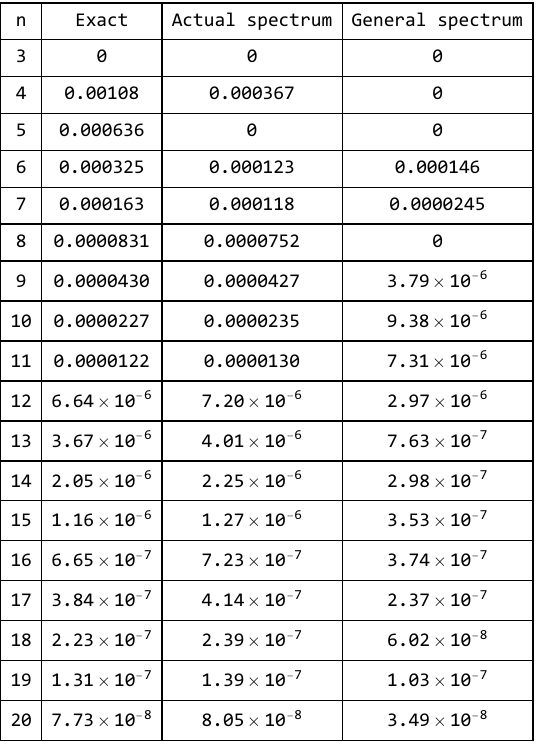} }}%
    \caption{Comparison of the $c_\ell^{(n)}$s on the leading and the subleading Regge trajectories between the amplitudes with the actual spectrum and the general spectrum for the open superstring in fig.(\ref{fig:dualitycontourplotCat1}).}
   \label{fig:GSleadingsubleadingopenstring}
\end{figure}

Here we will show that we can run our numerics using a general integer-spaced spectrum with a higher-spin cutoff and reproduce similar results for the open superstring amplitude in fig.(\ref{fig:dualitycontourplotCat1}) as well as for the black star amplitude in fig.(\ref{fig:dualitycontourplotCat2}). To demonstrate this we use the following general integer spaced spectrum with a higher spin cutoff: at mass level $n$ the spectrum contains $0\leq \ell \leq n-1$ in steps of $1$. Note that the higher-spin cutoff used here matches with that of the open superstring but we have removed the restriction that only even (odd) spins contribute when $n$ is odd (even). We use $N_{\text{max}}=30$ in $D=10$ and use the same grid of $\lambda$-constraints as the open superstring. The tolerance we use is $T=10^{-9}$. We show our results below. In fig.(\ref{fig:GSPlotopenstring}) we plot our results for the open superstring and find that both the actual spectrum and the general spectrum almost reproduce the open superstring (the general spectrum has slower convergence). In the two tables in fig.(\ref{fig:GSleadingsubleadingopenstring}) we compare the results for the leading and the subleading Regge trajectories for the open superstring  with the exact answer for both the actual spectrum and the general spectrum. We show that although the leading trajectory matches well with the exact answer for both the actual and the general spectrum, the subleading trajectory differs from the amplitude with the general spectrum. These discrepancies are compensated for by the new $c_\ell^{(n)}$s in the general spectrum. Some of the first few of these new $c_\ell^{(n)}$s are shown in table \ref{tab:newcelssopenstring}. It is important to emphasise that at the level of the plots, the amplitudes appear almost indistinguishable.

\begin{figure}[H]
    \centering
    \subfloat[\centering Black star: $s_2=-0.1$]{{\includegraphics[width=8cm]{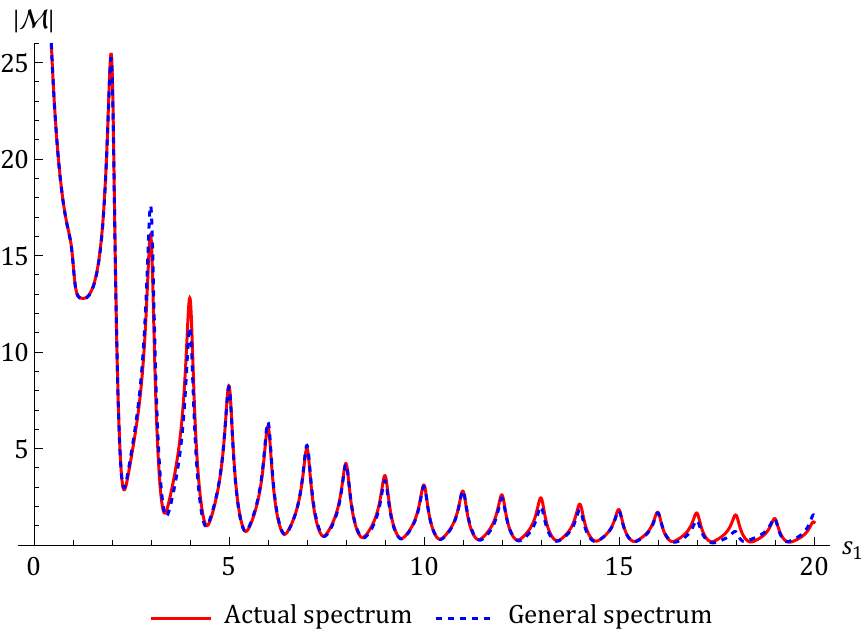} }}%
    \qquad
    \subfloat[\centering Black star: $s_2=1.1$]{{\includegraphics[width=8cm]{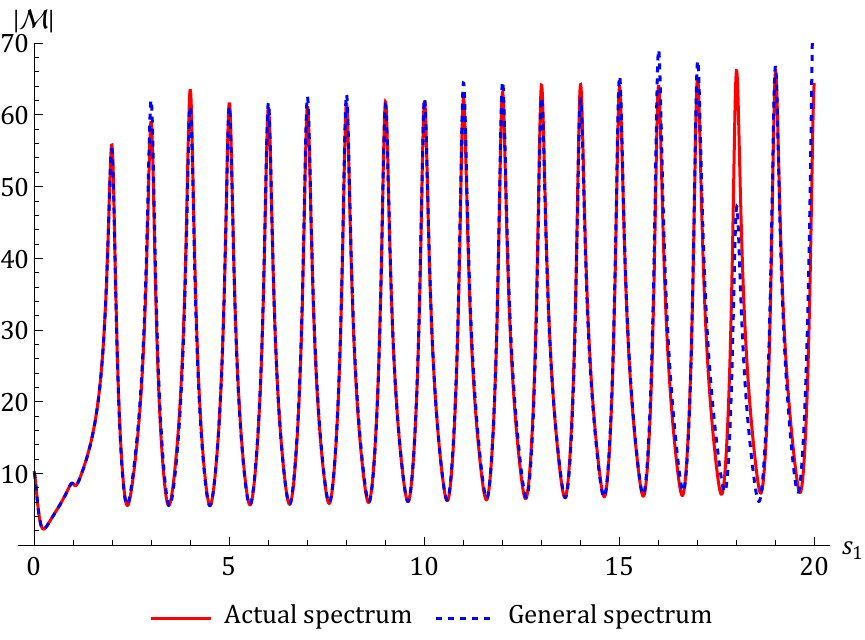} }}%
    \caption{Comparison between the amplitudes with the actual spectrum and the general spectrum for the black star in fig.(\ref{fig:dualitycontourplotCat2}).}
   \label{fig:GSPlotblackstar}
\end{figure}

\begin{figure}[H]
    \centering
    \subfloat[\centering Black star: Leading Regge trajectory]{{\includegraphics[width=7cm]{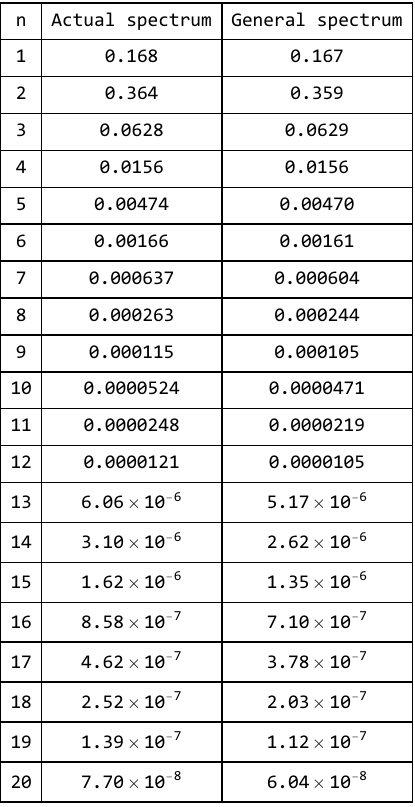} }}%
    \qquad
    \subfloat[\centering Black star: Subleading Regge trajectory]{{\includegraphics[width=7cm]{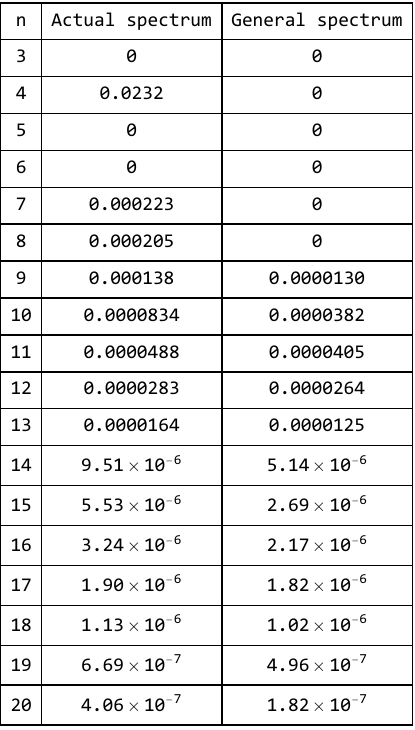} }}%
    \caption{Comparison of the $c_\ell^{(n)}$s on the leading and the subleading Regge trajectories between the amplitudes with the actual spectrum and the general spectrum for the black star in fig.(\ref{fig:dualitycontourplotCat2}).}
   \label{fig:GSleadingsubleadingblackstar}
\end{figure}

Next, we focus on the black star amplitude in fig.(\ref{fig:dualitycontourplotCat2}) and perform similar checks both with the actual spectrum and the general spectrum. The results for the black star are shown in fig.(\ref{fig:GSPlotblackstar}). Again we find very good match between between the amplitudes with the actual spectrum and the general spectrum. In fig.(\ref{fig:GSleadingsubleadingblackstar}) we compare the $c_\ell^{(n)}$s on the leading and the subleading Regge trajectories. We find a nice match between the two amplitudes for the leading Regge trajectory whereas the subleading Regge trajectory shows discrepancies between the two amplitudes. These discrepancies are being compensated by the new $c_\ell^{(n)}$s. The first few of them for the black star amplitude are given in table \ref{tab:newcellsblackstar}.
\begin{table}[hbt!]
    \centering
    \begin{tabular}{|c|c|c|c|c|c|c|} 
        \hline
        $c_{1}^{(3)}=0.000559$ & $c_{4}^{(6)}=0.0000893$ & $c_{5}^{(7)}=0.0000743 $ & $c_{6}^{(8)}=0.0000422$ & $c_{7}^{(9)}=0.0000202$ \\ 
        \hline
        $c_{8}^{(10)}=9.9\times10^{-6}$ & $c_{9}^{(11)}=5.4\times10^{-6}$ & $c_{10}^{(12)}=3.05\times10^{-6}$ & $c_{11}^{(13)}=1.65\times10^{-6}$ & $c_{12}^{(14)}=8.48\times10^{-7}$ \\ 
        \hline
    \end{tabular}
    \caption{The first few new $c_\ell^{(n)}$s for the open superstring amplitude using the general spectrum.}
    \label{tab:newcelssopenstring}
\end{table}
\begin{table}[hbt!]
    \centering
    \begin{tabular}{|c|c|c|c|c|c|c|} 
        \hline
        $c_{1}^{(3)}=0.0243$ & $c_{4}^{(6)}=0.000225$ & $c_{5}^{(7)}=0.000216 $& $c_{6}^{(8)}=0.000138$ & $c_{7}^{(9)}=0.0000686$ \\ 
        \hline
        $c_{8}^{(10)}=0.0000303$& $c_{9}^{(11)}=0.0000141$ & $c_{10}^{(12)}=7.52\times10^{-6}$ & $c_{11}^{(13)}=4.28\times10^{-6}$ &$c_{12}^{(14)}=2.29\times10^{-6}$\\        
        \hline
    \end{tabular}
    \caption{The first few new $c_\ell^{(n)}$s for the black star amplitude using the general spectrum.}
    \label{tab:newcellsblackstar}
\end{table}

\bibliographystyle{utphys}
\bibliography{Dispersion}

\end{document}